\documentclass[
 reprint,
 amsmath,amssymb,
 aps,
pra,
]{revtex4-2}
\usepackage{graphicx}
\usepackage{float}
\usepackage{dcolumn}
\usepackage{bm}
\usepackage{bbm}
\usepackage{braket}
\usepackage{xcolor}
\usepackage[T1]{fontenc}
\usepackage{enumerate,enumitem}
\usepackage{booktabs}
\usepackage{physics}
\usepackage{amsmath}
\usepackage{appendix}
\usepackage{hyperref}

\begin{document}

\title{Spin Kerr-cat qubits}

\author{Z.~M.~McIntyre$^{1}$}
\email{zoe.mcintyre@unibas.ch}
\author{Daniel Loss$^{1,2,3,4}$}
\affiliation{${}^{1}$Department of Physics, University of Basel, Klingelbergstrasse 82, 4056 Basel, Switzerland}
\affiliation{${}^{2}$Physics Department, King Fahd University of Petroleum and Minerals, 31261, Dhahran, Saudi Arabia}
\affiliation{${}^{3}$Quantum Center, KFUPM, Dhahran, Saudi Arabia}
\affiliation{${}^{4}$RDIA Chair in Quantum Computing}

\date{\today}

\begin{abstract}
    The use of noise-robust qubit encodings provides a way of extending the lifetime of quantum information at the hardware level. In this work, we introduce the spin Kerr-cat encoding, which leverages a sweet spot near a clock transition in the spectrum of quadrupolar nuclei (having spin $I\geq 1$) to achieve a suppression of charge and magnetic noise leading to qubit dephasing. The basis states of the spin Kerr-cat qubit are given by the two lowest levels of a $\mathbb{Z}_2$-symmetric nuclear-spin Hamiltonian and are well approximated by spin cat states for sufficiently high asymmetry parameters. For nuclei in highly asymmetric electric-field gradients, the nuclear spin could therefore be initialized in the spin-Kerr-cat subspace using SU(2) rotations and measurement rather than full SU($2I+1$) control. We calculate the dephasing time of the spin Kerr-cat qubit under a model of $1/f$ noise, as well as relaxation of the qubit due to breaking of the $\mathbb{Z}_2$ symmetry by charge-noise-induced fluctuations of the quadrupolar tensor. Using measured parameters for antimony (${}^{123}\mathrm{Sb}$) donors in silicon, we estimate that second-scale $T_2^*$ times could be achieved with this encoding, with bitflips and leakage occurring on significantly longer timescales. We propose a two-qubit gate acting on spin Kerr-cat qubits and electron-spin qubits, enabled by adiabatically moving an electron on and off the donor potential. Based on numerical simulations, we estimate that with an enhancement of measured quadrupolar couplings by a factor of ${\approx}  2$, a gate fidelity exceeding $99\%$ could be achieved for spin Kerr cats encoded in ${}^{123}\mathrm{Sb}$, with a gate time under $100$ ns. By entangling the nuclear spin with an electron spin in this way, the spin Kerr-cat qubit could then be measured using a Pauli spin blockade readout of the electron. 
\end{abstract}

\maketitle

\section{Introduction}

Protecting quantum information from decoherence is a central challenge for quantum computing. Among solid-state qubits, nuclear spins in silicon stand out for their exceptionally long coherence times~\cite{steger2012quantum,saeedi2013room}. The use of nuclear spins for quantum computing dates back to Kane's proposal for encoding information in the nuclear spins of phosphorous donors (${}^{31}\mathrm{P}$) in silicon~\cite{kane1998silicon}, following an earlier proposal by Loss and DiVincenzo for using electron spins in quantum dots to store and manipulate quantum information~\cite{loss1998quantum}. 
While most work on nuclear spins in silicon has focused on phosphorous donors~\cite{pla2013high,fricke2021coherent,mkadzik2022precision,reiner2024high}, which have spin $I=1/2$, there has also been significant experimental progress surrounding the use of high-spin donors ($I\geq 1$) for quantum information processing~\cite{asaad2020coherent,fernandez2024navigating,yu2025schrodinger}. Relative to ${}^{31}\mathrm{P}$, the larger Hilbert space of spins with $I\geq 1$ provides new opportunities for fundamental studies of nonclassicality~\cite{mourik2018exploring,vaartjes2025certifying}, as well as new prospects for encoding quantum information~\cite{morello2020donor,gross2021designing} and running quantum algorithms~\cite{leuenberger2003grover}.

Nuclear spins exhibit a strong noise bias, with lifetimes ($T_1$) typically far exceeding coherence times ($T_2^*$). For single ionized phosphorous donors (${}^{31}\mathrm{P}^+$) in natural silicon, nuclear-spin dephasing times of $T_2^*=3.3$ ms have been measured, and lifetimes on the order of $10{-}100$ s have been attributed to repeated ionization during readout, with the true $T_1$ time of the ionized donor likely far longer~\cite{pla2013high}. The use of isotopically purified silicon leads to considerable improvements in the $T_2^*$ time of ${}^{31}\mathrm{P}^+$ nuclear spins, with $T_2^*=0.6$ s having been measured in Ref.~\cite{muhonen2014storing}. Measurements of single antimony-123 (${}^{123}\mathrm{Sb}$) nuclear spins in isotopically purified silicon, having spin length $I=7/2$, found $T_2^*$ times on the order of tens of milliseconds and immeasurably long $T_1$ times believed to exceed tens of seconds~\cite{yu2025schrodinger}.

In this work, we present a nuclear-spin qubit encoding designed to enhance the dephasing time $T_2^*$ of quantum information stored in nuclear spins. This enhancement comes from tuning the qubit splitting to a charge-noise sweet spot in the nuclear-spin spectrum, where the first derivative of the qubit splitting with respect to variations in the quadrupole coupling vanishes. The sweet spot is located in the vicinity of a clock transition, characterized by a vanishing first derivative of the transition frequency with respect to magnetic-field fluctuations. Clock transitions originate in the field of atomic physics, where hyperfine transitions are used as time and frequency standards and must consequently remain stable against such fluctuations. Nuclear-spin clock transitions in rare-earth dopants have been studied for use in optical quantum memories~\cite{longdell2006characterization,mcauslan2012reducing}, and clock transitions in neutral bismuth donors have been proposed as the basis of donor-bound electron-spin qubits with reduced sensitivity to environmental ${}^{29}\mathrm{Si}$ spins~\cite{wolfowicz2013atomic}. Flip-flop qubits~\cite{tosi2017silicon}, encoded in the electron-nuclear states of a single donor, also leverage a (second-order) clock transition to achieve a reduced sensitivity to electrical noise, with $T_2^*=4$ $\mu$s having been measured experimentally~\cite{savytskyy2023electrically}.

The qubit introduced in this work is defined by the two lowest eigenstates of the nuclear-spin Hamiltonian, given to a good approximation by spin cat states. The definite $\hat{I}_z$ parity of these qubit basis states is ensured by an overall $\mathbb{Z}_2$ symmetry of the nuclear-spin Hamiltonian, realized through proper alignment of the applied magnetic field in relation to the electric-field-gradient (EFG) tensor at the site of the nuclear spin. In this respect, the \textit{spin Kerr-cat qubit} presented here is the spin analog of the Kerr-cat qubits used in circuit QED~\cite{puri2017engineering,grimm2020stabilization}, which are encoded in the lowest eigenstates of squeezed Kerr-nonlinear oscillators, given by cat states of fixed photon-number parity.  A  distinction worth noting, however, is that the Kerr-cat encoding does not suppress errors due to the dominant error source affecting superconducting resonators (photon loss), which leads to bitflips rather than phase flips. By contrast, the spin Kerr-cat encoding is designed to suppress the dominant error source affecting spin qubits.

Spin cat states have been considered as the logical basis states of quantum error correcting codes capable of detecting errors due to dephasing~\cite{gross2024hardware,omanakuttan2024fault}. These spin cat states can be generated (as non-eigenstates) in nuclear spins with quadrupolar nonlinearities~\cite{gross2024hardware, gupta2024robust}, as has been demonstrated experimentally with ${}^{123}\mathrm{Sb}$ donors in silicon and measured spin-cat coherence times of tens of milliseconds~\cite{yu2025schrodinger}. However, whereas quantum error correction seeks to correct errors after they have occurred, an alternate and ultimately complementary approach, considered here, is to passively suppress errors at the hardware level through the choice of qubit encoding. As will be seen, this strategy on its own could enhance the decoherence time of encoded quantum information by up to two orders of magnitude with no active quantum error correction or dynamical decoupling.

\enlargethispage{\baselineskip}

The layout of this article is as follows: In Sec.~\ref{sec:quadrupole}, we provide an overview of quadrupolar interactions between high-spin nuclei and electric-field gradients, and we explain how the applied magnetic field can be oriented to ensure the required $\mathbb{Z}_2$ symmetry of the nuclear-spin Hamiltonian. In Sec.~\ref{sec:spin-kerr-cat}, we introduce the spin Kerr-cat qubit before quantifying in Sec.~\ref{sec:decoherence} the enhancement in dephasing time obtained by operating at lower field strengths, tuned to a sweet spot in the vicinity of a clock transition in the nuclear-spin spectrum. We also analyze a relaxation mechanism due to charge noise, resulting from coupling between the nuclear spin and electric-field gradients. In Sec.~\ref{sec:operation}, we explain how single-qubit gates, initialization, and readout could be performed, as well as an electron-spin-mediated two-qubit gate compatible with spin shuttling. The paper concludes with an outlook in Sec.~\ref{sec:conclusion}.

\section{Nonlinearity due to quadrupole interactions}\label{sec:quadrupole}

Nuclei with spins $I\geq 1$ have an electric quadrupole moment arising from their non-spherical charge distribution. This quadrupole moment interacts with electric-field gradients (EFGs) at the site of the nucleus~\cite{cohen1957quadrupole}. For donors in silicon, establishing a large EFG at the position of the nucleus requires breaking the tetrahedral symmetry of the bulk silicon crystal, as can be achieved through lattice strain~\cite{franke2015interaction,pla2018strain}. In the device used in Ref.~\cite{asaad2020coherent}, the strain was attributed to the different thermal contraction of silicon and aluminum upon cooling to cryogenic temperatures.   

In the presence of a spatially varying electric potential $V(\bm{r})$, the nonzero quadrupole moment of the nuclear spin leads to a quadrupole interaction that can be written in terms of the second-rank quadrupole tensor $\bm{Q}$ as
\begin{equation}\label{quadrupole-cartesian}
    H_{\mathrm{q}}=\hat{\bm{I}}'^\top\bm{Q}\hat{\bm{I}}'=\sum_{\alpha,\beta\in \{x',y',z'\}} Q_{\alpha\beta}\hat{I}_\alpha\hat{I}_\beta,
\end{equation}
where $\{x',y',z'\}$ are Cartesian axes defined in some lab frame, $\hat{\bm{I}}'=(\hat{I}_{x'},\hat{I}_{y'},\hat{I}_{z'})$ is a vector of spin-$I$ nuclear-spin operators, and where $Q_{\alpha\beta}$ is given by
\begin{equation}
    Q_{\alpha\beta}=\frac{eq\mathcal{V_{\alpha\beta}}}{2I(2I-1)}.
\end{equation}
Here, $e$ is the elementary charge, $q$ is the electric quadrupole moment of the nucleus (quantifying the deviation of its charge distribution from spherical symmetry), and $\mathcal{V}_{\alpha\beta}=\partial^2 V/\partial\alpha\partial\beta$ is an element of the EFG tensor $\bm{\mathcal{V}}$. Coupling between the EFG and the quadrupole moment of the nucleus enables electrical control of the nuclear spin via nuclear electric resonance~\cite{bloembergen1961linear}, providing an important mechanism for coherently manipulating nuclear spins in quantum devices using radio-frequency (RF) electric fields rather than nuclear magnetic resonance (NMR). Nuclear electric resonance has been demonstrated for an ensemble of nuclear spins in GaAs~\cite{ono2013coherent}. It has also been demonstrated for a single ${}^{123}\mathrm{Sb}$ donor in silicon by modulating nearby gate voltages \cite{asaad2020coherent}.

\begin{figure}
    \centering
    \includegraphics[width=\linewidth]{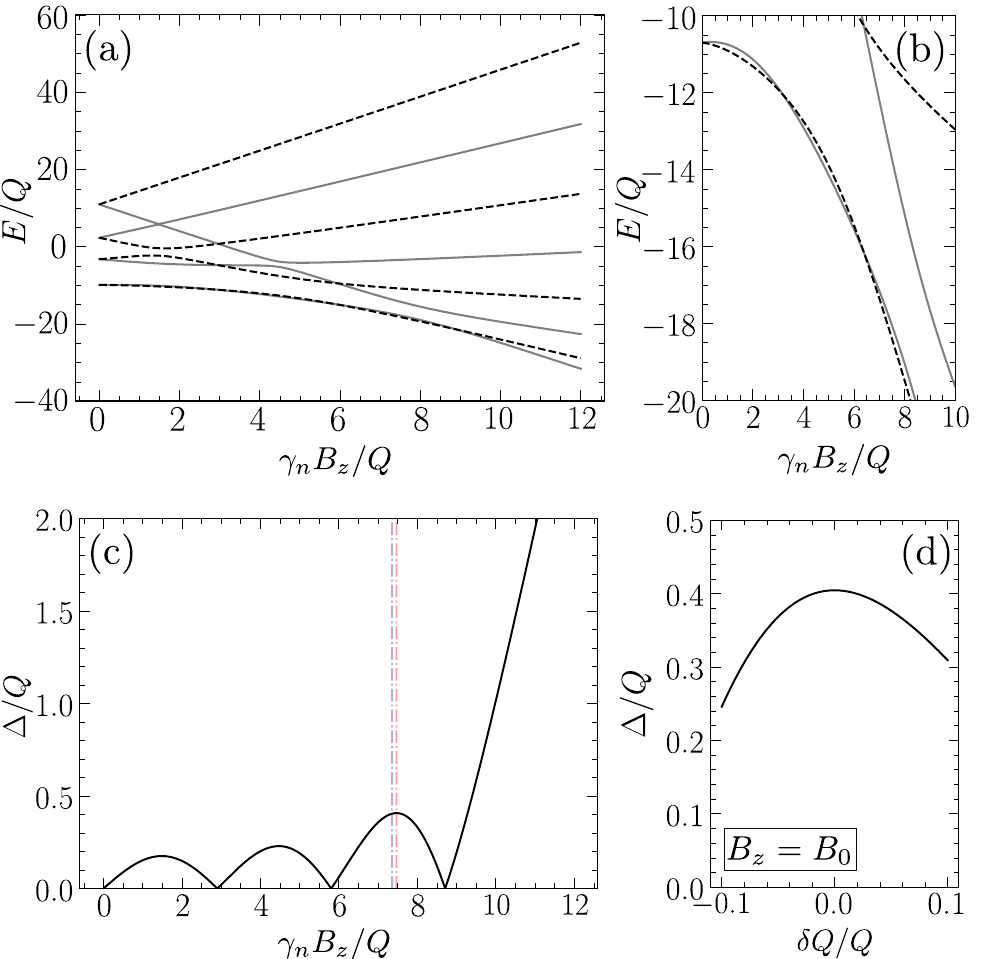}
    \caption{(a) Spectrum of $H$ for $I=7/2$ and $\eta=0.75$ as a function of the applied magnetic field $B_z$. The energies of even- and odd-parity states are plotted with dotted black and solid grey lines, respectively. Avoided crossings between same-parity states are a consequence of the two-spin-flip terms in the quadrupole Hamiltonian, which cannot couple states of different parities. (b) Zoomed-in version of (a), focusing on crossings in the lowest $\hat{I}_z$-parity doublet. (c) Energy difference $\Delta$ between the two states belonging to the lowest $\hat{I}_z$-parity doublet. A clock transition, where $\Delta$ is first-order insensitive to magnetic noise, is indicated by the dot-dashed red line. The blue dot-dashed line indicates a charge-noise sweet spot where $\Delta$ is first-order insensitive to variations $\delta Q$ in the quadrupole coupling. (d) The splitting $\Delta$ at the sweet-spot magnetic field $B_0$, plotted as a function of $\delta Q$. }
    \label{fig:spectrum}
\end{figure}

The EFG tensor is real and symmetric and can therefore be diagonalized by an orthogonal transformation. In the resulting principal axis system (PAS) $\{x,y,z\}$, the tensor is diagonal. Denoting this diagonal tensor $\bm{D_0}$, we have $\bm{R_0}^\top\bm{\mathcal{V}}\bm{R_0}=\bm{D_0}$, where $\bm{R_0}$ is an orthogonal matrix whose columns are the eigenvectors of $\bm{\mathcal{V}}$, and where we choose by convention a basis ordering in which the diagonal entries of $\bm{D_0}$ satisfy $\vert D_{xx}\vert\leq \vert D_{yy}\vert\leq \vert D_{zz}\vert$. In this PAS, the quadrupole interaction is given by~\cite{slichter2013principles}
\begin{equation}\label{quadrupole-hamiltonian}
    H_{\mathrm{q}}=\frac{Q}{2}\left[3\hat{I}_z^2+\frac{\eta}{2}\left(\hat{I}_+^2+\hat{I}_-^2\right)-\hat{\bm{I}}^2\right],
\end{equation}
where $\bm{\hat{I}}=\bm{R_0}^\top\hat{\bm{I}}'=(\hat{I}_x,\hat{I}_y,\hat{I}_z)$ is the transformed vector of nuclear-spin operators, $\hat{\bm{I}}^2=\hat{\bm{I}}\cdot\hat{\bm{I}}$, $Q= eq D_{zz}/[2I(2I-1)]$, and where we have introduced an asymmetry parameter $\eta\in [0,1]$ given by
\begin{equation}
    \eta=\frac{D_{xx}-D_{yy}}{D_{zz}}.
\end{equation}
The restriction of $\eta$ to the interval $[0,1]$ follows from the choice of basis ordering together with the fact that $\mathrm{Tr}\:\bm{D_0}=0$. The fact that $\bm{\mathcal{V}}$ is traceless and symmetric is itself a consequence of $V(\bm{r})$ satisfying the Laplace equation $\nabla^2 V=0$.

The principal axes $x$, $y$, and $z$ of the quadrupole tensor, as well as the parameters $Q$ and $\eta$ characterizing the quadrupole interaction in the PAS, can all be found through spectroscopic measurements while sweeping the direction of the applied field $B$~\cite{mourik2018exploring}. For a magnetic field $B_z$ applied along the $z$-axis, the full Hamiltonian $H$ of the nuclear spin is given by
\begin{equation}\label{hamiltonian}
    H=-\gamma_n B_z \hat{I}_{z}+H_{\mathrm{q}},
\end{equation}
where $\gamma_n$ is the gyromagnetic ratio of the nuclear spin. Notably, the Hamiltonian $H$ [Eq.~\eqref{hamiltonian}] has a $\mathbbm{Z}_2$ symmetry generated by $\hat{\Pi}=e^{-i\pi \hat{I}_{z}}$:
\begin{equation}\label{Z2-symmetry}
    e^{i\phi\hat{I}_z}He^{-i\phi\hat{I}_z}=H,\quad\phi=0,\pi.
\end{equation}
This $\mathbbm{Z}_2$ symmetry is absent for arbitrary magnetic-field orientations and immediately implies that the eigenstates of $H$ will be $\hat{I}_z$-parity eigenstates, i.e., simultaneous eigenstates of $\hat{\Pi}$. The reduction of the Hamiltonian symmetry from $U(1)$ to $\mathbbm{Z}_2$, where the former would correspond to having Eq.~\eqref{Z2-symmetry} hold for any $\phi \in [0,2\pi)=U(1)$, is a physical consequence of the nuclear spin being located at a point in the EFG without axial symmetry ($\eta\neq 0$). 

For $\eta=0$, the eigenstates of $H$ [Eq.~\eqref{hamiltonian}] are given simply by the eigenstates $\ket{I,m}$ of $\hat{I}_z$, independent of the strength of the magnetic field. For nonzero $\eta$, $H_{\mathrm{q}}$ generally leads to hybridization between $\hat{I}_z$ eigenstates differing by pairs of spin flips. The effects of such hybridization are, however, small in the high-field regime $\gamma_n B_z\gg Q$, where terms in $H_\mathrm{q}$ that do not commute with $\hat{I}_z$ can be neglected in a secular approximation. In this regime, the eigenstates of the full Hamiltonian are once again given by eigenstates of $\hat{I}_z$ to a good approximation, and the energy differences between eigenstates all disperse linearly as a function of the applied field. Since two-spin-flip processes are energetically suppressed for large magnetic-field strengths, the main consequence of the quadrupole coupling in this regime is the introduction of anharmonicity $\propto Q$ into the otherwise evenly spaced spectrum; this anharmonicity can then be used to address individual nuclear-spin transitions~\cite{asaad2020coherent,fernandez2024navigating}. The term $\propto Q\hat{I}_z^2$ that survives the secular approximation has the same form as the one-axis twisting interactions responsible for spin squeezing in atomic physics, which can be used to create nonclassical states for quantum-enhanced Ramsey interferometry with atomic ensembles~\cite{pezze2018quantum}. 
Free evolution under one-axis twisting  interactions can also be used to generate non-stationary nuclear-spin cat states in the high-field regime $\gamma_nB_z\gg Q$~\cite{gupta2024robust,yu2025schrodinger}.

For lower field strengths and nonzero $\eta$, the effects of the two-spin-flip terms in the quadrupole Hamiltonian $H_{\mathrm{q}}$ cannot be neglected. In this regime, the eigenstates of $H$ are not simply $\hat{I}_z$ eigenstates, and eigenenergies disperse nonmonotonically as a function of the applied magnetic field [Fig.~\ref{fig:spectrum}(a)]. For a spin-$I$ nuclear spin with half-integer $I$, the spectrum for $B_z=0$ consists of $I+1/2$ degenerate $\hat{I}_z$-parity doublets, i.e.~pairs of states with opposite $\hat{\Pi}$ eigenvalues. These degeneracies are a manifestation of Kramers' theorem: For $B_z=0$, the Hamiltonian $H=H_{\mathrm{q}}$ is time-reversal invariant, giving $[H,\hat{T}]=0$, where here, $\hat{T}$ is the time-reversal operator. For half-integer spin, the property $\hat{T}^2=-1$ of the time-reversal operator then implies that for every eigenstate $\ket{\psi}$ of $H$ with eigenenergy $E$, the time-reversed state $\hat{T}\ket{\psi}$ is a distinct eigenstate of $H$ having the same energy $E$. Since $\ket{\psi}$ is a simultaneous parity eigenstate ($\hat{\Pi}\ket{\psi}=\pm i\ket{\psi}$), the fact that $[\hat{T},\hat{\Pi}]=0$ finally implies that $\hat{T}\ket{\psi}$ must have the opposite $I_z$ parity.

For $B_z\neq 0$, time-reversal symmetry is broken and Kramers' theorem no longer applies, lifting the degeneracies observed at $B_z=0$. The dispersion of eigenenergies as the field strength is increased will exhibit avoided crossings between same-parity states, while states of opposite parity may cross. In particular, there are $I{-}1/2$ field strengths where the lowest $\hat{I}_z$-parity doublet is degenerate [Fig.~\ref{fig:spectrum}(c)] prior to the onset of the high-field regime discussed above where all eigenenergies disperse linearly with $B_z$. Similar degeneracies between opposite-parity states have been shown to be a consequence of interference between Berry phases~\cite{leuenberger2001spin,leuenberger2002quantum}. 

As we now show, the lowest $\hat{I}_z$-parity doublet can be used to define a qubit with a dephasing time that significantly exceeds the dephasing times obtained in the regime $\gamma_n B_z\gg Q$. This enhancement in the dephasing time is a result of operating at a charge-noise sweet spot in the vicinity of a clock transition [Fig.~\ref{fig:spectrum}(c)-(d)], where the splitting $\Delta$ between the qubit basis states is first-order insensitive to variations in the quadrupole coupling and more robust against fluctuations in the applied magnetic field. We focus in this work on the spin lengths $I=7/2$ and $I=9/2$, both of which are experimentally relevant to silicon-based quantum computing architectures. Antimony-123 (${}^{123}\mathrm{Sb}$) nuclear spins have length $I=7/2$, a gyromagnetic ratio of $\gamma_n=5.55$ MHz/T, and have been used to demonstrate significant advances in the quantum control of high-spin donors in silicon~\cite{asaad2020coherent,fernandez2024navigating,yu2025schrodinger}. Bismuth-209 (${}^{209}\mathrm{Bi}$) nuclear spins, having a gyromagnetic ratio of 6.96 MHz/T, have a spin length of $I=9/2$ and are another possible high-spin Group-V donor spin~\cite{morley2010initialization,wolfowicz2013atomic}.

\section{Spin Kerr-cat qubits}\label{sec:spin-kerr-cat}

Based on symmetry arguments, the eigenstates of $H$ must consist of linear combinations of $\hat{I}_z$ eigenstates having a common parity ($\hat{\Pi}$) eigenvalue. In this section, we show that the two lowest eigenstates of $H$, which we take to define a qubit, are given to a good approximation by spin cat states. These spin cat states are parity eigenstates by construction and are given by the symmetric and anti-symmetric linear combinations of two spin coherent states that are maximally distinguishable in their azimuthal coordinate, with an equal projection along the $\hat{I}_z$ axis. Spin coherent states can be produced through rotations $e^{-i \alpha \hat{\bm{n}}\cdot \hat{\bm{I}}}$ of the nuclear spin, which form a representation of the special unitary group $\mathrm{SU}(2)$ on the spin's Hilbert space. Given the ability to prepare the fully polarized reference spin state $\ket{I,I}$, this fact will ultimately allow the nuclear spin to be initialized in the spin Kerr-cat subspace through operations typically considered natural or ``straightforward'' for large single spins~\cite{gross2021designing}.  Gates, initialization, and readout of spin Kerr-cat qubits will be discussed in detail in Sec.~\ref{sec:operation}.

Single-qubit pure states can be associated with points $(\theta,\phi)$ on the Bloch sphere, obtainable by rotating some reference state, e.g.~$\ket{0}$, first by angle $\theta$ about the $y$ axis, then by angle $\phi$ about the $z$ axis. Notably, all single-qubit pure states can be reached via such operations, which effectively treat the qubit state like a classical vector being rotated in real space. Such states also saturate the uncertainty relation $\Delta S_x\Delta S_y\geq \vert \langle S_z\rangle\vert/2$, making them the most ``classical'' quantum states of a two-level system. For a spin with more than two levels ($I\geq 1$), it is no longer true that all pure states can be reached via two-axis control. Spin coherent states~\cite{radcliffe1971some,arecchi1972atomic} form a subset of the set of pure states and are the higher-dimensional generalization of pure single-qubit states, constituting the most ``classical'' states of larger spins. Like single-qubit pure states, spin coherent states saturate the uncertainty relation and can be parameterized through a pair of rotation angles according to
\begin{equation}\label{spin-coherent-state}
    \ket{\Theta,\phi}=e^{-i\phi\hat
    I_z}e^{-i\Theta\hat{I}_y}\ket{I,I},
\end{equation}
where $\phi\in[0,2\pi)$, $\Theta\in[0,\pi]$, and where $\ket{I,I}$ is the fully polarized spin state satisfying $\hat{I}_z\ket{I,I}=I\ket{I,I}$. In the limit $I\rightarrow\infty$, spin coherent states recover continuous-variable coherent states $\ket{\alpha}$ (satisfying $a\ket{\alpha}=\alpha\ket{\alpha}$, where $a$ is a bosonic annihilation operator). This connection can be made more transparent by defining the spin displacement operator $D(\xi)=e^{\xi \hat{I}_--\xi^*\hat{I}_+}$, in terms of which $\ket{\Theta,\phi}=e^{-i\phi I}D(e^{i\phi}\Theta/2)\ket{I,I}$.  In the limit $I\gg 1$, the spin coherent state $\ket{\Theta,\phi}$ is then the spin analog of the coherent state $\ket{\alpha}$ with $\alpha=\sqrt{2I}e^{i\phi}\tan{(\Theta/2)}$. This may be seen, for instance, by considering that under the Holstein-Primakoff transformation, $\hat{I}_-\rightarrow \sqrt{2I}a^\dagger$ \cite{radcliffe1971some}.

We next define the spin cat states
\begin{equation}\label{spin-cat}
    \ket*{\Theta,\phi}_\pm =\mathcal{N}_\pm(\Theta) \left(\ket{\Theta,\phi}\pm e^{-i\pi I}\ket{\Theta,\phi-\pi}\right),
\end{equation}
where $\mathcal{N}_\pm(\Theta)=\left[2(1\pm \gamma(\Theta))\right]^{-1/2}$ with $\gamma(\Theta)=\cos^{2I}{\Theta}$. In terms of $\hat{I}_z$ eigenstates, the spin coherent state $\ket{\Theta,\phi}$ is given by
\begin{equation}\label{fock-decomposition}
    \ket{\Theta,\phi}=\left(1+\vert \zeta\vert^2\right)^{-I}\sum_{n=0}^{2I}\sqrt{\binom{2I}{n}}\zeta^ne^{-i\phi I}\ket{I,I-n},
\end{equation}
where $\zeta=e^{i\phi}\tan{(\Theta/2)}$, and where the phase $e^{-i\phi I}$ ensures that under a $2\pi$ rotation about $\hat{I}_z$ ($\phi\rightarrow\phi+2\pi$), spin coherent states for half-integer spin acquire a global phase of $-1$. On the basis of Eq.~\eqref{fock-decomposition}, it is then readily apparent that the spin cat states $\ket*{\Theta,\phi}_\pm$ defined in Eq.~\eqref{spin-cat} are by construction parity eigenstates, as required by the overall $\mathbbm{Z}_2$ symmetry of the nuclear-spin Hamiltonian $H$ [Eq.~\eqref{Z2-symmetry}]: $\hat{\Pi}\ket*{\Theta,\phi}_\pm=\pm e^{-i\pi I} \ket*{\Theta,\phi}_\pm$. 

While the spin cat states $\ket{\Theta,\phi}_\pm$ are maximally distinguishable in their azimuthal coordinate, it is more common in the literature to define spin cat states as linear combinations of antipodal spin coherent states. These include $\ket{I,I}\pm \ket{I,-I}$ and rigid rotations thereof. Spin cat states of this form have been considered as the logical basis states of quantum error correcting codes that leverage the larger Hilbert space of spins with $I\geq 1$ to detect and correct errors~\cite{gross2024hardware,omanakuttan2024fault}. Logical basis states in the error-correcting code of Ref.~\cite{gross2024hardware}, which protects against dephasing, have been generated with ${}^{123}\mathrm{Sb}$ donors and were found to have coherence times of tens of milliseconds~\cite{yu2025schrodinger}.  Other codes have also been proposed to correct phase-flip errors~\cite{pirandola2008minimal,chiesa2020molecular,chiesa2021embedded,lim2025demonstrating} and rotation errors on spins~\cite{gross2021designing,lim2023fault}, with encoded states given by carefully chosen superpositions of $\hat{I}_z$ eigenstates. 

Although the approach taken here also aims to protect against dephasing, it differs from schemes based on quantum error correction in the sense that the qubit eigenstates are chosen so that the qubit encoding itself mitigates dephasing at the hardware level. This can result in a significant increase in the qubit dephasing time without active error correction, potentially by up to two orders of magnitude. In this respect, our approach is reminiscent of the main motivation (i.e.~suppressing phase-flip errors) for using Kerr-cat qubits defined in the states of Kerr-nonlinear resonators subjected to two-photon driving~\cite{puri2017engineering,grimm2020stabilization}. For driving at twice the resonator frequency, the resonator undergoes a bifurcation transition in which the phase-space origin becomes unstable and the Fock vacuum ceases to be the resonator ground state. The two degenerate ground states of the driven system constitute the basis states of the Kerr-cat qubit, given by the symmetric and anti-symmetric linear combinations of the finite-amplitude coherent states $\ket{\pm\alpha}$ centered about the two potential minima in phase space. With this choice of qubit encoding, phase-flip errors are exponentially suppressed in $\vert \alpha\vert^2$ due to a suppression of tunneling between the two minima. Given the predominance of bitflips over phase-flips, Kerr-cat qubits could then be used in combination with quantum error-correcting codes optimized for asymmetric noise channels~\cite{darmawan2021practical,putterman2025hardware}.

\begin{figure}
    \centering
    \includegraphics[width=\linewidth]{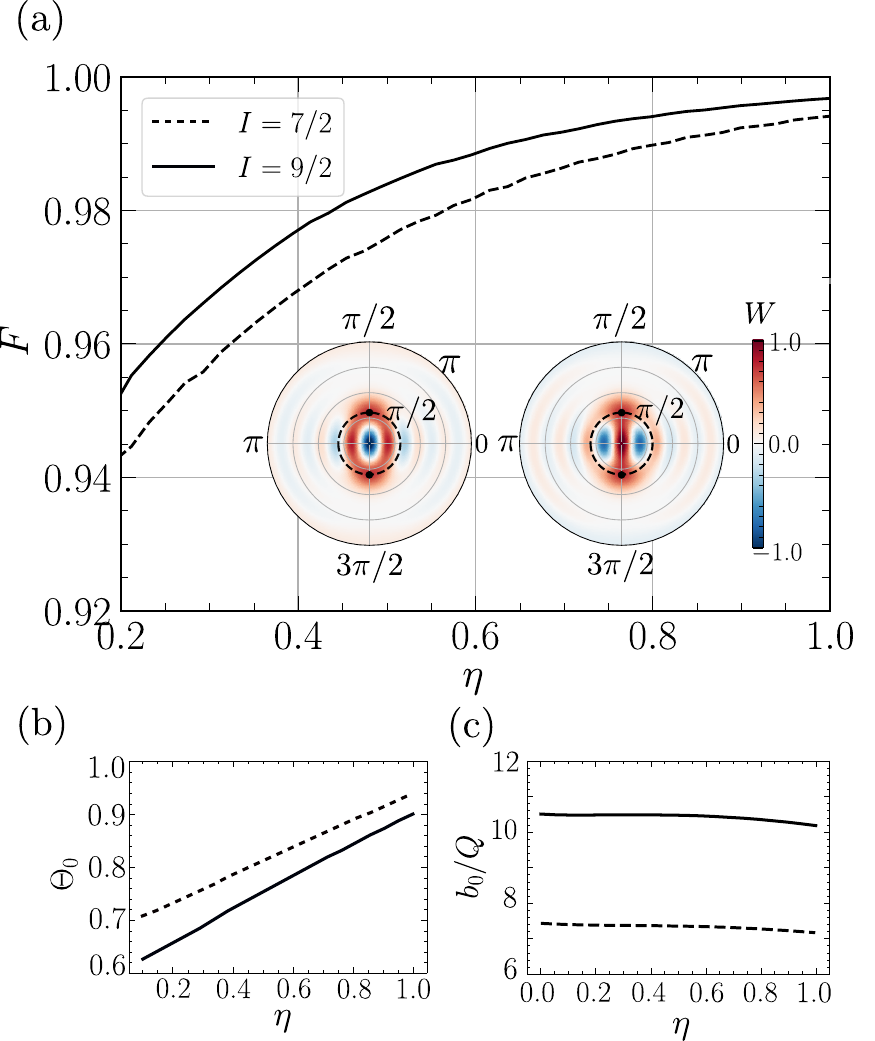}
    \caption{(a) Fidelity $F=\mathrm{max}_\Theta \bar{f}(\Theta)$ as a function of the asymmetry parameter $\eta$, where $\bar{f}(\Theta)$ is the average fidelity of the two lowest eigenstates of $H$ relative to the spin cat states $\ket{\Theta,\pi/2}_\sigma$ of the same parity. Inset: Azimuthal equidistant projections of the spin Wigner functions $W(\theta,\phi)$ of the spin Kerr-cat basis states for $I=7/2$ and $\eta=1$. The dotted black circle is a contour at $\theta=\Theta_0=0.954$, while the black dots indicate the ``heads'' of the cats at $\pm\pi/2$. (b) Polar angle $\Theta_0=\mathrm{argmax} \bar{f}(\Theta)$ defining the spin Kerr-cat encoding as a function of $\eta$. The legend is the same as in (a). (c) Sweet-spot Zeeman energy $b_0=\gamma_n B_0$ at which the first derivative $\partial \Delta/\partial Q$ of the qubit splitting with respect to $Q$ vanishes (indicated by a blue dot-dashed line in Fig.~\ref{fig:spectrum}). }
    \label{fig:cat-fidelities}
\end{figure}

While bitflips due to photon loss already constitute the main error source for superconducting resonators, with the Kerr-cat encoding serving to increase the noise bias, nuclear spins typically feature the opposite noise bias, with errors due to dephasing dominating against those due to relaxation~\cite{pla2013high,muhonen2014storing}. Hence, in contrast to the Kerr-cat encoding, the spin Kerr-cat encoding combats the dominant error source affecting spin systems. The name of the encoding comes from the fact that in the limit $I\gg 1$, the nuclear-spin Hamiltonian maps onto the Hamiltonian of a squeezed Kerr oscillator via the Holstein-Primakoff transformation, which maps spin operators to bosonic operators according to $\hat{I}_z=I-a^\dagger a$ and $\hat{I}_-\simeq \sqrt{2I}a^\dagger$. In the context of this transformation, the terms $\hat{I}_z^2$ and $\hat{I}_+^2+\hat{I}_-^2$ appearing in $H_{\mathrm{q}}$ [Eq.~\eqref{quadrupole-hamiltonian}] bear a qualitative resemblance to the Kerr nonlinearity $(a^\dagger a)^2$ and two-photon squeezing $ a^2+a^{\dagger 2}$ needed to define a Kerr-cat qubit. 
For nuclear spins, however, we emphasize that no external driving is required---the nonlinearities needed to define the qubit are produced entirely by quadrupolar interactions, which are generically present for spins with length $I\geq 1$ in the presence of a non-uniform electric field. 

Although a spin length of $I=9/2$ is not sufficient for quantitative agreement with results obtained under the Holstein Primakoff transformation, we nonetheless find that the low-energy states of $H$ can be described as superpositions of spin coherent states, in much the same way that the low-energy states of a squeezed Kerr oscillator are given by coherent-state superpositions: Recalling that the eigenstates of $H$ are necessarily $\hat{\Pi}$ eigenstates, we take the spin cat states given in Eq.~\eqref{spin-cat} as an ansatz. We then vary $\Theta$ and numerically evaluate
\begin{align}
    F&=\max_{\Theta}\bar{f}(\Theta),\\
    \Theta_0&=\mathrm{argmax}\:\bar{f}(\Theta),\label{theta0}
\end{align}
where $\bar{f}(\Theta)$ is the average fidelity of the two lowest eigenstates of $H$ relative to the spin cat states $\ket{\Theta,\pi/2}_\pm$:
\begin{equation}
    \bar{f}(\Theta)=\frac{1}{2}\left(\vert \langle 1\vert \Theta,\pi/2\rangle_-\vert^2+\vert \langle 0\vert \Theta,\pi/2\rangle_+\vert^2\right).
\end{equation}
The states $\ket{0},\ket{1}$ were obtained numerically with the applied magnetic field tuned to a sweet spot where the first derivative of the qubit splitting with respect to the quadrupole coupling $Q$ vanishes (indicated by the vertical blue line in Fig.~\ref{fig:spectrum}). We henceforth denote by $B_0$ the value of the magnetic field at this point. The sweet spot near the last local maximum in Fig.~\ref{fig:spectrum}(c) yielded the highest fidelities for the spin lengths of $I=7/2$ and $I=9/2$ considered here, and the results of the optimization for $F$ and $\Theta_0$ are shown in Figs.~\ref{fig:cat-fidelities}(a-b). Fidelities $F\gtrsim 99\%$ can be achieved for a range of asymmetry values $\eta$, with the quality of the cat-state description increasing for both larger $I$ and larger $\eta$.

The basis states $\ket{0},\ket{1}$ of the spin Kerr-cat qubit may then be approximated as 
\begin{align}
\begin{aligned}\label{spin-kerr-cat}
    &\ket{1}\simeq\ket{\Theta_0,\pi/2}_-,\\
    &\ket{0}\simeq\ket{\Theta_0,\pi/2}_+.
\end{aligned}
\end{align}
While the cat-state description is not necessary for the purpose of defining a qubit, it will play an important role in the procedure (given in Sec.~\ref{sec:initialization}) for initializing the nuclear spin in the spin-Kerr-cat subspace using $\mathrm{SU}(2)$ rotations and $\hat{I}_z$-parity measurements. For this initialization procedure, the fidelity $F$ then sets the state-preparation fidelity. (For lower values of $\eta$, initialization fidelities could be improved by allowing full $\mathrm{SU}(2I+1)$ control of the nuclear spin, demonstrated for instance in Ref.~\cite{fernandez2024navigating}.) The states $\ket{0},\ket{1}$ can be visualized through the spin Wigner function $W_\rho(\theta,\phi)=\mathrm{Tr}\{\rho \hat{w}(\theta,\phi)\}$, where $\rho$ is the density matrix of the nuclear spin and $\hat{w}(\theta,\phi)$ is the Wigner operator~\cite{stratonovich1957distributions}. Similar to the standard Wigner function used for continuous-variable systems, the spin Wigner function is a phase-space quasiprobability distribution for which negative values indicate nonclassical features in $\rho$~\cite{davis2021wigner}. As shown in Fig.~\ref{fig:cat-fidelities}, the spin Kerr-cat basis states exhibit the signature pattern of interference fringes characteristic of cat states in both finite-dimensional and continuous-variable systems.

Since tuning the qubit to the sweet spot requires a certain ratio between $b_0=\gamma_n B_0$ and $Q$ [Fig.~\ref{fig:spectrum}(c)], the magnetic field $B_0$ depends on both $Q$ and the nuclear gyromagnetic ratio $\gamma_n$. As shown in Fig.~\ref{fig:cat-fidelities}(c), this ratio is given by $b_0/Q\approx 7$ for $I=7/2$ and by $b_0/Q\approx 10$ for $I=9/2$. For $\gamma_n=5.55$ MHz/T (corresponding to the nuclear gyromagnetic ratio of ${}^{123}\mathrm{Sb}$, for which $I=7/2$), this translates to an applied magnetic field of $B_0\approx 120$ mT for $Q=100$ kHz. For ${}^{209}\mathrm{Bi}$ nuclear spins, for which $\gamma_n=6.96$ MHz/T and $I=9/2$, the same value of $Q$ gives $B_0\approx 144$ mT. 

\section{Decoherence}\label{sec:decoherence}

Since nuclear spins exhibit a strong bias in their dominant error source, with relaxation times typically far exceeding dephasing times, the spin Kerr-cat encoding is designed specifically to increase the dephasing time $T_2^*$ of quantum information stored in a nuclear-spin degree of freedom. In this section, we first analyze the errors resulting from an EFG-mediated relaxation channel, before quantifying the potential enhancement in dephasing time enabled by this encoding. The relaxation channel in question is the result of the nucleus' nonzero quadrupole moment allowing the nuclear spin to couple to charge noise in the environment, which can cause fluctuations in the EFG tensor $\bm{\mathcal{V}}$. As shown in Sec.~\ref{sec:relaxation}, we find that, assuming realistic two-level fluctuator (TLF) densities in the amorphous $\mathrm{SiO}_2$ layer, relaxation and leakage errors induced by charge noise are unlikely to dominate over dephasing provided the donor implantation depth is sufficiently large ($\approx 10$ nm).

\subsection{Relaxation}\label{sec:relaxation}

Since the nuclear spin encoding the spin Kerr-cat qubit couples to electric-field gradients through the quadrupole interaction, noise impacting the EFG at the position of the nuclear spin will lead to an additional decoherence channel not present for spins with $I=1/2$~\cite{franke2017multiple}. The $\mathbb{Z}_2$ symmetry of the nuclear-spin Hamiltonian, which ensures that the eigenvalue of $\hat{\Pi}$ is a good quantum number, relies on aligning the applied magnetic field with the principal axis of the EFG tensor. Charge noise can therefore break this intended symmetry and ultimately couple states of different $\hat{I}_z$ parity.  For small variations in the alignment of the magnetic field with respect to the principal axis of the unperturbed EFG tensor, these terms can be treated in perturbation theory and give rise to  transitions between the spin Kerr-cat basis states, as well as to leakage out of the spin Kerr-cat subspace. 

In this section, we provide a formalism for characterizing these errors before estimating the transition rate between spin Kerr-cat basis states due to the presence of TLFs in the vicinity of the nuclear spin. We estimate that for realistic donor-implantation depths and TLF densities in $\mathrm{SiO}_2$, bitflip errors $\ket{0}\leftrightarrow \ket{1}$ due to charge noise will occur on a timescale exceeding 300 s for $Q=100$ kHz (estimated conservatively based on the nearest-neighbor TLF distance acting as a lower bound for 90\% of random TLF distributions). While this section focuses on relaxation due to charge noise, we have also considered relaxation due to phonons, which couple to the nuclear spin via the gradient-elastic tensor (Appendix~\ref{sec:phonons}). The effects of phonons were found to be negligibly small relative to those of charge noise. 

With the quadrupole tensor $\bm{Q}$ expressed in a lab-frame coordinate system $\{x',y',z'\}$, the quadrupole interaction $H_{\mathrm{q}}$ is given by Eq.~\eqref{quadrupole-cartesian}. As explained in Sec.~\ref{sec:quadrupole} above, this Hamiltonian can be re-written in the PAS of the EFG tensor $\bm{\mathcal{V}}$ through an orthogonal transformation $\bm{R_0}^\top \bm{\mathcal{V}}\bm{R_0}=\bm{D_0}$, and in this PAS, $H_{\mathrm{q}}$ is given by Eq.~\eqref{quadrupole-hamiltonian}. Charge noise can lead to variation or drift of the PAS due to variations $\mathcal{V}_{\alpha\beta}\mapsto \mathcal{V}_{\alpha\beta}+\delta \mathcal{V}_{\alpha\beta}$ of the electric potential at the site of the nuclear spin. Such noise will perturb the EFG tensor by an amount $\bm{\delta \mathcal{V}}$, and consequently the orthogonal transformation required to diagonalize it, which we now write as
\begin{equation}\label{diagonalize-delta}
    \bm{R_\delta}^\top (\bm{\mathcal{V}}+\bm{\delta \mathcal{V}})\bm{R_\delta}=\bm{D_\delta}\equiv \bm{D_0}+\bm{\delta D_0}.
\end{equation}
Here, the orthogonal matrix $\bm{R_\delta}$ and diagonal tensor $\bm{D_\delta}$ are defined such that $\bm{R_\delta}=\bm{R_0}$ and $\bm{D_\delta}=\bm{D_0}$ for $\bm{\delta \mathcal{V}}=\bm{0}$. 

In general, a magnetic field aligned with the $\hat{z}$ axis in the PAS of $\bm{\mathcal{V}}$ will not be aligned exactly with the $\hat{z}$ axis in the PAS of $\bm{\mathcal{V}}+\bm{\delta \mathcal{V}}$, but rather, along some different orientation specified by $\bm{R_\delta}^{\top} \bm{R_0}\hat{z}$. This misalignment will lead to $\mathbb{Z}_2$-symmetry-breaking transversal ($\propto \hat{I}_x,\hat{I}_y$) terms originating from the nuclear Zeeman interaction. To quantify the misalignment $\bm{R_\delta}^{\top} \bm{R_0}\hat{z}$ of the magnetic field in the PAS of $\bm{\mathcal{V}}+\bm{\delta \mathcal{V}}$, we transform $\bm{\delta \mathcal{V}}$ to the eigenbasis of $\bm{\mathcal{V}}$ and denote the resulting matrix $\bm{\Omega}$:
\begin{equation}\label{delta-matrix}
    \bm{\Omega}=\bm{R_0}^\top \bm{\delta \mathcal{V}} \bm{R_0}.
\end{equation}
We parameterize the orthogonal matrix $\bm{R_\delta}$ needed to diagonalize $\bm{\mathcal{V}}+\bm{\delta \mathcal{V}}$ as 
\begin{equation}
    \bm{R_\delta}=\bm{R_0}e^{\bm{S_\delta}},
\end{equation}
where orthogonality of $\bm{R_\delta}$ requires that $\bm{S_\delta}$ be skew symmetric: $\bm{S_\delta}^\top=-\bm{S_\delta}$. From Eq.~\eqref{diagonalize-delta}, it then follows that 
\begin{equation}\label{bch}
    \bm{D_0}+\bm{\Omega}=e^{\bm{S}_\delta}\bm{D_\delta} e^{-\bm{S_\delta}}.
\end{equation}

In principle, the matrix $\bm{S_\delta}$ could be systematically solved for, order-by-order, by expanding the right-hand side of Eq.~\eqref{bch} with the Baker-Campbell-Hausdorff expansion. We will, however, restrict our attention to leading-order corrections due to $\bm{\delta \mathcal{V}}$, giving
\begin{equation}\label{delta-solution}
    \bm{\Omega}= \bm{\delta D_0}+[\bm{S_\delta}, \bm{D_0} ]+ O(\bm{\delta \mathcal{V}}^2),
\end{equation}
where $[A,B]=AB-BA$ denotes the commutator of $A$ and $B$. Since $\bm{\delta D_0}$ and $[\bm{S_\delta},\bm{D_0}]$ are purely diagonal and off-diagonal, respectively, Eq.~\eqref{delta-solution} allows us to characterize the leading-order corrections to the eigenvalues and eigenvectors of $\bm{\mathcal{V}}$ due to $\bm{\delta \mathcal{V}}$. In particular, we can solve for $\bm{\Omega}$ element-wise, giving
\begin{align}
    &\Omega_{ij}=(D_{jj}-D_{ii})S_{ij},\quad i\neq j,\label{delta-ij}\\
    &\Omega_{ii}=\delta D_{ii},\label{delta-ii}
\end{align}
where $D_{ii}$ are the unperturbed eigenvalues of $\bm{D_0}$ ($i=x,y,z$), and where $\delta D_{ii}$ denotes the $(i,i)$ element of $\bm{\delta D_0}$. Having solved for $\bm{S_\delta}$ (whose diagonal elements are necessarily zero following from $\bm{S_\delta}^\top=-\bm{S_\delta}$), vectors written in the PAS of $\bm{\mathcal{V}}$ can now straightforwardly be rotated into the PAS of $\bm{\mathcal{V}}+\bm{\delta \mathcal{V}}$. In particular, the direction of a magnetic field oriented along $\hat{z}$ in the PAS of $\bm{\mathcal{V}}$ can be written in the PAS of $\bm{\mathcal{V}}+\bm{\delta \mathcal{V}}$ as
\begin{align}
    \bm{R_\delta}^{\top}\bm{R_0}\hat{z}&=e^{-\bm{S_\delta}}\hat{z}\label{orthonormal}\\
    &= (\mathbbm{1}-\bm{S_\delta})\hat{z}+O(\bm{\delta \mathcal{V}}^2).\label{transform}
\end{align}
Substituting the matrix elements given in Eq.~\eqref{delta-ij} into Eq.~\eqref{transform} then gives [to $O(\bm{\delta \mathcal{V}}^2)$]
\begin{equation}\label{coordinate-transform}
    \bm{R_\delta}^\top\bm{R_0}\hat{z}\simeq  \hat{z}- \beta_x\hat{x}-\beta_y\hat{y},
\end{equation}
where
\begin{equation}\label{beta-alpha}
    \beta_i=\frac{\Omega_{i z}}{D_{zz}-D_{ii}},\quad i=x,y.
\end{equation}
Since $\bm{\mathcal{V}}$ is traceless, we have $D_{zz}+D_{yy}+D_{xx}=0$, and by convention, we have also chosen a basis ordering such that $\vert D_{zz}\vert \geq \vert D_{yy}\vert \geq \vert D_{xx}\vert$. The denominators in Eq.~\eqref{beta-alpha} cannot be zero subject to the above constraints except in the trivial case where $D_{ii}=0$ for all $i=x,y,z$.

Although one could, in principle, evaluate $e^{-\bm{S}_\delta}\hat{z}$ to retain orthonormality in Eq.~\eqref{coordinate-transform}, the results of doing so are only accurate to leading order in $\bm{S_\delta}$ unless $\bm{S_\delta}$ incorporates higher-order corrections in $\bm{\delta \mathcal{V}}$. The leading-order treatment considered here can be understood geometrically as follows: For small variations in the orientation of the magnetic field relative to the principal axis of $\bm{\mathcal{V}}+\bm{\delta \mathcal{V}}$, the projection of the magnetic field along the $\hat{x}$ and $\hat{y}$ axes will depend on the sine of some small angle, yielding linear corrections at leading order [cf.~Eq.~\eqref{coordinate-transform}]. By contrast, the reduction of the magnetic field along the $\hat{z}$ axis scales like the cosine of this same small angle, yielding corrections only at quadratic order.

In the PAS of $\bm{\mathcal{V}}+\bm{\delta \mathcal{V}}$, a magnetic field oriented along the principal axis of the \textit{unperturbed} EFG tensor will give rise to transverse terms $\propto\hat{I}_x,\hat{I}_y$ in the nuclear Zeeman interaction [cf.~Eq.~\eqref{coordinate-transform}]. Noise leading to variations in the EFG tensor will consequently break the intended $\mathbbm{Z}_2$ symmetry of $H$ [Eq.~\eqref{Z2-symmetry}], which, in the PAS of $\bm{\mathcal{V}}+\bm{\delta \mathcal{V}},$ now takes the following form for an applied magnetic field of strength $B_0$:
\begin{equation}\label{transverse-field}
    H=-\gamma_n B_0 \hat{I}_z- \bm{\delta b}_\perp\cdot \bm{\hat{I}} +H_{\mathrm{q}}.
\end{equation}
Here, $\bm{\delta b}_\perp=\gamma_n B_0(\beta_x,\beta_y,0)$ is a transverse magnetic field that depends on the matrix elements of $\bm{\delta \mathcal{V}}$ in the eigenbasis of $\bm{\mathcal{V}}$ [Eq.~\eqref{delta-matrix}]. Since $\bm{\delta \mathcal{V}}$ can vary, $\bm{\delta b}_\perp$ is generally a time-dependent quantity. The quadrupole interaction $H_{\mathrm{q}}$ in Eq.~\eqref{transverse-field} is given by Eq.~\eqref{quadrupole-hamiltonian}, but with $Q\mapsto Q+\delta Q$ and $\eta\mapsto \eta+\delta \eta$. The shifts $\delta Q$ and $\delta \eta$ result from the first-order corrections in the eigenvalues of $\bm{D_0}$ due to $\bm{\delta \mathcal{V}}$, which again depend on $\bm{\Omega}$ via Eq.~\eqref{delta-ii}. 

The $\mathbbm{Z}_2$-symmetry-breaking Zeeman term $\bm{\delta b}_\perp(t)\cdot\hat{\bm{I}}$ can be treated in first-order perturbation theory under the assumption that the noise producing $\bm{\delta \mathcal{V}}$ is weak. With $H_0=-\gamma_n B_0\hat{I}_z+H_\mathrm{q}$ denoting the unperturbed Hamiltonian (having eigenstates $\ket{m}$ of definite parity associated with eigenvalues $\epsilon_m$), the interaction-picture evolution of an initial eigenstate $\ket{m}$ can be expressed in the basis of $H_0$ eigenstates as 
\begin{equation}
    \ket{m}_t=\sum_{m'}c_{m'}(t)\ket{m'}.
\end{equation}
The evolution of $c_{m}(t)$ is governed by the Schr\"odinger equation,  
\begin{equation}\label{schrodinger}
    i\dot{c}_m(t)=\sum_n \tilde{V}_{mn}(t) c_n(t),
\end{equation}
where $\tilde{V}_{mn}(t)=e^{i \epsilon_{mn} t}V_{mn}(t)$ denotes a matrix element of the interaction-picture perturbation with respect to the unperturbed eigenstates $\ket{m}$ and $\ket{n}$. Here, $\epsilon_{mn}=\epsilon_m-\epsilon_n$ and $V_{mn}(t)=\sum_{j=x,y}\delta b_j(t)\langle m\vert \hat{I}_j\vert n\rangle$.

For concreteness, we take as our initial state the ground state $\ket{0}$ of the spin Kerr-cat qubit, given by the lowest-energy eigenstate of $H_0$. To first order in perturbation theory, valid for weak noise, we can assume that $c_0(t)=1-\delta(t)$ with $\vert\delta(t)\vert\ll 1$, in which case Eq.~\eqref{schrodinger} simplifies to $\dot{c}_m(t)\simeq -i \tilde{V}_{m0}(t)$ for $m\neq 0$. The probability $\vert c_{m}(t)\vert^2$ of the nuclear spin undergoing a transition from $\ket{0}$ to $\ket{m\neq 0}$ after some time $t$ is then given by
\begin{equation}\label{transition-rates}
    \vert c_{m}(t)\vert^2\simeq \int_0^t\int_0^t dt_1dt_2\: e^{-i\epsilon_{m0}(t_1-t_2)}V_{m0}^*(t_1)V_{m0}(t_2).
\end{equation}
Since $\hat{I}_{x}$ and $\hat{I}_y$ only couple states of opposite parity, $\mathbbm{Z}_2$-symmetry-breaking noise will only drive transitions between opposite-parity states within first-order perturbation theory. 

Rather than evaluate $\langle m\vert \hat{I}_j\vert 0\rangle$ explicitly for all relevant $m$, we will simply bound the transition probability given above by noting that for a spin-$I$ system, these matrix elements must satisfy $\langle m\vert \hat{I}_j\vert 0\rangle\leq I$ independent of $m$. Furthermore, since fluctuations of $\bm{\delta b}_\perp$ originate from fluctuations of the magnetic-field orientation in the PAS of $\bm{\mathcal{V}}+\bm{\delta \mathcal{V}}$ [cf.~Eq.~\eqref{coordinate-transform}], fluctuations in $\delta b_x$ and $\delta b_y$ are correlated and have a common spectral content. Under the assumption that the noise producing $\bm{\delta \mathcal{V}}$ is stationary, averaging $\vert c_{m}(t)\vert^2$ over realizations of the noise allows us to express the average transition probability $\langle \vert c_{m}(t)\vert^2\rangle$ in terms of the self- and cross-spectral densities of $\delta b_x(t)$ and $\delta b_y(t)$: 
\begin{align}\label{average-transition-prob}
    &\langle\vert c_{m}(t)\vert^2\rangle\lesssim I^2\int\frac{d\omega}{2\pi}J_\perp(\omega)F_{m}(\omega,t),\\
    &J_\perp(\omega)=\sum_{i,j=x,y}J_{ij}(\omega),
\end{align}
where $F_{m}(\omega,t)=\vert \int_0^tdt'e^{-i(\epsilon_{m0}+\omega)t'}\vert^2$ and $J_{ij}(\omega)=\int dt\: e^{i\omega t}\langle \delta b_i(t)\delta b_j\rangle$. The inequality in Eq.~\eqref{average-transition-prob} follows from the bound on the matrix elements of $\hat{I}_{x,y}$ discussed above.

In the long-time limit, $F_{m}(\omega,t)$ becomes increasingly peaked at $\omega=-\epsilon_{m0}$: $\lim_{t\rightarrow\infty}F_{m}(\omega,t)=2\pi t\delta(\epsilon_{m0}+\omega)$. This recovers Fermi's golden rule, where the transition (in this case, excitation) probability depends on the spectral density $J_\perp$ evaluated at $-\epsilon_{m0}$:
\begin{equation}\label{fermi-golden}
    \langle \vert c_{m}(t)\vert^2\rangle\lesssim \Gamma_{m0} t,\quad\Gamma_{m0}=I^2J_\perp(-\epsilon_{m0}),\quad t\rightarrow \infty.
\end{equation}
To quantify the transition rate $\Gamma_{10}^{-1}$ from the ground to the excited state of the spin Kerr-cat qubit, it then remains to estimate the typical size of the spectral density $J_\perp$ at $-\Delta=\epsilon_0-\epsilon_1$ for a realistic model of charge noise.

\subsubsection{Noise due to a single TLF}\label{sec:relax-tlf}

To this end, we first characterize the perturbation $\bm{\delta \mathcal{V}}$ to the EFG tensor resulting from a single two-level fluctuator (TLF) located at an average distance $r$ from the position $\bm{r}_0$ of the nuclear spin, here taken to define the origin. We treat the TLF as a point charge with charge $e$ that undergoes random displacements $\bm{\delta r}$ from its average position. This model could describe, for instance, an electron randomly tunneling between two metastable configurations. 

The dominant contribution to $\bm{\delta\mathcal{V}}$ originates from the electric-field response tensor, which characterizes the impact of an applied electric field on the EFG due to the distortion of bond orbitals~\cite{armstrong1961linear}. A ``direct'' contribution coming from the EFG tensor of the TLF itself is negligibly small in comparison (see Appendix \ref{sec:direct-tlf}). In this section, the lab-frame coordinate system $\{x',y',z'\}$ is taken to correspond to the silicon crystallographic axes. For a crystal with tetrahedral symmetry, like bulk silicon, the change $\bm{\delta\mathcal{V}}$ in the EFG tensor due to a perturbation $\bm{\delta E}=(\delta E_{x'},\delta E_{y'}, \delta E_{z'})$ in the electric field of the TLF is given by
\begin{equation}\label{response-tensor}
    \underline{\bm{\delta\mathcal{V}}}=\begin{pmatrix}
        \bm{0}\\
        R_{14}\mathbbm{1}_3
    \end{pmatrix}\bm{\delta E},
\end{equation}
where $\mathbbm{1}_3$ is a $3\times 3$ identity matrix, and where $\underline{\bm{\delta\mathcal{V}}}$ is a vector containing the six independent elements of $\bm{\delta \mathcal{V}}$:
\begin{equation}
    \underline{\bm{\delta\mathcal{V}}}=\begin{pmatrix}
         \delta\mathcal{V}_{x'x'} & \delta\mathcal{V}_{y'y'} & \delta\mathcal{V}_{z'z'} & \delta\mathcal{V}_{y'z'} & \delta\mathcal{V}_{x'z'} & \delta\mathcal{V}_{x'y'}
    \end{pmatrix}^\top.
\end{equation}
The size of the tensor element $R_{14}$ has been estimated as $R_{14}\approx 10^{12}\;\mathrm{m}^{-1}$ based on measurements of the Rabi frequencies of nuclear-electric resonance transitions in antimony~\cite{asaad2020coherent}. 

At the location of the nuclear spin, the electric field due to the TLF is given simply by $\bm{E}_{\mathrm{TLF}}(\bm{r}_0)=(ke/r^2)\bm{\hat{r}}$, where $k$ is Coulomb's constant and $\bm{\hat{r}}$ is a unit vector pointing from the TLF to the nuclear spin. The notation $\bm{\nabla}'$ denotes the gradient expressed in the crystallographic coordinate system $\{x',y',z'\}$. If the point charge is displaced from its original position by some small vector $\bm{\delta r}$, then the perturbation $\bm{\delta E}$ to the electric field at the site of the nuclear spin is given to leading order by
\begin{align}
    \bm{\delta E}&=(\bm{\delta r}\cdot\bm{\nabla'})\bm{E}_{\mathrm{TLF}}(\bm{r}_0)\\
    &=\frac{ke}{r^3}\left[\bm{\delta r}-3(\bm{\hat{r}}\cdot \bm{\delta r})\bm{\hat{r}}\right].\label{dipole-tensor}
\end{align}
From Eqs.~\eqref{response-tensor} and \eqref{dipole-tensor}, the elements of $\bm{\delta\mathcal{V}}$ therefore scale according to
\begin{equation}
    \delta\mathcal{V}_{i'j'}\simeq R_{14}\frac{ke}{r^3}\delta r,\quad i\neq j,
\end{equation}
where $\delta r$ is the magnitude of the TLF displacement, and where we have neglected the geometric dependence coming from the dot product $\bm{\hat{r}}\cdot\bm{\delta r}$.

The typical size of the transverse magnetic-field components in the PAS of $\bm{\mathcal{V}}+\bm{\delta\mathcal{V}}$ depends on both the strength $B_0$ of the applied magnetic field and the dimensionless prefactors $\beta_i$ given in Eq.~\eqref{beta-alpha}. For asymmetry parameters $\eta \geq 0.5$, the magnitude of the denominator of $\beta_i$ is $\lvert D_{zz}-D_{ii}\vert\geq \vert D_{zz}\vert\propto Q$ for both $i=x,y$. The numerator of $\beta_i$ depends on the matrix elements $\Omega_{iz}$ of $\bm{\delta\mathcal{V}}$ in the eigenbasis (PAS) of $\bm{\mathcal{V}}$.  Since $\vert\Omega_{i z}\vert\lesssim \vert\delta \mathcal{V}_{i'j'}\vert$, following from the fact that Eq.~\eqref{delta-matrix} describes an orthogonal transformation, we then have
\begin{equation}\label{beta-alpha-bound}
    Q\vert\beta_i\vert\lesssim \frac{1}{h}\bigg\vert\frac{ke^2q}{2I(2I-1)}\frac{R_{14}\delta r}{r^3}\bigg\vert,
\end{equation}
where we have restored a dependence on Planck's constant for a comparison in Hz.
Group-V donors in silicon have electric quadrupole moments $q$ with a magnitude of roughly half a barn (1 barn $=10^{-28}$ $\mathrm{m}^2$)~\cite{morello2020donor}.  Taking $\delta r=.1\:\mathring{\mathrm{A}}$ as the typical displacement of a point charge when tunneling over a single atomic bond~\cite{golding1979intrinsic,martinis2005decoherence}, Eq.~\eqref{beta-alpha-bound} then gives
\begin{equation}\label{comparison}
    Q\vert \beta_i\vert \lesssim 10^{-23}\:r^{-3}\mathrm{m}^3\:\mathrm{Hz}.
\end{equation}

If we assume that the donor is implanted at a distance of $z=10$ nm~\cite{adshead2025isotopically} below a layer of $\mathrm{SiO}_2$ having a TLF density of $n_{\mathrm{TLF}}=10^{14}/\mathrm{m}^2$ (realized in state-of-the-art devices~\cite{elsayed2024low}), then based on the statistics of random (Poissonian) TLF distributions, the probability of the closest TLF being at a distance of $r\geq 20$ nm is $e^{-\pi n_{\mathrm{TLF}}(r^2-z^2)}\approx 0.9$, with the peak of the probability distribution occurring at $r=40$ nm. Taking $\delta \mathcal{V}_{i'j'}$ to set the scale of the diagonal elements $\Omega_{ii}$ of $\bm{\Omega}$ as well, which ultimately control variations in $Q$ and $\eta$ [cf.~Eq.~\eqref{delta-ii}], Eq.~\eqref{comparison} then lets us estimate that for $r=40$ nm ($r=20$ nm), we have $\delta Q\approx 0.06$ Hz ($0.5$ Hz) and $\delta \eta\sim   10^{-7}$ ($\sim 10^{-6}$). 

In the long-time limit, the transition rate from state $\ket{0}$ to $\ket{m\neq 0}$ [cf.~Eq.~\eqref{fermi-golden}] depends on the spectral density $J_\perp(\omega)$ evaluated at $\epsilon_{0m}=\epsilon_0-\epsilon_m$. Since the orthogonal transformation $\bm{R_0}$ preserves the typical strength of the noise, we will assume for simplicity that $\bm{R_0}$ rotates $\bm{\delta\mathcal{V}}$ such that all of the noise is along the $x$ axis in the PAS of $\bm{\mathcal{V}}+\bm{\delta\mathcal{V}}$, giving $\Omega_{yz}=0$ and $\Gamma_{m0}=I^2 J_{xx}(\epsilon_{0m})$. The spectral density $J_{xx}$ itself depends on the autocorrelation function $\langle \delta b_x(t)\delta b_x\rangle$ of the magnetic field along $\hat{x}$, whose time dependence is inherited from the $\mathring{\mathrm{A}}$-scale displacements of the point charge. We model the time dependence of $\delta r(t)$ as a random telegraph signal: $\delta r(t)=\delta r\times \xi(t)$ with $\xi(t)=\pm 1/2$. The autocorrelation function of $\xi(t)$ is exponential, $\langle \xi(t)\xi\rangle=(1/4)e^{-2\kappa \vert t\vert}$, with $\kappa$ denoting the switching rate between the two values of $\pm 1/2$~\cite{cywinski2008enhance}. It consequently follows that
\begin{equation}\label{gamma-m0}
    \Gamma_{m0}=I^2\left(\gamma_n B_0 \beta_x\right)^2 \frac{\kappa}{\epsilon_{0m}^2+4\kappa^2}<\frac{I^2(\gamma_n B_0 \beta_x)^2}{2 \vert \epsilon_{0m}\vert},
\end{equation}
where $\vert \beta_x\vert$ is bounded by Eqs.~\eqref{beta-alpha-bound} and \eqref{comparison}, and where the second inequality holds independent of the value of $\kappa$.  

The smallest value of $\epsilon_{m0}$ is on the order of $(0.1{-}1)Q$ (Fig.~\ref{fig:spectrum}) and corresponds to the splitting $\Delta=\epsilon_1-\epsilon_0$ between the basis states of the spin Kerr-cat qubit, while the rate of leakage out of the qubit subspace is suppressed in comparison by larger energy gaps. Taking $I=7/2$, $\gamma_n B_0\approx 10Q$ (Fig.~\ref{fig:cat-fidelities}), and $2\epsilon_{10}\approx Q$, we then have
\begin{equation}\label{estimate-gamma-10}
    \Gamma_{10} < \left[1.5\times10^{-22}r^{-3} \mathrm{m}^3\right]^2\frac{\mathrm{Hz}^2}{Q}.
\end{equation}
For $r=20$ nm and $Q=100$ kHz, Eq.~\eqref{estimate-gamma-10} gives a bitflip time of $\Gamma_{10}^{-1}>280$ s due to charge noise. The strong polynomial dependence $\propto r^{-6}$
of $\Gamma_{10}$ on the distance $r$ between the TLF and nuclear spin does however lead to a rapid reduction of $\Gamma_{10}^{-1}$ as $r$ is reduced: For a separation of $r=15$ nm, we instead have $\Gamma_{10}^{-1}> 50$ s. Since we are treating the noise classically, the noise spectral density is symmetric for positive and negative frequencies. The transition rate $\Gamma_{01}$ for the opposite process, $\ket{1}\rightarrow\ket{0}$, is therefore equal to $\Gamma_{10}$.

\subsubsection{Generalization to multiple TLFs}

The generalization to several TLFs follows straightforwardly, owing to the fact that electric fields add vectorially. In particular, we now have 
\begin{equation}\label{delta-V-many-TLF}
    \underline{\bm{\delta\mathcal{V}}}=\begin{pmatrix}
        \bm{0}\\
        R_{14}\mathbbm{1}_3
    \end{pmatrix}\sum_j\bm{\delta E}_j,
\end{equation}
where $\bm{\delta E}_j$ is the perturbation to the electric field due to the $j^{\mathrm{th}}$ TLF, located at position $\bm{r}_j$. The matrix elements of $\bm{\Omega}$ controlling the size $\propto \beta_i$ of the transverse magnetic-field components in the PAS of $\bm{\mathcal{V}}+\bm{\delta \mathcal{V}}$ [Eq.~\eqref{beta-alpha}] can be obtained from Eq.~\eqref{delta-V-many-TLF} and will generally weight each TLF differently depending on its position relative to the nuclear spin. In general, however, $\vert\beta_i\vert$ can be estimated by neglecting these geometry-dependent weights, giving
\begin{align}
    \vert\beta_i\vert&= \sum_j\vert\beta_{\mathrm{TLF}_j}\vert,\\
    Q\beta_{\mathrm{TLF}_j}&\approx \frac{1}{h}\frac{ke^2q}{2I(2I-1)} \frac{R_{14}\delta r}{r_j^3}.
\end{align}
Similar to the case of a single TLF, we assume that the noise is due to random jumps in the position of each TLF, described by $\delta r_j(t)=\delta r\times \xi_j(t)$ with $\xi_j(t)=\pm 1/2$. With $\langle \xi_j(t)\xi_j\rangle=(1/4) e^{-2\kappa_j \vert t\vert}$ giving the autocorrelation function of $\xi_j(t)$ in terms of the switching rate $\kappa_j$ of the $j$th TLF, the transition rate $\Gamma_{10}$ can then be written as a sum over the spectral density of each TLF,
\begin{equation}
    \Gamma_{10}\approx I^2(\gamma_n B_0)^2 \sum_j \beta_{\mathrm{TLF}_j}^2 \frac{\kappa_j}{\epsilon_{01}^2+4\kappa_j^2}.
\end{equation}

We can conservatively neglect the $j$-dependence of $\beta_{\mathrm{TLF}_j}$ by setting all $\beta_{\mathrm{TLF}_j}$ equal to $\beta_{\mathrm{max}}=\max_j\beta_{\mathrm{TLF}_j}$. A $1/f$ noise spectrum can then be obtained by replacing the sum over TLFs by an integral over switching rates $\kappa_j$, taken to be log-uniform distributed~\cite{cywinski2008enhance}. This procedure recovers the same overall scaling as the case of a single TLF [Eq.~\eqref{gamma-m0}], with the transition rate scaling like $\beta_{\mathrm{max}}^2$. Since $\beta_{\mathrm{max}}^2$ scales like $r_{\mathrm{min}}^{-6}$, where $r_{\mathrm{min}}$ is the distance to the nearest TLF, the size of $\Gamma_{10}$ will fall off rapidly with increasing $r_{\mathrm{min}}$ and will ultimately be set predominantly by this TLF (and possibly by a few others at a comparable distance). As a result, the bounded estimate for $\Gamma_{10}$ derived for the case of a single TLF [Eq.~\eqref{estimate-gamma-10}] provides a reasonable estimate for the transition rate obtained in the presence of a more complicated configuration of TLFs, up to a potential constant prefactor corresponding to the number of TLFs located at the same minimum distance $r_{\mathrm{min}}$ from the nuclear spin.

\subsection{Dephasing}

Qubit dephasing originates from variations in the qubit splitting $\Delta$ and the resulting uncertainty in the relative phase acquired by superposition states. For a qubit splitting $\Delta(b_z,Q)$ set by $b_z=\gamma_n B_z$ and $Q$ [Fig.~\ref{fig:spectrum}(c)], the dynamics due to pure dephasing can be obtained from the Hamiltonian
\begin{equation}
    H_{\mathrm{qubit}}=\frac{\Delta (b_z,Q)}{2}\hat{\tau}_z,
\end{equation}
where $\hat{\tau}_z=\ketbra{1}-\ketbra{0}$ is a Pauli operator acting in the qubit subspace. The shorter dephasing times measured for ${}^{123}\mathrm{Sb}$ nuclear spins relative to ${}^{31}\mathrm{P}$ spins in a comparable device suggest that noise of electrical origins may play an appreciable role in the dephasing of nuclei with $I\geq 1$~\cite{franke2017multiple,asaad2020coherent}. The impacts of charge noise can be mitigated at the level of the qubit encoding by setting the external magnetic field to the charge-noise sweet spot, $b_z=b_0=\gamma_nB_0$, where the first derivative $\partial_Q\Delta$ of the qubit splitting with respect to fluctuations in $Q$ is equal to zero. Although magnetic-field fluctuations can still impact the qubit splitting at linear order, their impact is reduced by the near-coincidence of the charge-noise sweet spot with the clock transition, at which $\partial_{b_z}\Delta=0$. For small variations of $b_z$ and $Q$ around their average values [$b_z(t)=b_0+\delta b_z(t)$, $Q(t)=Q_0+\delta Q(t)$], the qubit splitting can be expanded in a Taylor series.  The variations $\delta\Delta(\bm{x})=\Delta(\bm{x})-\Delta(\bm{x}_0)$ in the qubit splitting can then be approximated as
\begin{align}
\begin{aligned}\label{taylor-exp}
    \delta\Delta(\bm{x})&\simeq \partial_{b_z}\Delta\big\vert_{\bm{x}_0}\delta b_z+\partial_Q\Delta\big\lvert_{\bm{x}_0}\delta Q+\frac{1}{2}\partial_Q^2\Delta\big\vert_{\bm{x}_0}\delta Q^2\\
    &+\frac{1}{2}\partial_{b_z}^2\Delta\big\vert_{\bm{x}_0}\delta b_z^2+\partial_Q\partial_{b_z}\Delta\big\vert_{\bm{x}_0}\delta b_z\delta Q,
\end{aligned}
\end{align}
where $\bm{x}=(b_z,Q)$ denotes a vector of parameters. 

In the regime $\gamma_n B_z\gg Q$, the eigenstates of $H$ are given to a good approximation by $\hat{I}_z$ eigenstates since the $U(1)$-symmetry-breaking terms $\hat{I}_\pm^2$ can reasonably be neglected in a secular approximation. In this regime, the splitting between all nuclear-spin eigenstates therefore increases linearly with both $b_z$ and $Q$. Taking the lowest two levels to define a qubit with splitting $\Delta(\bm{x})$, the decay of the qubit coherence would then be dominated by the linear terms in Eq.~\eqref{taylor-exp}. The first derivatives of $\Delta$ relevant to dephasing at the sweet spot and in the high-field regime are then $\partial_Q\Delta(\bm{x}_0)=0$ and $\partial_{b_z}\Delta(\bm{x}_0)< 1$ for $B_z=B_0$, and $\partial_{b_z}\Delta(\bm{x}_0)=\partial_Q\Delta(\bm{x}_0)=1$ for $\gamma_n B_z\gg Q$. In the remainder of this section, we calculate the dephasing time obtained for $1/f$ noise in both the high-field regime and at the sweet spot. Contributions due to sub-leading terms in Eq.~\eqref{taylor-exp} will be neglected in each case.

Variations of $b_z$ and $Q$ lead to a decay of the average off-diagonal elements of the qubit density matrix, described by the coherence factor~\cite{cywinski2008enhance}
\begin{equation}
    C(t)=\bigg\langle\frac{\langle \hat{\tau}_+\rangle_t}{\langle\hat{\tau}_+\rangle_0}\bigg\rangle,
\end{equation}
where $\langle\hat{\tau}_+\rangle_t=\mathrm{Tr}\{\rho(t)\hat{\tau}_+\}$, $\hat{\tau}_+=\ketbra{1}{0}$, and where $\rho$ is the qubit density matrix written in a frame rotating at $\Delta(\bm{x}_0)$. The angle brackets $\langle\cdot\rangle$ (with no subscript) denote an average over noise realizations. The coherence factor can be written in terms of a cumulant expansion,
\begin{align}
    C(t)&=\langle e^{i\phi(t)}\rangle\\
    &=\mathrm{exp}\bigg\{\sum_{n=1}^\infty \frac{i^n}{n!}\Lambda_n(t)\bigg\},\label{cumulant-expansion}
\end{align}
where here, $\Lambda_n(t)$ is the $n$th cumulant of the acquired phase $\phi(t)=\int_0^t dt'\delta \Delta (t')$. For zero-mean Gaussian noise, the dynamics of $C(t)$ are set entirely by the variance $\Lambda_2$, while for non-Gaussian noise, the inclusion of higher cumulants may be needed to accurately capture the full decay curve to arbitrarily long times. Notably, quadratic fluctuations $\delta Q^2$ are generically non-Gaussian even when $\delta Q$ is itself normally distributed. However, as will be shown in this section, the dominant contribution to the spin-Kerr-cat dephasing time (as estimated on the basis of measured quantities) will likely arise from the linear susceptibility of $\Delta$ to variations $\delta b_z$ due to magnetic noise. Under the assumption that $\delta b_z$ is Gaussian distributed, we therefore truncate the cumulant expansion at second order, giving  $C(t)\simeq e^{i\Lambda_1(t)-\frac{1}{2}\Lambda_2(t)}$.  

For stationary noise, the variance $\Lambda_2(t)$ of the accumulated phase $\phi(t)$ can be written as 
\begin{align}
    \Lambda_2(t)=\int\frac{d\omega}{2\pi} F(\omega,t)S_\Delta(\omega),\label{variance}
\end{align}
where $F(\omega,t)=\big\vert\int_0^t dt' e^{i\omega t'}\big\vert^2$ is the usual filter function for free-induction decay, and where $S_\Delta(\omega)$ is the spectral density of the noise coupling to the qubit. Under the assumption that $\delta b_z$ and $\delta Q$ are uncorrelated, $S_\Delta(\omega)$ is given by the sum of the spectral densities of variations in $\Delta(\bm{x})$ due to $\delta b_z$ and $\delta Q$. In the regime $\gamma_n B_z\gg Q$, we then have
\begin{equation}\label{spectral-density-high-field}
    S_\Delta(\omega)=S_b(\omega)+S_Q(\omega),\quad \gamma_n B_z\gg Q,
\end{equation}
where $S_b(\omega)=\int dt e^{i\omega t}\langle \delta b_z(t)\delta b_z\rangle$ is the noise spectral density of $\delta b_z$, and where $S_Q(\omega)$ is defined analogously. 

For $B_z=B_0$, fluctuations in $\delta Q$ affect the qubit splitting only at quadratic order. In this case, the spectral density $S_\Delta(\omega)$ can be evaluated by applying Isserlis's theorem to the autocorrelation function of $\delta Q^2$, giving
\begin{equation}\label{spectral-density}
    S_\Delta(\omega)=(\Delta')^2S_b(\omega)+\frac{1}{4\pi}(\Delta'')^2(S_Q\star S_Q)(\omega),\quad B_z=B_0,
\end{equation}
where we have introduced the shorthand $\Delta'=\partial_{b_z}\Delta\vert_{\bm{x}_0}$, $\Delta''=\partial_Q^2\Delta\vert_{\bm{x}_0}$, and where $(S_Q\star S_Q)(\omega)=\int d\Omega S_Q(\omega-\Omega)S_Q(\Omega)$ is a convolution of $S_Q$ with itself. Equation \eqref{spectral-density} holds regardless of the form of $S_Q(\omega)$, provided $\delta Q(t)$ is Gaussian-distributed and stationary, and provided fluctuations in $b_z$ and $Q$ are independent. As a side note, we showed in Sec.~\ref{sec:relaxation} that charge noise can alter the orientation of the EFG's principal axis system with respect to the applied magnetic field, and that this can produce fluctuating magnetic-field components in the principal axis system of the perturbed EFG tensor. At leading order, these fluctuating field components are transversal [cf.~Eq.~\eqref{transverse-field}]. The field fluctuations leading to qubit dephasing, however, are those that preserve the $\mathbbm{Z}_2$ symmetry of the nuclear-spin Hamiltonian, i.e.~that couple to $\hat{I}_z$.  As discussed below Eq.~\eqref{beta-alpha}, longitudinal fluctuations in $b$ due to charge noise enter at quadratic order in some small angle characterizing the misorientation of the external magnetic field relative to the principal axis of the perturbed EFG tensor. Consequently, the assumption that $\delta b_z$ and $\delta Q$ are uncorrelated will likely hold to a good approximation provided the nuclear spin also couples to ``direct'' sources of magnetic noise, sources of which include hyperfine coupling to residual ${}^{29}\mathrm{Si}$ spins and dipolar coupling to paramagnetic impurities like silicon dangling bonds (``$P_b$ centers'') at the Si/$\mathrm{SiO}_2$ interface~\cite{lemke1978dangling,de2007dangling}. 

Due to its widespread relevance~\cite{galperin2006non,schriefl2006decoherence,paladino20141}, we now focus on the case of $1/f$ noise. In particular, we assume that fluctuations in both $b_z$ and $Q$ are described by the spectral density $S_\beta(\omega)=\nu_\beta/\vert\omega\vert$ ($\beta=b,Q$) for frequencies $\omega_{\mathrm{IR}}\leq\vert \omega\vert\leq\omega_{\mathrm{UV}}$ falling in between the infrared and ultraviolet cutoffs $\omega_{\mathrm{IR}}$ and $\omega_{\mathrm{UV}}$. In practice, $\omega_{\mathrm{IR}}=2\pi/T_m$ is set by the total time $T_m$ over which the noise is measured, while $\omega_{\mathrm{UV}}$ is set by the fastest time scale in the environment. For the free-induction decay considered in this section, qubit dephasing is insensitive to the value of $\omega_{\mathrm{UV}}$ due to spectral weight in $F(\omega,t)$ being concentrated at low frequencies.

We first evaluate the variance $\Lambda_2(t)$ [Eq.~\eqref{variance}] with the spectral density of the qubit-frequency variations given by $S_\Delta(\omega)=S_b(\omega)+S_Q(\omega)$ [Eq.~\eqref{spectral-density-high-field}]. This gives~\cite{makhlin2004dissipative,fehse2023generalized}
\begin{align}\label{variance-1/f}
 \Lambda_2(t)&\simeq  \frac{\nu_{\mathrm{tot}}}{\pi}t^2 \ln{[(\omega_{\mathrm{IR}}t)^{-1}]},
\end{align}
where $\nu_{\mathrm{tot}}=\nu_b+\nu_Q$, valid up to logarithmic corrections that are suppressed for $\ln{(1/\omega_{\mathrm{IR}}t)}>1$. We define $T_2^*$ to be the time at which $\vert C(T_2^*)\vert=1/e$, or equivalently, at which $\Lambda_2(T_2^*)/2=1$. Rearranging lets us to solve for the noise amplitude $\nu_{\mathrm{tot}}$, giving
\begin{equation}\label{nu}
    \nu_{\mathrm{tot}}\simeq \frac{2\pi}{(T_2^*)^2\ln{\left(\frac{1}{\omega_{\mathrm{IR}}T_2^*}\right)}}.
\end{equation}
This expression allows us to estimate $\nu_{\mathrm{tot}}$ based on the measured $T_2^*$ time and the total measurement time $T_m=2\pi/\omega_{\mathrm{IR}}$. It can also be inverted to solve for $T_2^*$: Through recursive substitution and up to the same logarithmic corrections [small for $\ln{(\sqrt{\nu_{\mathrm{tot}}/2\pi}/\omega_{\mathrm{IR}})}> 1$], we then find that~\cite{makhlin2004dissipative,yang2019achieving}
\begin{equation}\label{T2s}
    (T_2^*)^{-1}\simeq \sqrt{\frac{\nu_{\mathrm{tot}}}{2\pi}\ln{\left(\frac{1}{\omega_{\mathrm{IR}}}\sqrt\frac{\nu_{\mathrm{tot}}}{2\pi}\right)}},\quad \gamma_n B_z\gg Q.
\end{equation}

Calculating a dephasing time $T_{2,\mathrm{cat}}^*$ for the spin Kerr-cat qubit requires that the convolution $(S_Q\star S_Q)(\omega)$ first be evaluated. We calculate this quantity
for $1/f$ noise in Appendix \ref{appendix-dephasing} and substitute the resulting expression for $S_\Delta(\omega)$ [Eq.~\eqref{spectral-density}] into Eq.~\eqref{variance}. This gives
\begin{align}\label{variance-1}
    \Lambda_2(t)&\simeq \frac{\nu_b(\Delta')^2}{\pi}t^2\ln\left(\frac{1}{\omega_\mathrm{IR}t}\right)+\frac{(\nu_Q \Delta'')^2}{2\pi^2}t^2\ln^2{\left(\frac{1}{\omega_{\mathrm{IR}}t}\right)},
\end{align}
valid up to logarithmic corrections. As before, we define $T_{2,\mathrm{cat}}^*$ to be the time at which $\Lambda_2(T_{2,\mathrm{cat}}^*)/2=1$. To solve for $T_{2,\mathrm{cat}}^*$, we introduce the variable $u=\mathrm{ln}[(\omega_{\mathrm{IR}}T_{2,\mathrm{cat}}^*)^{-1}]$. From Eq.~\eqref{variance-1}, we then obtain an equation for $u$ given by
\begin{equation}\label{u-sol}
    u\simeq \frac{1}{2}\ln{u}+\frac{1}{2}\ln{\left[\frac{\nu_b(\Delta')^2+(\nu_Q\Delta'')^2u/2\pi}{2\pi\omega_{\mathrm{IR}}^2}\right]}.
\end{equation}
The approximate equality in Eq.~\eqref{u-sol} was inherited from Eq.~\eqref{variance-1}; no other approximations have been made at this point. 

With $\Delta$ scaling like $\Delta(Q,b_z)=Qg(x=b_z/Q)$, where $g(x)$ is a dimensionless function of $x=b_z/Q$, the derivatives appearing in Eq.~\eqref{u-sol} are given by $\Delta'=g'(x_0)$ and $\Delta''=x_0^2g''(x_0)/Q$ with $x_0=b_0/Q$. Existing measurements of the dephasing times of ${}^{123}\mathrm{Sb}$ nuclear spins~\cite{asaad2020coherent,fernandez2024navigating,yu2025schrodinger} are performed in the high-field regime $\gamma_nB_z\gg Q$, where the effects of magnetic and charge noise are difficult to separate. However, the significantly shorter dephasing times obtained for ${}^{123}\mathrm{Sb}$ nuclear spins compared to ${}^{31}\mathrm{P}$ nuclear spins in similar devices~\cite{asaad2020coherent,muhonen2014storing} (for which $Q=0$ since $I=1/2$) would tend to suggest that charge noise is the limiting error source for quadrupolar nuclear spins. We can therefore estimate the noise amplitude $\nu_Q$ ($\nu_b$) by taking $T_2^*=50$ ms ($T_2^*=0.5$ s) in Eq.~\eqref{nu}, consistent with the typical dephasing times obtained for ${}^{123}\mathrm{Sb}$~\cite{asaad2020coherent, fernandez2024navigating, yu2025schrodinger} and ${}^{31}\mathrm{P}$ nuclear spins~\cite{asaad2020coherent,muhonen2014storing}. Assuming a total integration time of $T_m=1$ h, this gives $\nu_b\approx 3.6\;\mathrm{s}^{-2}$ and $\nu_Q\approx 270\;\mathrm{s}^{-2}$. (Of course, this procedure assumes that the dephasing time of phosphorus provides a representative measure of the magnetic noise affecting donor spins in real devices, i.e.~that high-spin nuclei will experience similar levels of magnetic noise.) Solving for $\Delta'$ and $\Delta''$ numerically, we then estimate that $\nu_b(\Delta')^2$ will typically exceed $(\nu_Q\Delta'')^2$ by roughly an order of magnitude. [Since $u\simeq \ln{(T_m/T_{2,\mathrm{cat}}^*)}$ grows only logarithmically with $T_m$, we take $u/2\pi\simeq 1$ in this comparison. A quantitative estimate based on numerically solving Eq.~\eqref{u-sol} is given below.] 

\begin{figure}
    \centering
    \includegraphics[width=\linewidth]{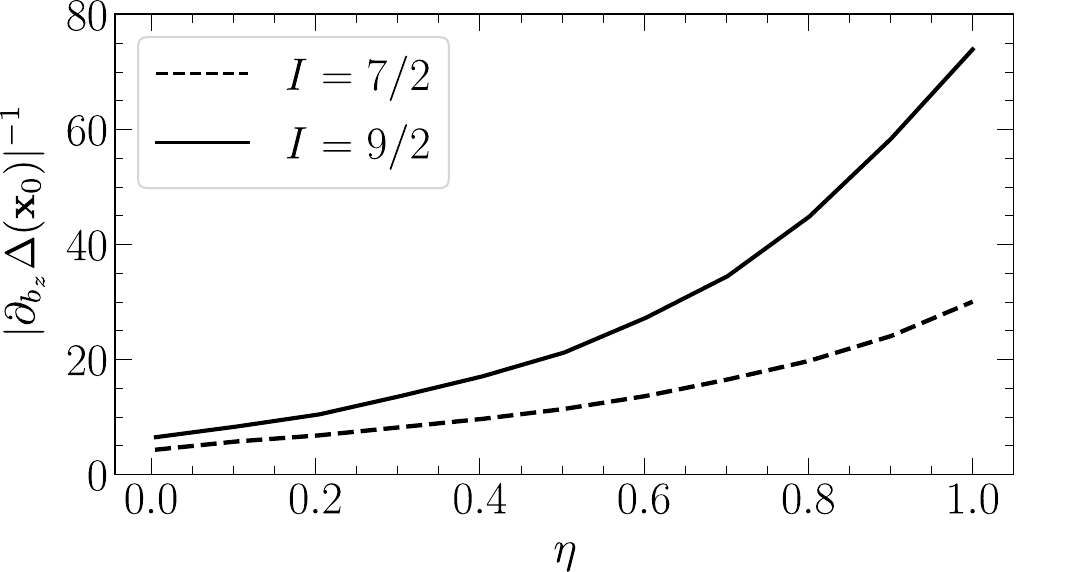}
    \caption{Inverse of the partial derivative $\partial_{b_z}\Delta(\bm{x}_0)$ of the qubit splitting with respect to the Zeeman energy $b_z=\gamma_n B_z$, evaluated at the charge-noise sweet spot [indicated by a dot-dashed blue line in Fig.~\ref{fig:spectrum}(c)]. }
    \label{fig:slope}
\end{figure}

Under the assumption that $\nu_b(\Delta')^2>(\nu_Q\Delta'')^2$, $T_{2,\mathrm{cat}}^*$ will be controlled principally by the term $\propto \nu_b(\Delta')^2$, with the largest sub-leading corrections exhibiting slow double-logarithmic growth $\propto\ln(\ln[\nu_b(\Delta')^2 / 2\pi\omega_{\mathrm{IR}}^2])$. An iterative solution can therefore be found by substituting $u\simeq (1/2)\ln{\left[\nu_b(\Delta')^2/2\pi\omega_\mathrm{IR}^2\right]}$ back into Eq.~\eqref{u-sol}, giving
\begin{equation}\label{T2ss}
    \left(T_{2,\mathrm{cat}}^*\right)^{-1}\simeq \sqrt{\frac{\nu_b(\Delta')^2}{2\pi}\ln{\Xi}}\left[1+\frac{(\nu_Q\Delta'')^2}{2\pi\nu_b(\Delta')^2}\ln{\Xi}\right]^{1/2},
\end{equation}
where $\Xi=(1/\omega_{\mathrm{IR}})\sqrt{\nu_b(\Delta')^2/2\pi}$. The second term in the square brackets describes the sub-leading corrections to $T_{2,\mathrm{cat}}^*$ originating from quadratic fluctuations $\delta Q^2$ in the qubit splitting. 

Comparing the leading term in Eq.~\eqref{T2ss} to Eq.~\eqref{T2s}, we find that the noise amplitude has effectively been reduced from $\nu_{\mathrm{tot}}$ to $\nu_b(\Delta')^2$. If we take $T_2^*({}^{31}\mathrm{P})=0.5$ s as being limited by $\nu_b$ only, then Eq.~\eqref{T2ss} predicts an enhancement relative to this dephasing time by a factor of $1/\vert \Delta'\vert=\vert \partial_{b_z}\Delta(\bm{x}_0)\vert^{-1}$ (Fig.~\ref{fig:slope}). On the basis of Fig.~\ref{fig:slope}, we therefore estimate that the dephasing time $T_{2,\mathrm{cat}}^*$ could reach ${\approx} 10$ s for $\eta=1$ and $I=7/2$. This in turn constitutes an improvement of roughly $200\times$ over the typical $T_2^*$ values measured for ${}^{123}\mathrm{Sb}$ in the high-field regime. For a more careful estimate, we can solve Eq.~\eqref{u-sol} numerically. Using the values of $\nu_b$ and $\nu_Q$ given previously, this yields an estimated dephasing time of $T_{2,\mathrm{cat}}^*\approx 10.6$ s based on Eq.~\eqref{u-sol}, and of $T_{2,\mathrm{cat}}^*\approx 12.3$ s when using the approximate solution given in Eq.~\eqref{T2ss}. 

Although we have focused on variations in $b_z$ and $Q$, the asymmetry parameter $\eta$ can also vary in principle. Such variations will generally affect $\delta\Delta$ at linear order, similar to variations in $b_z$. However, based on the TLF concentration assumed in Sec.~\ref{sec:relaxation}, we estimated variations $\delta \eta\approx 10^{-6}{-}10^{-7}$, which, for $\eta\gtrsim 0.75$, translate to variations of $\vert\delta\Delta\vert< 0.01$ Hz. Since this would translate to dephasing on a timescale of ${\approx} 100$ s, we expect the dominant linear contribution to originate from magnetic noise, consistent with the calculations presented above.

\section{Operation}\label{sec:operation}

In this section, we provide concrete strategies for the control and readout of spin Kerr-cat qubits. State preparation, readout, and two-qubit gates could be realized by leveraging the hyperfine interaction between the nuclear spin and an electron that is controllably moved on and off the donor. Single-qubit control, to which we now turn our attention, can be implemented in the absence of the electron using techniques well established in NMR.

\subsection{Single-qubit control}\label{sec:single-qubit}

Realizing arbitrary single-qubit rotations requires the ability to drive rotations about two orthogonal axes, here denoted by capital letters, $\hat{Z}$ and $\hat{X}$, to distinguish them from the principal axes of the EFG tensor. Initialization of the nuclear spin in the spin Kerr-cat subspace will be discussed in Sec.~\ref{sec:initialization}. For a nuclear-spin already initialized in the qubit subspace, rotations about $\hat{Z}$ can be realized through free evolution under the nuclear-spin Hamiltonian $H$, of which the spin Kerr-cat basis states are eigenstates. 

Rotations about a second, non-collinear axis could be realized through standard NMR techniques by applying an AC magnetic field at a frequency resonant with the qubit splitting, $B_{\mathrm{ac}}(t)=B_\perp\cos{(\Delta t+\varphi)}$. In practice, an AC magnetic field can be delivered using a broadband on-chip antenna, as is standard for both NMR and electron spin resonance (ESR). Ideally, the field would oscillate along the $y$ axis in the PAS of the quadrupolar tensor, but for generality we assume that it instead drives the nuclear spin about some different axis $\hat{n}_\perp=(\cos{\theta},\sin{\theta},0)$ in the $xy$ plane. The Hamiltonian of the driven spin is then given by 
\begin{equation}\label{nmr-hamiltonian}
    H=-\gamma_n B_0\hat{I}_z+H_\mathrm{q}+\gamma_n B_\perp\cos{(\Delta t+\varphi)}\hat{I}_\perp,
\end{equation}
where $\hat{I}_\perp=\cos{(\theta)}\hat{I}_x+\sin{(\theta)}\hat{I}_y$. With this configuration, the oscillating field can only drive transitions between certain eigenstates, namely those of opposite $\hat{I}_z$ parity. This includes the basis states $\ket{0}$ and $\ket{1}$ of the spin Kerr-cat qubit. 

Projecting into the qubit subspace and transforming to a frame rotating with the qubit splitting, we then obtain the rotating-frame Hamiltonian 
\begin{equation}\label{rwa-1}
    \tilde{H}\simeq \frac{\gamma_n B_\perp}{2}\left(e^{-i\varphi}I_{10} \hat{\tau}_++\mathrm{h.c.}\right),
\end{equation}
valid for $\gamma_n B_\perp\ll \Delta$ within a rotating-wave approximation. Here, $\hat{\tau}_+=\ketbra{1}{0}$ is a spin-raising operator acting within the qubit subspace, and $I_{10}=\langle 1\vert \hat{I}_\perp\vert 0\rangle$ is a matrix element that ultimately sets the Rabi frequency $\Omega_{\mathrm{R}}$. Although the true nuclear-spin eigenstates are not exactly equal to spin-cat states, we can nonetheless estimate the size of $\Omega_{\mathrm{R}}$ by evaluating matrix elements of $\hat{I}_x$ and $\hat{I}_y$ with respect to the spin cat states in Eq.~\eqref{spin-kerr-cat}. This procedure gives
\begin{equation}\label{rwa-2}
    \tilde{H}\simeq \Omega_\mathrm{R}\left(\hat{\tau}_x\cos{\vartheta}+\hat{\tau}_y\sin{\vartheta}\right),
\end{equation}
where $\vartheta=\varphi-\mathrm{arg}\:I_{10}$, and where $\Omega_{\mathrm{R}}=(\gamma_n B_\perp I/2) f(\theta)$ with
\begin{align}
    &f(\theta)=\frac{\tan{\Theta_0}}{\sqrt{1-\gamma^2}}\left[\gamma^2\cos^2{\theta}+\cos^2{\Theta_0}\sin^2{\theta}\right]^{1/2}.
\end{align}
Here, as before, $\gamma(\Theta_0)=\cos^{2I}{\Theta_0}$ comes from the overlap of the two spin coherent states involved in the definition of $\ket{\Theta_0,\pi/2}_\pm$. The strength $\Omega_\mathrm{R}$ of the Rabi drive depends on the orientation of the rotation axis $\hat{n}_\perp$ in the $xy$ plane of the quadrupole tensor, reaching a maximum value set by $f(\pi/2)\simeq  \sin{\Theta_0}$ for driving about $\hat{y}$ and a minimum set by $f(0)\simeq\gamma \tan{\Theta_0}$ for driving about $\hat{x}$.  Since $\Theta_0$ assumes a maximum value of $\Theta_0\approx 0.9$ (Fig.~\ref{fig:cat-fidelities}), a factor of $\sin{0.9}\approx 0.8$ does not result in a significant suppression of the AC magnetic field, whereas for $\Theta_0\approx 0.9$ and $I=7/2$, the suppression due to $f(0)\approx 0.045$ is appreciable and should be avoided if fabrication allows it.

Since $\Delta$ is on the order of the quadrupole coupling $Q$ (Fig.~\ref{fig:spectrum}), the strength of the AC field is restricted to values satisfying $\gamma_n B_\perp I\ll Q$. For $Q=100$ kHz, this translates to a timescale $\gg 10\:\mu$s for Rabi oscillations. (Notably, existing strategies for addressing individual nuclear-spin transitions are similarly limited by the size of $Q$~\cite{fernandez2024navigating}.) As an alternative to free evolution, rotations about $\hat{Z}$ could also be implemented virtually by changing the phase $\varphi$ of the AC magnetic field~\cite{mckay2017efficient,gupta2024robust,yu2025schrodinger}.

\subsection{Coupling to an electron spin}\label{sec:coupling-e-spin}

\begin{figure}
    \centering
    \includegraphics[width=\linewidth]{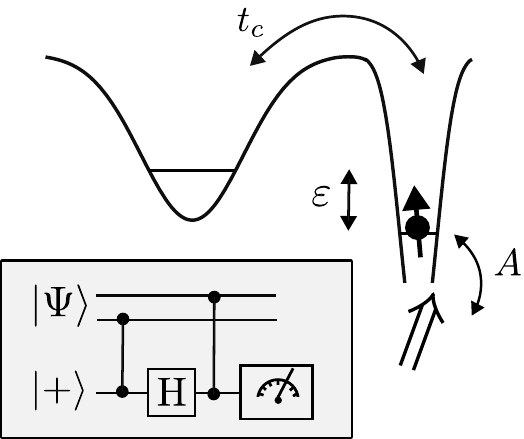}
    \caption{Schematic of an electron tunneling from a quantum-dot potential to the confinement potential of a donor spin. When bound to the donor, the electron spin interacts with the nuclear spin via the hyperfine interaction. An entangling gate logically equivalent to a CZ gate, up to single-spin rotations, can be generated by allowing the electron and nuclear spins to interact for a total time $T$. Box: Circuit representation of an electron-spin-mediated two-qubit gate between spin Kerr-cat qubits. The output is logically equivalent to applying a CZ gate to the two qubits initially in state $\ket{\Psi}$. }
    \label{fig:setup}
\end{figure}

Further control of spin Kerr-cat qubits---including two-qubit gates, initialization, and readout---can be achieved with operations involving ancillary electron-spin qubits. Donor-bound electrons can themselves be operated as qubits~\cite{morello2010single,fricke2021coherent,geng2025high}, or used as ancillas to read out the nuclear spin through spin-state-dependent tunneling~\cite{pla2013high}. In this case, however, we imagine that the electron is initially confined by a gate-defined quantum dot, but can be made to tunnel on and off of the donor potential through control over the dot-donor detuning (Fig.~\ref{fig:setup}). 

We take the Hamiltonian $H_{\mathrm{c}}$ describing the electron's charge degree of freedom to be given by
\begin{equation}\label{charge-hamiltonian}
    H_{\mathrm{c}}=\frac{\varepsilon}{2}\hat{\nu}_z+t_c\hat{\nu}_x,
\end{equation}
where $\hat{\nu}_z$ and $\hat{\nu}_x$ are Pauli matrices acting in the electron's orbital subspace, spanned by the orbital ground states of the quantum dot and donor. Here, $t_c$ is a tunnel coupling between the dot and donor, while $\varepsilon$ denotes the dot-donor detuning, which can be controlled through voltages applied to the gates defining the quantum-dot potential. (The valley excited state is neglected in this model, as are any donor excited states~\cite{mielke2021nuclear}.) When the electron is confined to the donor, the electron and nuclear spins interact via the hyperfine interaction $(A/2)\hat{\bm{\sigma}}\cdot\hat{\bm{I}}$, where $A$ is the strength of the hyperfine coupling and $\hat{\bm{\sigma}}$ is a vector of Pauli matrices acting on the electron's spin degree of freedom. We can therefore write the interaction $H_{\mathrm{HF}}$ between the electron and nuclear spins as~\cite{mielke2021nuclear}
\begin{equation}\label{hf-1}
    H_{\mathrm{HF}}=\frac{A}{4}\hat{\bm{\sigma}}\cdot\hat{\bm{I}}(\mathbbm{1}+\hat{\nu}_z),
\end{equation}
where $(1/2)(\mathbbm{1}+\hat{\nu}_z)$ is a projector onto the donor's orbital ground state. For concreteness, we assume that $A>0$. Generalizing to $A<0$ is straightforward.

The electron can be loaded on and off the donor potential by controlling the dot-donor detuning $\varepsilon$, with $\varepsilon\gg t_\mathrm{c}$ ($-\varepsilon\gg t_\mathrm{c}$) corresponding to the electron being localized on the dot (donor). Under an adiabatic sweep of $\varepsilon$ (requiring that $\dot{\varepsilon}\ll t_\mathrm{c}^2$), we can treat the orbital dynamics within an adiabatic approximation and assume that an electron initially located on the dot [$\varepsilon(0)\gg t_\mathrm{c}$] will remain in the instantaneous ground state of $H_\mathrm{c}$. The strength of the hyperfine interaction then acquires a time dependence originating from the overlap of the instantaneous lower orbital eigenstate $\ket{-(t)}$ with the donor ground state (Appendix~\ref{sec:time-dep-hyperfine}):
\begin{equation}\label{hf-2}
    A(t)=\frac{A}{2}\langle{-}(t)\vert (\mathbbm{1}+\hat{\nu}_z)\vert {-}(t)\rangle. 
\end{equation}
Restricting the dynamics to the lower orbital eigenstate is valid provided $A\ll t_\mathrm{c}$, in which case the hyperfine coupling is too weak to drive Landau-Zener transitions. In this regime, all of the orbital dynamics are captured by $A(t)$, allowing us to focus on the spin dynamics with the understanding that the orbital wavefunction is given simply by $\ket{-(t)}$ up to the usual dynamical and Berry's phases.

In the presence of an applied magnetic field $B_0$, the Zeeman splitting of the electron is $\gamma_e B_0$, where $\gamma_e=28$ GHz/T is the electron gyromagnetic ratio. Under the assumption that $\gamma_e B_0\gg  A$, we can treat the hyperfine interaction in a secular approximation and retain only its Ising-like component: $H_{\mathrm{HF}}\simeq (A/2)\hat{\sigma}_z\hat{I}_z$. For an electron with spin $\ket{\uparrow}$ ($\ket{\downarrow}$), the nuclear spin consequently experiences an effective magnetic field of strength $\gamma_n B_0-A/2$ ($\gamma_nB_0+A/2$). We further assume that $\vert \gamma_n B_0\pm A\vert \gg Q$ so that in the presence of a hyperfine-coupled electron, the quadrupole interaction $H_{\mathrm{q}}$ \textit{can} be treated in a secular approximation. Since the nuclear Zeeman splitting $\gamma_n B_0$ is not large enough on its own to justify neglecting the off-diagonal terms $\propto \eta$ in $H_{\mathrm{q}}$, satisfying this inequality requires that $ A \gg \gamma_nB_0$, yielding the hierarchy of energy scales
\begin{equation}\label{hierarchy}
    \gamma_n B_0\ll  A\ll \gamma_e B_0.
\end{equation}

In view of this hierarchy, we take the Hamiltonian $H_{\mathrm{tot}}$ of the electron spin and nuclear spin to be given by
\begin{align}
    &H_{\mathrm{tot}}=H_0+V,\label{Htot}\\
    &H_0=-\gamma_n B_0\hat{I}_z+Q\hat{I}_z^2+\frac{\gamma_e }{2}B_0\hat{\sigma}_z,\label{H0}\\
    &V=\frac{1}{2}A(t)\hat{\sigma}_z\hat{I}_z.\label{V}
\end{align}
In practice, the hyperfine coupling $A$ will rise and fall over a finite timescale corresponding to the time required to traverse the charge anticrossing adiabatically. We can treat the dynamics generated by $H_{\mathrm{tot}}$ in a sudden approximation (with the non-secular terms in $H$ ``turning off'' instantaneously) under the assumption that the time required to load the electron onto the donor is short compared to the timescale $Q^{-1}$ over which the spin Kerr-cat qubit evolves in the absence of the electron. Gates on hopping spins in silicon have been realized with interdot charge transfers occurring on nanosecond timescales~\cite{unseld2025baseband}. Since $Q^{-1}=10$ $\mu$s for $Q=100$ kHz, this simplification is likely well justified for realistic experimental parameters.

For ${}^{123}\mathrm{Sb}$ donors in bulk silicon, the strength of the hyperfine interaction is $A=101.52$ MHz, while $A=96.584$ MHz has been measured for a donor in a silicon nanoelectronic device~\cite{fernandez2024navigating}. The hyperfine coupling strength is much stronger for ${}^{209}\mathrm{Bi}$ and is given by $A=1.475$ GHz for bulk silicon. Since operating near the clock transition requires lower magnetic-field strengths (on the order of 100 mT), the second inequality in Eq.~\eqref{hierarchy} will likely not be satisfied for ${}^{209}\mathrm{Bi}$ if the electron is allowed to tunnel fully onto the donor as assumed here. For the remainder of this section, we therefore focus on realistic parameters for ${}^{123}\mathrm{Sb}$. 

\subsubsection{Two-qubit gates}

Entanglement between the electron and nuclear spins can be generated via the hyperfine interaction: In an interaction picture defined with respect to $H_0$, the time-dependent hyperfine interaction $A(t)$ can be used to generate the unitary operation
\begin{equation}\label{U-CROT}
    U_{\mathrm{CR}}=e^{-i\frac{\pi}{2}\hat{\sigma}_z\hat{I}_z},
\end{equation}
obtained by allowing the electron to interact with the nuclear spin for a total time $T$ satisfying $\int_0^Tdt\:A(t)=\pi$. This unitary applies a phase flip to the electron spin conditioned on the nuclear spin being in an even-parity state (up to a known basis change for the electron spin):
\begin{align}
    &U_{\mathrm{CR}}\ket{0}\ket{\pm}_x=e^{-i\frac{\pi}{2}\hat{I}_z}\ket{0}S\ket{\mp}_x,\label{cz-1}\\
    &U_{\mathrm{CR}}\ket{1}\ket{\pm}_x=e^{-i\frac{\pi}{2}\hat{I}_z}\ket{1}S\ket{\pm}_x.\label{cz-2}
\end{align}
Here, $\ket{\pm}_x$ are the $\pm 1$ eigenstates of $\hat{\sigma}_x$, and $S$ is an $S$ gate on the electron spin, in terms of which $\ket{\pm}_y=S\ket{\pm}_x$. Up to single-spin rotations, $U_{\mathrm{CR}}$ is therefore logically equivalent to a controlled-Z (CZ) gate between the spin Kerr-cat qubit and electron spin. 

Equations \eqref{cz-1}-\eqref{cz-2} hold for arbitrary $\eta$ and ultimately proceed from the $\mathbbm{Z}_2$ symmetry of the nuclear-spin eigenstates, which dictates that $\hat{\Pi}\ket{s}=i(-1)^s\ket{s}$ for $I=7/2$, where $\hat{\Pi}=e^{-i\pi\hat{I}_z}$ is the parity operator and $s=0,1$.  To keep the nuclear spin in the spin Kerr-cat subspace, the extra rotation about $\hat{I}_z$ should be canceled following the gate. This can be achieved by allowing the nuclear spin to evolve in the presence of a hyperfine-coupled ancilla electron held in, e.g., $\ket{\downarrow}$. If several two-qubit gates must be performed in quick succession, then the unwanted rotations could also be left uncompensated until the end of the gate sequence since $e^{-2\pi i \hat{I}_z}$ simply contributes a global phase of $-1$.  

Since the Holstein-Primakoff transformation maps $\hat{I}_z\mapsto I-a^\dagger a$, $U_{\mathrm{CR}}$ is the spin analog of the two-qubit gates used to entangle bosonic cat qubits and ancillary readout qubits~\cite{sun2014tracking,rosenblum2018fault,putterman2025hardware}. Such gates leverage a dispersive coupling of the form $\hat{\sigma}_z a^\dagger a$, which can similarly be used to impart a phase flip to the readout qubit conditioned on the cat qubit having an odd photon-number parity, $e^{-i\pi a^\dagger a}=-1$. For stabilized cat qubits (of which Kerr-cat qubits are an example), evolution due to the two-photon drive or dissipation providing the stabilization mechanism does not commute with evolution due to the dispersive coupling, requiring that the stabilization mechanism be turned off during the entangling gate~\cite{putterman2025hardware}. In the present case, the two-spin-flip terms for which $[\hat{I}_\pm^2,\hat{I}_z]\neq 0$ are rendered effectively negligible during the entangling operation by the large nuclear Zeeman splitting ${\sim}  A\gg Q$ coming from the presence of the hyperfine-coupled electron. The free-evolution dynamics of the nuclear spin during the entangling gate therefore consists of rigid rotations about the $\hat{I}_z$ axis. 

In general, both the size of $Q$ and the orientation of the principal axis may be affected by the presence of the hyperfine-coupled electron; in Ref.~\cite{fernandez2024navigating}, for instance, the quadrupolar splitting between $\hat{I}_z$ eigenstates (in the regime $\gamma_n B_z\gg Q$) was found to differ by 8 kHz for an ionized versus neutral ${}^{123}\mathrm{Sb}$ donor. Changes in the principal axis or in the value of the quadrupolar coupling are rendered negligible on the timescale of the entangling dynamics by the large effective magnetic field introduced by the hyperfine coupling, which suppresses all terms in the quadrupole Hamiltonian apart from the term $\propto \hat{I}_z^2$. The hierarchy of energy scales given in Eq.~\eqref{hierarchy} also ensures that the quadrupolar nonlinearity $Q\hat{I}_z^2$ does not lead to significant distortions of the nuclear-spin state due to one-axis twisting~\cite{gupta2024robust} over timescales $T\ll Q^{-1}$, with the same argument holding even if $Q$ is altered by the presence of the electron.

\begin{figure}
    \centering
    \includegraphics[width=\linewidth]{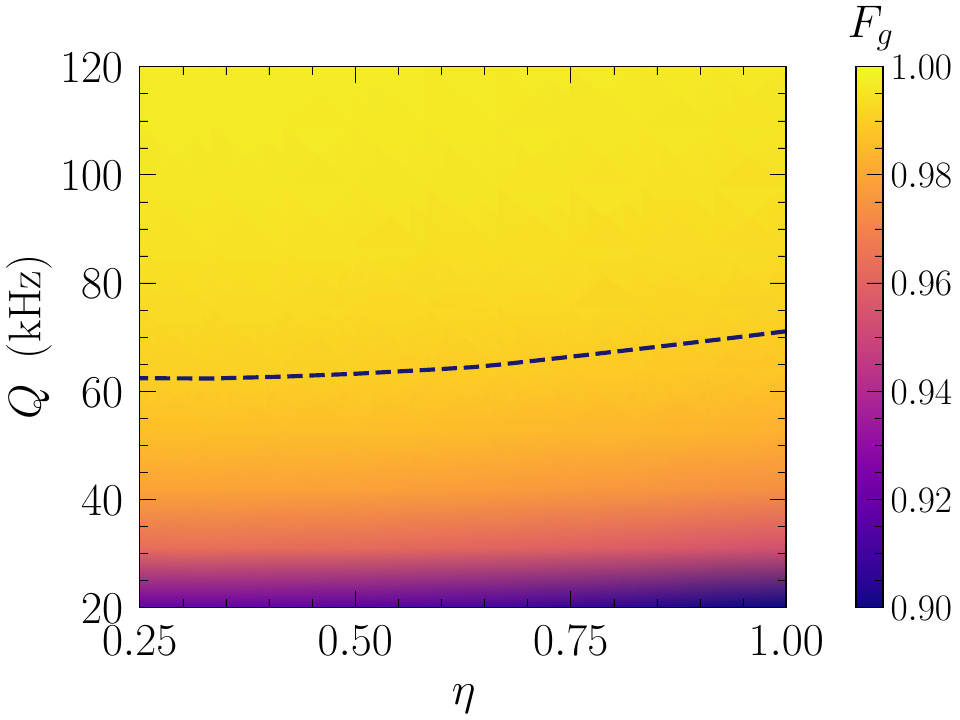}
    \caption{Average gate fidelity $F_g$ of the two-qubit gate $U_{\mathrm{CR}}$, accounting for errors due to non-secular terms in the hyperfine and quadrupole interactions. The black dashed line indicates a contour at which $F_g=0.99$. }
    \label{fig:gate-fid}
\end{figure}

To quantify residual errors due to the secular approximations made in writing Eqs.~\eqref{Htot}-\eqref{V}, we simulate the dynamics of the spin Kerr cat and electron spin under the full Hamiltonian $H_{\mathrm{exact}}$ obtained by including nonsecular terms in both the quadrupole and hyperfine interactions:
\begin{align}
    H_{\mathrm{exact}}&=H+\frac{\gamma_e}{2}B_0\hat{\sigma}_z+\frac{1}{2}A(t)\hat{\bm{\sigma}}\cdot \hat{\bm{I}},\\
    H&=-\gamma_n B_0\hat{I}_z+H_\mathrm{q}.
\end{align}
The magnetic-field strength $B_0$ is determined by the two parameters $Q$ and $\eta$ characterizing the quadrupole Hamiltonian $H_\mathrm{q}$ [Fig.~\ref{fig:cat-fidelities}(c)]. We take $A=101.5$ MHz and  $\gamma_n=5.55$ MHz to be the hyperfine coupling strength and gyromagnetic ratio of the ${}^{123}\mathrm{Sb}$ nuclear spin, and for simplicity, we model the time dependence of $A(t)$ as a trapezoidal ramp from $A_{\mathrm{min}}=0$ to $A_{\mathrm{max}}=A/2$ and back to $A_{\mathrm{min}}=0$, with symmetric rise and fall times of $T/4$ and a plateau of duration $T/2$, where $T=8\pi/(3A)$. This choice satisfies $\int_0^Tdt\:A(t)=\pi$ as required and yields a total interaction time of $T\simeq82$ ns.

Starting from some initial state $\ket{\psi}_0$, we evaluate the final time-evolved state $\ket{\psi_{\mathrm{true}}}=\ket{\psi}_{T}$ of the electron spin and nuclear spin by integrating the Schr\"odinger equation up to a final time $t=T$. We then compute the fidelity of $\ket{\psi_{\mathrm{true}}}$ relative to the ideal final state $\ket{\psi_{\mathrm{ideal}}}$ obtained by applying $U_{\mathrm{CR}}$ [Eq.~\eqref{U-CROT}] to $\ket{\psi}_0$, accounting for the free-evolution phases acquired by the electron spin. (Recall that $U_{\mathrm{CR}}$ was given in an interaction picture defined with respect to the decoupled Hamiltonian $H_0$.) The free evolution of the nuclear spin can also be compensated by accounting for precession of $\ket{0}$ and $\ket{1}$ at the spin-Kerr-cat qubit splitting, but the effects on the fidelity are minimal since $A\gg Q$. The results in Fig.~\ref{fig:gate-fid} are not visibly affected in the case where free evolution of the nuclear spin during the gate remains uncompensated.

Finally, we compute an average two-qubit gate fidelity $F_g$ by averaging the state fidelity obtained in the manner described above over the two-qubit Haar measure $d\psi$:
\begin{equation}
    F_g=\int d\psi\:\vert \langle \psi_{\mathrm{true}}\vert \psi_{\mathrm{ideal}}\rangle\vert^2.
\end{equation}
The average gate fidelity is plotted as a function of $Q$ and $\eta$ in Fig.~\ref{fig:gate-fid}. Values of $F_g=99\%$ (indicated by the dashed contour) are attainable provided $Q$ can be made sufficiently large. Although the quadrupole splittings measured with ${}^{123}\mathrm{Sb}$ to date~\cite{asaad2020coherent,fernandez2024navigating,yu2025schrodinger} do not reach 60 kHz and do not correspond exactly to the quadrupole coupling $Q$, an enhancement of the electric-field gradient by a factor of $\approx 2$ would likely be sufficient for achieving $99\%$ fidelity. High gate fidelities can also be achieved for values of $\eta$ for which the spin-cat approximation does not hold to a high degree of accuracy. However, since the spin-cat approximation forms the basis of the initialization procedure given in Sec.~\ref{sec:initialization}, high asymmetry parameters $\eta\gtrsim0.75$ remain desirable.

Since $A(t)$ depends on the overlap of the instantaneous orbital state $\ket{-(t)}$ with the donor orbital ground state [Eq.~\eqref{hf-2}], charge noise will generally degrade the gate fidelity $F_g$ shown in Fig.~\ref{fig:gate-fid}. In terms of the dot-donor detuning $\varepsilon$ and tunnel coupling $t_\mathrm{c}$, this overlap can be written as $A(t)=(A/4)\left[1-\varepsilon(t)/\Omega_{\mathrm{c}}(t)\right]$, where $\Omega_\mathrm{c}(t)=\sqrt{\varepsilon^2(t)+4t_c^2}$ is the instantaneous splitting of the effective charge qubit. Since charge noise couples to electrostatic potentials, we assume that charge noise results predominantly in variations $\varepsilon(t)\rightarrow\varepsilon(t)+\delta\varepsilon(t)$ of the detuning ramp. Such variations will lead to over- or under-rotations relative to the target operation by some random angle $\delta\theta$ sampled from a distribution. Under the assumption that the detuning noise $\delta\varepsilon(t)\simeq \delta \varepsilon$ is quasistatic over the duration $T$ of the gate with variance $\langle\delta\varepsilon^2\rangle=\sigma_\varepsilon^2$, the standard deviation $\sigma_{\delta\theta}$ of this distribution can be expressed as
\begin{equation}\label{rotation-error}
    \sigma_{\delta\theta}\simeq\frac{A\sigma_{\varepsilon}}{4}\int_0^Tdt\:[\Omega_{\mathrm{c}}(t)]^{-1},
\end{equation}
valid to leading order in $\delta \varepsilon/t_c$. 

Given some target profile $A(t)$, the corresponding ramp in $\varepsilon(t)$ can be found as $\varepsilon(t)=t_c[1-2\lambda(t)]/[\lambda(t)\bar{\lambda}(t)]^{1/2}$, where $\lambda(t)=2A(t)/A$ and $\bar{\lambda}(t)=1-\lambda(t)$.  For the trapezoidal $A(t)$ used above, $\int_0^Tdt\:[\Omega_\mathrm{c}(t)]^{-1}\approx 0.2 T/t_\mathrm{c}$ with $T\approx 82$ ns. Assuming $t_\mathrm{c}=2.5$ GHz and $\sigma_{\varepsilon}=1\:\mu$eV~\cite{mielke2021nuclear}, Eq.~\eqref{rotation-error} then gives $\sigma_{\delta\theta}\simeq 0.04$. Since the gate infidelity scales quadratically in $\delta\theta$, this in turn corresponds to a typical gate infidelity of 0.16\% due to charge noise. The infidelity can likely be reduced by choosing a different pulse shape, which was not optimized in this work.

In a realistic setup, it is likely that spin Kerr-cat qubits will not interact significantly due to a small magnetic-dipole coupling between nearest-neighbor nuclei. Two-qubit gates between spin Kerr-cat qubits could then be realized by entangling the spin Kerr cats with a mobile electron that can hop on and off the donor potentials:  Given $U_{\mathrm{CR}}$, a two-qubit gate could be realized by sequentially entangling an electron with each nuclear spin, and then measuring the spin of the electron (Fig.~\ref{fig:setup}). The fidelity of such a gate can be estimated as $(2F_g-1)-\epsilon_{\mathrm{s}}-\epsilon_{\mathrm{m}}$, where $\epsilon_{\mathrm{s}}$ and $\epsilon_{\mathrm{m}}$ are errors impacting the electron during shuttling and readout. High-fidelity spin shuttling is now routinely achieved in silicon; in Ref.~\cite{de2025high}, for instance, spin-state-preserving shuttling over distances of 10 $\mu$m was achieved in 200 ns and with fidelities of 99.5\%. Although readout fidelities tend to lag behind gate fidelities in spin-qubit devices, readout fidelities of 99\% have been demonstrated using, e.g., gate-based reflectometry~\cite{connors2020rapid}. 

In an architecture where spin Kerr cats are instead used as data qubits in a quantum error-correcting code, two-qubit gates between spin Kerr cats are not needed. The parity checks needed for syndrome extraction could be performed using the two-qubit gate $U_{\mathrm{CR}}$, for which the parameter regime needed for 99\% fidelity was given in Fig.~\ref{fig:gate-fid}. Arbitrary connectivity could then be achieved by routing the electron-spin ancilla qubits as desired, up to the distances permitted by the electron-spin coherence time.

\subsubsection{Readout}\label{sec:readout}

Nuclear-spin readout is most commonly realized by allowing an electron to tunnel onto the donor, flipping its state conditioned on the state of the nuclear spin, and then using energy-selective tunneling~\cite{elzerman2004single} to read out the state of the electron. This approach requires energetically well-separated electron-spin states, as may be obtained in the presence of a large magnetic field ($\gtrsim 1$ T in Ref.~\cite{morello2010single}). 

Since the spin Kerr-cat encoding operates at lower field strengths (Sec.~\ref{sec:spin-kerr-cat}), an approach more amenable to the regime considered here is to take advantage of the hyperfine interaction between the nuclear spin and an electron that hops on and off the donor potential [Eq.~\eqref{V}]. By preparing the electron in a known spin state, e.g.~$\ket{+}_x$, a single-shot readout of the spin Kerr-cat qubit could be realized using the entangling operation $U_{\mathrm{CR}}$ introduced in Sec.~\ref{sec:coupling-e-spin}: Given some state $\ket{\psi}=\alpha\ket{0}+\beta\ket{1}$ of the spin Kerr-cat qubit, $U_{\mathrm{CR}}$ can be used to map the $\hat{I}_z$ parity of the nuclear spin onto the $Y$-basis eigenvalue of the electron spin according to
\begin{equation}
    U_{\mathrm{CR}}\ket{\psi}\ket{+}_x= e^{-i\frac{\pi}{2}\hat{I}_z}(\alpha\ket{0}\ket{-}_y+\beta\ket{1}\ket{+}_y).
\end{equation}
A $Y$-basis measurement of the electron then amounts to a $Z$-basis measurement of the spin Kerr-cat qubit. Following such a measurement, free evolution of the nuclear spin in the presence of a donor-bound electron held in state, e.g., $\ket{\downarrow}$ could be used to compensate the rotation about $\hat{I}_z$ that takes the nuclear spin out of the spin Kerr-cat subspace.

\subsubsection{Initialization}\label{sec:initialization}

Given some arbitrary initial state, initialization of the nuclear spin in the spin Kerr-cat subspace can be realized in three steps: (1) preparation of the fully polarized spin state $\ket{I,I}$, (2) rotation about the $\hat{I}_y$ axis to prepare the spin coherent state $\ket{\Theta_0,0}$, and (3) measurement of the nuclear spin's $\hat{I}_z$ parity. This procedure assumes that the spin-cat approximation to the true nuclear-spin eigenstates holds to a high fidelity ($\gtrsim 99\%$, cf.~Fig.~\ref{fig:cat-fidelities}), in which case initialization can be performed with $\mathrm{SU}(2)$ control and measurement only. If the asymmetry parameter $\eta$ is too low for achieving high initialization fidelities using this procedure, then state preparation could be improved by combining steps (1)-(3) with further finetuning enabled by $\mathrm{SU}(2I+1)$ control. 

For an electron bound to the donor, the electron-spin splitting depends on the projection of the nuclear spin along the $\hat{I}_z$ axis [Eq.~\eqref{Htot}]. By preparing the electron in state $\ket{\downarrow}$ and applying an ESR pulse that is resonant conditioned on a particular $\hat{I}_z$ eigenstate, a measurement of the electron can then distinguish that eigenstate from the rest of the nuclear-spin Hilbert space~\cite{pla2013high,asaad2020coherent}. Cycling through all possible ESR frequencies then enables a projective measurement of the nuclear spin in the $\hat{I}_z$ basis. Following this initial measurement, the fully polarized state $\ket{I,I}$ could be prepared in the presence of a donor-bound electron using successive $\pi$ pulses between adjacent levels~\cite{asaad2020coherent,fernandez2024navigating,yu2025schrodinger}.  Recently, an adaptive readout protocol~\cite{vaartjes2026maximizing} has demonstrated improvements in the fidelity of such measurements by reducing the impact of deviations from the ideal case of quantum non-demolition readout [resulting from nonsecular corrections to Eq.~\eqref{Htot}]. 

Given $\ket{I,I}$, the nuclear spin should next be rotated around the $\hat{I}_y$ axis to produce the spin coherent state
\begin{equation}
    \ket{\Theta_0,0}=\frac{1}{2}\sum_{\sigma=\pm} \mathcal{N}_\sigma^{-1}(\Theta_0)\ket{\Theta_0,0}_\sigma.
\end{equation}
This operation could be performed in the presence of a donor-bound electron~\cite{fernandez2024navigating,yu2025schrodinger}, or for an ionized donor using a multi-frequency generalization of the NMR driving discussed in Sec.~\ref{sec:single-qubit}. In the latter case, the Hamiltonian $H$ of the nuclear spin is given by Eq.~\eqref{nmr-hamiltonian} with the AC magnetic field replaced by $B_{\mathrm{ac}}(t)=B_\perp\sum_{i}\cos{(\omega_i t)}$. With $\ket{m}$ and $\epsilon_m$ denoting the eigenstates and eigenvalues of the undriven spin, we transform $H$ to the generalized rotating frame~\cite{leuenberger2003grover,gupta2024robust}, defined by the unitary $U_{\mathrm{rot}}=\mathrm{diag}(e^{-i\epsilon_{2I+1}t},\dots, e^{-i\epsilon_0t})$, giving
\begin{equation}\label{gen-rot}
    \tilde{H}(t)=\gamma_n B_\perp\sum_{i}\cos{(\omega_i t)}\sum_{mn}I_{mn}e^{i\epsilon_{mn}t}\ketbra{m}{n}.
\end{equation}
Here, $I_{mn}=\langle m\vert \hat{I}_\perp\vert n\rangle$ is a matrix element that is nonzero only for eigenstates of opposite parity, and  $\epsilon_{mn}=\epsilon_m-\epsilon_n$. Noting that $\hat{I}_\perp=\sum_{mn}I_{mn}\ketbra{m}{n}$, the drive frequencies $\omega_i$ should be chosen to match  the $(I+1/2)^2$ transition frequencies for which $I_{mn}\neq 0$. Within a rotating-wave approximation valid for $\gamma_n B_\perp\ll \Delta=\mathrm{min}_i\:\omega_i$, Eq.~\eqref{gen-rot} can then be approximated as
\begin{equation}
    \tilde{H}\simeq \frac{\gamma_n B_\perp}{2}\hat{I}_\perp.
\end{equation}

If  $\hat{I}_\perp=\hat{I}_y$, then the ability to prepare $\ket{\Theta_0,0}$ follows directly. However, for an arbitrary orientation of the AC magnetic field in the $xy$ plane, the state $\ket{I,I}$ will instead be rotated into a different spin coherent state given by
\begin{equation}
    \tilde{U}(t)\ket{I,I}\simeq\ket{\Theta(t),\pi/2-\theta},
\end{equation}
where $\tilde{U}(t)=e^{-i \tilde{H}t}$ and $\Theta(t)=\gamma_n B_\perp t$. A misalignment $\theta\neq \pi/2$ due to driving about $\hat{n}\neq \hat{y}$ can be corrected by allowing the spin to precess in the presence of a donor-bound electron in state $\ket{\uparrow}$ for a time $t_0$ satisfying $(1/2)\int_0^{t_0}dtA(t)=\theta-\pi/2$. The final spin coherent state is then $\ket{\Theta_0,0}$ as intended.

The final $\hat{I}_z$ parity measurement can be realized using the strategy laid out in Sec.~\ref{sec:readout}: An electron initially in state $\ket{+}_x$ is loaded on and off the donor potential, so that the electron and nuclear spins undergo the entangling gate $U_{\mathrm{CR}}$. (If an axis correction was required in the previous step due to having $\hat{n}\neq\hat{y}$, then the electron used to rotate the spin coherent state can be left on the donor. In this case, a Hadamard can be used to flip the electron state from $\ket{\uparrow}\mapsto\ket{+}_x$ at time $t=t_0$.) This interaction yields the entangled state
\begin{align}
\begin{aligned}
    U_{\mathrm{CR}}\ket{\Theta_0,0}\ket{+}_x&=\lambda\ket{\Theta_0,\pi/2}_+\ket{-}_y\\
    &+\sqrt{1-\lambda^2}\ket{\Theta_0,\pi/2}_-\ket{+}_y,
\end{aligned}
\end{align}
where $\lambda=\sqrt{[1+\gamma(\Theta_0)]/2}$. A measurement of the electron in the $\ket{\pm}_y$ basis then projects the nuclear spin onto an (approximate) spin Kerr-cat basis state [cf.~Eq.~\eqref{spin-kerr-cat}]. Since the outcome is probabilistic, the preparation of a fixed initial state may require feedback in the form of an $X$ gate applied whenever the post-measurement state is of the wrong parity. This gate can be applied in software (i.e.,~without physical driving and in zero time) via a Pauli frame update. 

Notably, the action of $U_{\mathrm{CR}}$ on $\ket{\Theta,\phi}_\pm$ is identical in form to that given in Eqs.~\eqref{cz-1}-\eqref{cz-2}. This follows directly from the fact that these states are also eigenstates of $\hat{\Pi}$. Since the spin cat states $\ket{\Theta_0,\pi/2}_\pm$ are only approximate eigenstates of the nuclear-spin Hamiltonian, the initialization fidelity is controlled by the fidelity of the spin-cat approximation (Fig.~\ref{fig:cat-fidelities}), which exceeds 99\% for $\eta\gtrsim 0.75$. For lower values of $\eta$, knowledge of the exact eigenstates (which can be obtained from exact diagonalization) could then be used to redistribute amplitude between $I_z$ eigenstates using $\mathrm{SU}(2I+1)$ control as needed.

\section{Discussion and Conclusion}\label{sec:conclusion}

In this work, we have presented a ``spin Kerr-cat'' encoding that leverages quadrupolar interactions appearing for high-spin nuclei in the presence of an electric-field gradient to achieve a reduced sensitivity to dephasing noise. For an external magnetic field applied along the principal axis of the quadrupolar tensor, the nuclear-spin Hamiltonian acquires a $\mathbb{Z}_2$ symmetry that makes the spin's $\hat{I}_z$ parity a good quantum number, analogous to the definite photon-number parity characterizing the basis states of the Kerr-cat qubits used in circuit QED. The spin Kerr-cat qubit is then given by the lowest parity doublet of the nuclear-spin Hamiltonian, with basis states well approximated by spin cat states for sufficiently high asymmetry parameters. For nuclei in highly asymmetric electric-field gradients, the cat-state approximation would then allow the nuclear spin to be initialized in the spin-Kerr-cat subspace to a high fidelity using projective measurements and SU(2) rotations rather than full SU($2I+1$) control.

We propose strategies for manipulation and readout based on parity-conditioned operations on electron spins. In particular, we find using realistic parameters for ${}^{123}\mathrm{Sb}$  that a two-qubit gate between a spin Kerr-cat qubit and an electron-spin qubit could be performed with 99\% fidelity given only modest improvements in the quadrupole coupling $Q$ relative to the values measured  to date~\footnote{Since the magnetic field in Refs.~\cite{asaad2020coherent,fernandez2024navigating, yu2025schrodinger} is not aligned with
the quadrupole tensor’s principal axis, the quadrupole
splittings reported in these works are not exactly equal
to $Q$. We nonetheless take the values reported here as
reasonable order-of-magnitude estimates for $Q$.}. Given the availability of such an entangling gate between nuclear and electron spins, an interesting future direction would be to consider the use of spin Kerr-cat qubits as quantum memories for electron-spin qubits, taking advantage of both the long decoherence times of nuclear spins and the faster manipulation times of electrons. 

We have calculated the spin Kerr-cat dephasing time $T_{2,\mathrm{cat}}^*$ under a model of $1/f$ noise, as applicable to charge-noise-induced fluctuations of the quadrupole tensor or to magnetic noise due to paramagnetic surface impurities. Based on measured parameters, we estimate that second-scale coherence times could likely be achieved. We have also analyzed a relaxation channel specific to this encoding, resulting from misalignment or drift of the magnetic field in the principal axis system of the electric-field gradient at the site of the nuclear spin. Such drift breaks the intended $\mathbb{Z}_2$ symmetry of the nuclear-spin Hamiltonian and can originate from random fluctuations in the locations of nearby two-level charge fluctuators. For realistic TLF densities and implantation depths, we estimate bitflip times exceeding a hundred seconds, with charge-noise-induced leakage out of the qubit subspace occurring on even longer timescales. 

Spectroscopic characterization, combined with a sweep of the magnetic-field direction, would be required in order to find the principal axis of the quadrupole tensor and to extract the values of $Q$ and $\eta$~\cite{mourik2018exploring}. While existing measurements performed in the high-field regime can be used to estimate the typical size of $Q$~\cite{asaad2020coherent,fernandez2024navigating,yu2025schrodinger}, the magnetic fields used in these experiments are not aligned with the $z$ axis, and the reported quadrupolar splittings therefore combine information about $Q$ and $\eta$. As shown in this work, the spin Kerr-cat encoding requires a high asymmetry parameter $\eta\gtrsim 0.75$ for the spin-cat approximation to the true nuclear-spin eigenstates to be quantitatively accurate, and consequently, for reaching high initialization fidelities with $\mathrm{SU}(2)$ rotations and measurement only. Finding strategies for controlling the value of $\eta$ is beyond the scope of this work.

In Ref.~\cite{asaad2020coherent}, the electric-field gradient at the location of the nuclear spin is attributed to lattice strain resulting from the different thermal contractions of silicon and aluminum upon cooling to cryogenic temperatures.  Since both the strength and direction of the applied magnetic field must be tuned in relation to the quadrupole tensor, scaling up this approach would require a high level of uniformity of the strain fields near different qubits, combined with precise control over donor placement. Non-uniformity of the strain field could, for instance, lead to misalignment of the magnetic field in relation to the principal axes of at least some of the spins' quadrupole tensors. Such misalignment would introduce intrinsic (i.e.~not noise-induced) relaxation errors whose size could be estimated using the formalism established in Sec.~\ref{sec:relaxation}, providing a direct link between qubit properties and tolerable levels of strain non-uniformity across different sites. Control over the electric-field gradient at the location of each qubit could in principle be achieved through strain engineering enabled by nanometer-scale piezoelectric actuators, which have been demonstrated experimentally in silicon-based heterostructures~\cite{dreher2011electroelastic}. Finally, we remark that the need to tune the local environment of each qubit in order to realize favorable properties is not a feature unique to the proposal presented here, and that sweet spots defined by magnetic-field orientations also exist for, e.g., hole spins, requiring the ability to tune system parameters so that several qubits can simultaneously be set at a sweet spot~\cite{bosco2021fully,bassi2025optimal,carballido2025compromise}.

\section*{Acknowledgments}
Z.~M.~M. thanks W.~A.~Coish for useful discussions. We also thank A.~Morello and V.~Mourik for insights on feasibility following the original version of this manuscript. This work was supported by the Swiss National Science Foundation, NCCR SPIN (Grant No.~225153). D.L.~acknowledges the Deanship of Research and the Quantum Center at KFUPM for the support received under Grant no.~CUP25102 and no.~INQC2600, respectively.

\appendix

\section{Relaxation due to phonons}\label{sec:phonons}

In this Appendix, we estimate the qubit bitflip rate due to coupling of the nuclear spin to phonons. In the context of this calculation, we take the lab-frame coordinate system $\{x',y',z'\}$ to correspond to the silicon crystallographic axes. The unprimed coordinate system corresponds as before to the principal axis system of the EFG tensor. 

Phonons couple to the nuclear spin through the gradient-elastic tensor~\cite{asaad2020coherent}, producing a perturbation $\bm{\delta \mathcal{V}}$ to the EFG tensor given by
\begin{equation}\label{elastic-gradient}
    \delta\mathcal{V}_{\alpha\beta}=\sum_{\gamma\delta}S_{\alpha\beta\gamma\delta} \delta\epsilon_{\gamma\delta},
\end{equation}
where $S_{\alpha\beta\gamma\delta}$ is an element of the gradient-elastic tensor and $\delta\epsilon_{\gamma\delta}$ is a perturbation to the strain tensor $\bm\epsilon$ resulting from the phonon-induced lattice displacement. For $\mathrm{T}_d$ tetrahedral symmetry, relevant to bulk silicon, the gradient-elastic tensor has only two nonzero elements $S_{11}$ and $S_{44}$, and we can write Eq.~\eqref{elastic-gradient} in a more compact form given by 
\begin{equation}\label{elastic-gradient-vector}
    \underline{\bm{\delta \mathcal{V}}}= \begin{pmatrix}
        \bm{S_1} & \bm{0}\\
        \bm{0} & \bm{S_4}
    \end{pmatrix}\underline{\bm{\delta \epsilon}}.
\end{equation}
Here, $\underline{\bm{\delta \mathcal{V}}}$ and $\underline{\bm{\delta \epsilon}}$ are vectorized versions of  $\bm{\delta\mathcal{V}}$ and $\bm{\delta\epsilon}$ that include only their six independent elements:
\begin{align*}
    \underline{\bm{\delta \mathcal{V}}}&=\begin{pmatrix}
        \delta \mathcal{V}_{x'x'} & \delta\mathcal{V}_{y'y'} & \delta\mathcal{V}_{z'z'} & \delta\mathcal{V}_{y'z'} & \delta\mathcal{V}_{x'z'} & \delta\mathcal{V}_{x'y'}
    \end{pmatrix}^\top,\\
    \underline{\bm{\delta \mathcal{\epsilon}}}&=\begin{pmatrix}
        \delta \mathcal{\epsilon}_{x'x'} & \delta\mathcal{\epsilon}_{y'y'} & \delta\mathcal{\epsilon}_{z'z'} & \delta\mathcal{\epsilon}_{y'z'} & \delta\mathcal{\epsilon}_{x'z'} & \delta\mathcal{\epsilon}_{x'y'}
    \end{pmatrix}^\top.
\end{align*}
The matrices $\bm{S_1}$ and $\bm{S_4}$ relate $\underline{\bm{\delta \mathcal{V}}}$ to the normal and shear strains, respectively, and are themselves given by
\begin{align}
    \bm{S_1}&=\begin{pmatrix}
        S_{11} & -S_{11}/2 & -S_{11}/2\\
        -S_{11}/2 & S_{11} & -S_{11}/2\\
        -S_{11}/2 & -S_{11}/2 & S_{11}
    \end{pmatrix},\\
    \bm{S_4}&=\begin{pmatrix}
        2S_{44} & 0 & 0\\
        0 & 2S_{44} & 0\\
        0 & 0 & 2S_{44}
    \end{pmatrix}.
\end{align}
Based on density functional theory calculations, the values of $S_{11}$ and $S_{44}$ were estimated to be $S_{11}=2.4\times 10^{22}\;V/\mathrm{m}^2$ and $S_{44}=6.1\times 10^{22}\;V/\mathrm{m}^2$~\cite{asaad2020coherent}.

In terms of the lattice displacement vector $\bm{u}(\bm{r})$, the elements $\delta\mathcal{\epsilon}_{\alpha\beta}$ of the strain tensor are given by
\begin{equation}\label{strain-tensor}
    \delta\epsilon_{\alpha\beta}=\frac{1}{2}\left(\frac{\partial u_{\alpha}}{\partial \beta}+\frac{\partial u_\beta}{\partial\alpha}\right)\bigg\vert_{\bm{r}=0},
\end{equation}
where the derivatives are evaluated at the location of the nuclear spin, here taken to correspond to the origin. The displacement vector can in turn be expressed in terms of phononic raising and lowering operators $b_{\bm{q},\lambda},b_{\bm{q},\lambda}^\dagger$ as
\begin{equation}\label{phonon-displacement}
    \bm{u}(\bm{r})=\sum_{\bm{q},\lambda}\sqrt{\frac{\hbar}{2\rho V\omega_{\bm{q},\lambda}}}\hat{e}_{\bm{q},\lambda}e^{i\bm{q}\cdot\bm{r}}(b_{\bm{q},\lambda}+b_{-\bm{q},\lambda}^\dagger), 
\end{equation}
where $\rho$ is the silicon mass density, $V$ is the volume of the crystal, $\hat{e}_{\bm{q},\lambda}$ is the polarization vector for a phonon in branch $\lambda$ having wavenumber $\bm{q}$, and where $\omega_{\bm{q},\lambda}=v_{\lambda}\vert \bm{q}\vert$ is the frequency of such a phonon, expressible in terms of a branch-dependent speed of sound $v_\lambda$. We neglect any contribution due to optical phonons under the assumption that they are frozen out.

The transition rate $\Gamma_{10}$ is controlled by the spectral density $J_\perp(-\Delta)$ of the transverse magnetic-field fluctuations in the principal axis system of the perturbed EFG tensor. The transverse field fluctuations $\delta b_{i}$ ($i=x,y$) are in turn given by $\delta b_{i}=\gamma_n B_0 \beta_i$, where 
\begin{equation}
    \beta_i=\frac{\Omega_{i z}}{D_{zz}-D_{ii}}.
\end{equation}
Here, $\Omega_{iz}=[\bm{R}_0^\top\bm{\delta\mathcal{V}}\bm{R}_0]_{iz}$ is a matrix element of $\bm{\delta \mathcal{V}}$ expressed in the principal axis system of the unperturbed EFG tensor $\bm{\mathcal{V}}$. These matrix elements  could be evaluated directly from Eq.~\eqref{elastic-gradient-vector}, provided the orientation of the principal axes relative to the crystallographic axes was known. However, since the offset in coordinate systems provides at most $O(1)$ corrections, we can estimate the typical size of $\Omega_{iz}$ by neglecting the orthogonal rotation by $\bm{R}_0$, and by taking the typical size $\delta\mathcal{V}$  of the perturbation to be given by the typical size of the gradient-elastic tensor ($\approx 10^{22}\; V/\mathrm{m}^2$) multiplied by the typical size $\delta\epsilon$ of the perturbation to the strain tensor:
\begin{equation}
    \delta \mathcal{V}\sim \left(10^{22}\;V/\mathrm{m}^2\right)\delta \epsilon.
\end{equation}
A reasonable estimate of $\delta\epsilon$ is obtained by neglecting the dependence of $\delta\epsilon_{\alpha\beta}$ on the orientation of the phonons' polarization vectors and propagation directions in the crystal coordinate system. At this level of approximation, the spatial derivatives in Eq.~\eqref{strain-tensor} all contribute a factor of $q$ coming from the plane-wave term in Eq.~\eqref{phonon-displacement}, giving
\begin{equation}
    \delta \epsilon\sim \sum_{\bm{q},\lambda}\sqrt{\frac{\hbar}{2\rho V\omega_{\bm{q},\lambda}}}q(b_{\bm{q},\lambda}+b_{-\bm{q},\lambda}^\dagger).
\end{equation}

Time dependence of $\delta \mathcal{V}$ is generated by the phonon Hamiltonian 
\begin{equation}
    H_{\mathrm{ph}}=\sum_{\bm{q},\lambda}\omega_{\bm{q},\lambda}b_{\bm{q},\lambda}^\dagger b_{\bm{q},\lambda}.
\end{equation}
Noise resulting from phonon-driven lattice displacements can therefore be treated as stationary under the assumption that the initial state of the phonon bath commutes with $H_{\mathrm{ph}}$, as would be the case for a thermal state. We assume such a thermal state here, taking
\begin{align}
    &\langle b_{\bm{q},\lambda}b_{\bm{q}',\lambda'}\rangle=0,\\
    &\langle b_{\bm{q},\lambda}^\dagger b_{\bm{q}',\lambda'}\rangle=\delta_{\bm{q},\bm{q}'}\delta_{\lambda,\lambda'}\bar{n}(\omega_{\bm{q},\lambda}),
\end{align}
where $\langle\cdot \rangle=\mathrm{Tr}\{(\cdot) \bar{\rho}_{\mathrm{ph}}\}$ indicates an average with respect to the thermal state $\bar{\rho}_{\mathrm{ph}}\propto e^{-H_{\mathrm{ph}}/k_BT}$, and where $\bar{n}(\omega)$ is the Bose-Einstein distribution.

Since $\vert D_{zz}-D_{ii}\vert\geq 2I(2I-1)Q\hbar /(eq)$, the typical size of the correlation function $\langle \delta b_i(t)\delta b_j\rangle$ is then given by
\begin{align}\label{correlation-function-phonon}
\begin{aligned}
    \langle \delta b_i(t)\delta b_j\rangle \lesssim C^2\sum_{\bm{q},\lambda}&g_{\lambda}(\bm{q})\big[e^{i\omega_{\bm{q},\lambda}t}\bar{n}(\omega_{\bm{q},\lambda})\\&+e^{-i \omega_{\bm{q},\lambda}t}(\bar{n}(\omega_{\bm{q},\lambda})+1)\big],
\end{aligned}
\end{align}
where 
\begin{align}
    &C=\frac{ \gamma_n B_0}{Q} \frac{eq}{2I(2I-1)\hbar} \times 10^{22}\;\mathrm{V/m}^2,\\
    &g_{\lambda}(\bm{q})=\frac{\hbar q^2}{2\rho \omega_{\bm{q},\lambda}V}.
\end{align}
For simplicity, we assume that all phonons propagate with the same velocity, $v=v_\lambda$. The sum over $\lambda$ then contributes an overall factor of 3, which we drop in the order-of-magnitude estimates below.

A bound on $J_\perp(\omega)$ can be obtained as the Fourier transform of Eq.~\eqref{correlation-function-phonon}.  We convert the sum over $\bm{q}$ into an integral over frequency, 
\begin{equation}
    \sum_{\bm{q}}\rightarrow\int_0^\infty d\omega\:D(\omega),
\end{equation}
where $D(\omega)=(V/2\pi^2) \omega^2/v^3$ is the phonon density-of-states in three dimensions. At $\omega=-\Delta$, the value of the spectral density is then given by
\begin{equation}
    J_\perp (-\Delta)\lesssim C^2 D(\Delta)g(\Delta)\bar{n}(\Delta),
\end{equation}
where
\begin{equation}
    g(\omega)=\frac{\hbar (\omega/v)^2}{2\rho \omega V}.
\end{equation}
Notably, the result is independent of the volume $V$ of the crystal. Taking $\gamma_n B_0=10 Q$, $q=10^{-28}\;\mathrm{m}^2$, $\rho=2300\;\mathrm{kg}/\mathrm{m}^3$, $v=6500\;\mathrm{m/s}$, $\Delta=50\;\mathrm{kHz}$, and a temperature of 10 mK, we obtain a value of $\Gamma_{10}\propto J_\perp(-\Delta)\lesssim 10^{-22}\;\mathrm{Hz}$. At this temperature, we have a thermal phonon occupation of $\bar{n}(\Delta)\approx 4200$. The relaxation rate $\Gamma_{01}\propto J(\Delta)$ is therefore effectively equal to $\Gamma_{10}$. The estimate obtained here suggests that phonons are unlikely to dominate over charge noise as a source of qubit relaxation.

\section{EFG tensor due to a point charge}\label{sec:direct-tlf}

In this Appendix, we calculate the change in the EFG tensor at the site of the nuclear spin due to the displacement of a TLF located at an average distance $r$ from the nuclear spin. In contrast to the scenario discussed in the main text, where the change $\bm{\delta E}$ in the electric field due to the TLF is transduced into a change in the EFG via the electric-field response tensor, we consider here the EFG tensor $\bm{\mathcal{V}}_{\mathrm{TLF}}$ due to the electric field $\bm{E}_{\mathrm{TLF}}(\bm{r}_0)$ itself. This tensor can also be perturbed by a displacement of the TLF.

If the TLF is displaced from its original position by an amount $\bm{\delta r}$, then the perturbation $\bm{\delta\mathcal{V}}$ to $\bm{\mathcal{V}}_{\mathrm{TLF}}$ is given to leading order by
\begin{equation}\label{direct-efg}
    \bm{\delta\mathcal{V}}=(\bm{\delta r}\cdot \bm{\nabla}')\bm{\mathcal{V}}_{\mathrm{TLF}},
\end{equation}
where $\bm{\nabla}'$ denotes the gradient in some lab-frame coordinate system $\{x',y',z'\}$. The complete geometric dependence of $\bm{\delta\mathcal{V}}$ could be evaluated in principle using the expression $\bm{E}_{\mathrm{TLF}}(\bm{r_0})=(ke/r^2)\bm{\hat{r}}$ for the electric field of a point charge. For simplicity, however, we will only estimate the typical size of $\bm{\delta\mathcal{V}}$. Since $\bm{\mathcal{V}}_{\mathrm{TLF}}$ involves derivatives of $\bm{E}_{\mathrm{TLF}}$, the elements $\delta\mathcal{V}_{i',j'}$ of $\bm{\mathcal{\delta V}}$ [Eq.~\eqref{direct-efg}] will scale roughly like
\begin{equation}
    \delta\mathcal{V}_{i'j'}\approx \delta r\frac{ke}{r^4},
\end{equation}
where $\delta r$ gives the magnitude of the TLF displacement. The typical size of $\beta_i$ will then scale like
\begin{equation}\label{comparison-direct}
    Q\vert \beta_i\vert\lesssim \frac{1}{h}\bigg\vert\frac{ke^2q}{2I(2I-1)}\frac{\delta r}{r^4}\bigg\vert\approx 10^{-36} r^{-4}\mathrm{m}^4\;\mathrm{Hz}.
\end{equation}
Equation \eqref{comparison-direct} can be compared with the estimate given in Eq.~\eqref{comparison} for the contribution due to the electric-field response tensor. The smaller prefactor in Eq.~\eqref{comparison-direct}, combined with the more aggressive falloff with distance, suggests that this effect will be negligible in comparison. For $r=20$ nm, for instance, Eq.~\eqref{comparison-direct} predicts variations $\delta Q$ on the order of $10$ $\mu$Hz.

\section{Evaluation of the spectral density}\label{appendix-dephasing}

We evaluate the convolution $(S_Q\star S_Q)(\omega)$ [cf.~Eq.~\eqref{spectral-density}] for the case where $S_Q(\omega)$ is given by
\begin{equation}\label{1/f-spec}
    S_Q(\omega)=\begin{cases}
    \frac{\nu}{\vert \omega\vert},&\quad\omega_{\mathrm{IR}}\leq\vert\omega\vert\leq\omega_{\mathrm{UV}}\\
    0,&\quad \text{otherwise}
    \end{cases}.
\end{equation}
Since the integral $\int d\omega \:S_Q(\omega)$ diverges, we have introduced infrared and ultraviolet cutoffs $\omega_{\mathrm{IR}}$ and $\omega_{\mathrm{UV}}$ controlling the frequency range over which $S_Q(\omega)$ is taken to be finite. The infrared cutoff $\omega_{\mathrm{IR}}=2\pi/T_m$ is determined by the total time $T_m$ over which the noise is measured, which ultimately limits the accessible Fourier components of the noise spectral density. In practice, the true noise spectrum would have to fall off faster than $\vert \omega\vert^{-1}$ at high frequencies in order for the variance $\sigma_Q^2=\langle \delta Q^2\rangle$ of the noise to be finite. However, the ultraviolet cutoff and the details of the high-frequency noise ultimately play no role in the free-induction decay since spectral weight in the filter function $F(\omega,t)$ is concentrated at low frequencies. We nevertheless include an explicit dependence on $\omega_{\mathrm{UV}}$ for clarity.

With $S_Q(\omega)$ given by Eq.~\eqref{1/f-spec}, we can write $(S_Q\star S_Q)$ as 
\begin{equation}\label{conv-def}
    (S_Q\star S_Q)(\omega)=\int d\Omega \:I_\omega(\Omega),
\end{equation}
where the integrand $I_\omega(\Omega)$ is nonzero only when both $S_Q(\Omega)$ and $S_Q(\omega-\Omega)$ are nonzero:
\begin{equation}
    I_\omega(\Omega)=\begin{cases}
        \frac{\nu^2}{\vert \Omega\vert\vert\omega-\Omega\vert},\quad &\omega_{\mathrm{IR}}\leq\vert \Omega\vert,\vert\omega-\Omega\vert\leq \omega_{\mathrm{UV}}\\
        0,\quad &\text{otherwise}
    \end{cases}.
\end{equation}
Since $(S_Q\star S_Q)(\omega)$ is an even function of $\omega$, we focus on evaluating Eq.~\eqref{conv-def} for $\omega\geq 0$. We divide the full range of integration $\Omega\in (-\infty,\infty)$ into three regions $R_1$, $R_2$, and $R_3$ where $I_\omega(\Omega)$ can assume finite values, given respectively by $(-\omega_{\mathrm{UV}},-\omega_{\mathrm{IR}})$, $(\omega_{\mathrm{IR}},\omega-\omega_{\mathrm{IR}})$, and $(\omega+\omega_{\mathrm{IR}},\omega_{\mathrm{UV}}$). The second region is defined only for $\omega\geq 2\omega_{\mathrm{IR}}$ since for $\omega\in [0,2\omega_{\mathrm{IR}})$, the would-be singularities in $S_Q(\Omega)$ and $S_Q(\omega-\Omega)$ overlap. Integrating $I_\omega(\Omega)$ over each of these regions then gives
\begin{align*}
    \int_{R_1} d\Omega\:I_\omega(\Omega)&=\frac{\nu^2}{\omega}\left[\ln{\left(\frac{\omega+\omega_{\mathrm{IR}}}{\omega_{\mathrm{IR}}}\right)}-\ln{\left(\frac{\omega_{\mathrm{UV}}+\omega}{\omega_{\mathrm{UV}}}\right)}\right],\\
    \int_{R_2} d\Omega\:I_\omega(\Omega)&=\frac{2\nu^2}{\omega}\ln{\left(\frac{\omega-\omega_{\mathrm{IR}}}{\omega_{\mathrm{IR}}}\right)},\\
    \int_{R_3} d\Omega\:I_\omega(\Omega)&=\frac{\nu^2}{\omega}\left[\ln{\left(\frac{\omega+\omega_{\mathrm{IR}}}{\omega_{\mathrm{IR}}}\right)}+\ln{\left(\frac{\omega_{\mathrm{UV}}-\omega}{\omega_{\mathrm{UV}}}\right)}\right].
\end{align*}

In the expression for $\Lambda_2(t)$ [Eq.~\eqref{variance}], the convolution $(S_Q\star S_Q)(\omega)$ is weighted by the filter function $F(\omega,t)=t^2\mathrm{sinc}^2(\omega t/2)$. This filter function  has most of its weight concentrated at low frequencies, effectively acting as a low-pass filter for $\vert\omega\vert\lesssim 1/t$. At high frequencies, $F(\omega,t)$ falls off like $\omega^{-2}$ and therefore suppresses the influence of higher-frequency components in $S_Q\star S_Q$, making the free-induction decay largely independent of the specific microscopic mechanisms bounding the total integrated spectral weight in actuality. Since only frequencies $\vert \omega\vert\lesssim 1/t\ll\omega_{\mathrm{UV}}$ contribute appreciably to the integral, we restrict our attention to frequencies $\omega\ll \omega_{\mathrm{UV}}$, giving
\begin{align}
\begin{aligned}\label{conv-2}
    (\Delta'')^2(S_Q\star S_Q)(\omega)&\simeq \frac{2(\nu\Delta'')^2}{\omega}\bigg[\ln{\left(\frac{\omega+\omega_{\mathrm{IR}}}{\omega_{\mathrm{IR}}}\right)}\\&+\Theta_{2\omega_{\mathrm{IR},}}(\omega)\ln{\left(\frac{\omega-\omega_{\mathrm{IR}}}{\omega_{\mathrm{IR}}}\right)}\bigg],
\end{aligned}
\end{align}
valid up to corrections $O((\nu\Delta'')^2/\omega_{\mathrm{UV}})$ that we henceforth neglect. In Eq.~\eqref{conv-2}, $\Theta_{\omega_0}(\omega)$ is a Heaviside function that is equal to 1 when $\omega-\omega_0\geq 0$ and equal to 0 otherwise. 

We denote by $\Lambda_{2,Q}(t)$ the contribution to $\Lambda_2(t)$ [Eq.~\eqref{variance}] coming from the term $\propto (\Delta'')^2$ in Eq.~\eqref{spectral-density}. Consistent with the low-pass-filter character of $F(\omega,t)$, we can evaluate $\Lambda_{2,Q}(t)$  by replacing the upper limit of integration  by $1/t$, giving
\begin{equation}
    \Lambda_{2,Q}(t)\simeq \frac{1}{4\pi}(\Delta'')^2t^2\int_{0}^{1/t}\frac{d\omega}{\pi} (S_Q\star S_Q)(\omega),
\end{equation}
where we have set $\mathrm{sinc}^2(x)\simeq 1$ for small $x$ and used the fact that $F(\omega,t)$ and $(S_Q\star S_Q)(\omega)$ are both even functions of $\omega$ to re-write the range of integration over positive frequencies only. Evaluating the integral then gives
\begin{align}
\begin{aligned}
    \Lambda_{2,Q}(t)&\simeq \frac{(\nu \Delta'')^2}{2\pi^2}t^2\bigg[\ln{\left(\frac{1}{\omega_{\mathrm{IR}}t}-1\right)\ln{\left(\frac{1}{\omega_{\mathrm{IR}}t}\right)}}\\
    &+\mathrm{Li}_2\left(1-\frac{1}{\omega_{\mathrm{IR}}t}\right)-\mathrm{Li}_2\left(\frac{-1}{\omega_{\mathrm{IR}}t}\right)+\frac{\pi^2}{12}\bigg],
\end{aligned}
\end{align}
where $\mathrm{Li}_2(x)$ denotes the polylogarithm function. No further approximations have been made at this stage. Taylor expanding in $\omega_{\mathrm{IR}}t\ll 1$ then allows us to write
\begin{equation}
    \Lambda_{2,Q}(t)\simeq \frac{(\nu \Delta'')^2}{2\pi^2} t^2\left[\ln^2\left(\frac{1}{\omega_{\mathrm{IR}}t}\right)+\frac{\pi^2}{12}+O((\omega_{\mathrm{IR}}t)^2)\right],
\end{equation}
where, provided $\ln{(1/\omega_{\mathrm{IR}}t)}\gg 1$, we are justified in neglecting the constant term $\pi^2/12$.

\section{Time-dependent hyperfine coupling under an adiabatic charge transfer}\label{sec:time-dep-hyperfine}

In this Appendix, we derive the time-dependent hyperfine coupling $A(t)$ of the electron spin and nuclear spin under an adiabatic sweep $\varepsilon(t)$ of the dot-donor detuning. The Hamiltonian describing the electron's charge degree of freedom is given by $H_\mathrm{c}(t)=\varepsilon(t)\hat{\nu}_z/2+t_c\hat{\nu}_x$ [Eq.~\eqref{charge-hamiltonian}], while the hyperfine coupling between the nuclear spin and donor-bound electron is taken to be described by $H_{\mathrm{HF}}=(A/4)(\hat{\bm{\sigma}}\cdot\hat{\bm{I}})(\mathbbm{1}+\hat{\nu}_z)$ [Eq.~\eqref{hf-1}], where $(\mathbbm{1}+\hat{\nu}_z)/2$ is a projector onto the orbital ground state of the donor potential.

Denoting by $\ket{\pm(t)}$ and $\lambda_\pm(t)$ the instantaneous eigenstates and eigenvalues of $H_{\mathrm{c}}(t)$, we transform the total charge Hamiltonian $H_{\mathrm{tot}}=H_\mathrm{c}+H_{\mathrm{HF}}$ to the instantaneous adiabatic eigenbasis via the unitary transformation
\begin{equation}
    \tilde{H}_\mathrm{tot}=U H_\mathrm{tot}U^\dagger-i U\dot{U}^\dagger,
\end{equation}
where $U=\sum_{s=\pm}\ketbra{s}{s(t)}$. We treat the evolution within an adiabatic approximation under the assumption that the usual adiabaticity condition holds, $\vert \langle {-}(t)\vert {\dot{+}}(t)\rangle\vert\ll 2t_\mathrm{c}$, giving $\tilde{H}_\mathrm{tot}\simeq \tilde{H}_{\mathrm{a}}=U H_\mathrm{tot}U^\dagger$ with 
\begin{equation}
    \tilde{H}_{\mathrm{a}}=\sum_{s=\pm}\lambda_s(t)\ketbra{s}+\frac{A}{2}(\hat{\bm{\sigma}}\cdot\hat{\bm{I}})\sum_{s,s'=\pm}\xi_{ss'}(t)\ketbra{s}{s'},
\end{equation}
where $\xi_{ss'}(t)=(1/2)\langle s(t)\vert (\mathbbm{1}+\hat{\nu}_z)\vert s'(t)\rangle$ is a matrix element of the donor orbital projector with respect to the instantaneous charge eigenstates.

Within a rotating-wave approximation valid for $A\ll t_\mathrm{c}$, we can neglect terms in the hyperfine interaction with $s\neq s'$. This translates to the assumption that the hyperfine interaction is not strong enough to drive Landau-Zener transitions. Projecting $\tilde{H}_\mathrm{a}$ onto the lower orbital level $\ket{-}$ then gives an effective Hamiltonian $H_{\mathrm{eff}}$ describing the spin and orbital degrees of freedom of an electron that remains in the lower instantaneous orbital eigenstate throughout the detuning sweep:
\begin{equation}
    H_{\mathrm{eff}}=\lambda_-(t)\ketbra{-}+\frac{A(t)}{2}(\hat{\bm{\sigma}}\cdot\hat{\bm{I}})\ketbra{-}{-}.
\end{equation}
The time dependence of the hyperfine coupling is given by $A(t)=A\xi_{--}(t)$ and is inherited from the time-dependent overlap of $\ket{-(t)}$ with the donor ground state.


\begin{thebibliography}{79}%
\makeatletter
\providecommand \@ifxundefined [1]{%
 \@ifx{#1\undefined}
}%
\providecommand \@ifnum [1]{%
 \ifnum #1\expandafter \@firstoftwo
 \else \expandafter \@secondoftwo
 \fi
}%
\providecommand \@ifx [1]{%
 \ifx #1\expandafter \@firstoftwo
 \else \expandafter \@secondoftwo
 \fi
}%
\providecommand \natexlab [1]{#1}%
\providecommand \enquote  [1]{``#1''}%
\providecommand \bibnamefont  [1]{#1}%
\providecommand \bibfnamefont [1]{#1}%
\providecommand \citenamefont [1]{#1}%
\providecommand \href@noop [0]{\@secondoftwo}%
\providecommand \href [0]{\begingroup \@sanitize@url \@href}%
\providecommand \@href[1]{\@@startlink{#1}\@@href}%
\providecommand \@@href[1]{\endgroup#1\@@endlink}%
\providecommand \@sanitize@url [0]{\catcode `\\12\catcode `\$12\catcode
  `\&12\catcode `\#12\catcode `\^12\catcode `\_12\catcode `\%12\relax}%
\providecommand \@@startlink[1]{}%
\providecommand \@@endlink[0]{}%
\providecommand \url  [0]{\begingroup\@sanitize@url \@url }%
\providecommand \@url [1]{\endgroup\@href {#1}{\urlprefix }}%
\providecommand \urlprefix  [0]{URL }%
\providecommand \Eprint [0]{\href }%
\providecommand \doibase [0]{https://doi.org/}%
\providecommand \selectlanguage [0]{\@gobble}%
\providecommand \bibinfo  [0]{\@secondoftwo}%
\providecommand \bibfield  [0]{\@secondoftwo}%
\providecommand \translation [1]{[#1]}%
\providecommand \BibitemOpen [0]{}%
\providecommand \bibitemStop [0]{}%
\providecommand \bibitemNoStop [0]{.\EOS\space}%
\providecommand \EOS [0]{\spacefactor3000\relax}%
\providecommand \BibitemShut  [1]{\csname bibitem#1\endcsname}%
\let\auto@bib@innerbib\@empty
\bibitem [{\citenamefont {Steger}\ \emph {et~al.}(2012)\citenamefont {Steger},
  \citenamefont {Saeedi}, \citenamefont {Thewalt}, \citenamefont {Morton},
  \citenamefont {Riemann}, \citenamefont {Abrosimov}, \citenamefont {Becker},\
  and\ \citenamefont {Pohl}}]{steger2012quantum}%
  \BibitemOpen
  \bibfield  {author} {\bibinfo {author} {\bibfnamefont {M.}~\bibnamefont
  {Steger}}, \bibinfo {author} {\bibfnamefont {K.}~\bibnamefont {Saeedi}},
  \bibinfo {author} {\bibfnamefont {M.~L.~W.}\ \bibnamefont {Thewalt}},
  \bibinfo {author} {\bibfnamefont {J.~J.~L.}\ \bibnamefont {Morton}}, \bibinfo
  {author} {\bibfnamefont {H.}~\bibnamefont {Riemann}}, \bibinfo {author}
  {\bibfnamefont {N.~V.}\ \bibnamefont {Abrosimov}}, \bibinfo {author}
  {\bibfnamefont {P.}~\bibnamefont {Becker}},\ and\ \bibinfo {author}
  {\bibfnamefont {H.-J.}\ \bibnamefont {Pohl}},\ }\bibfield  {title} {\bibinfo
  {title} {Quantum information storage for over 180 s using donor spins in a
  {S}i-28 “semiconductor vacuum”},\ }\href@noop {} {\bibfield  {journal}
  {\bibinfo  {journal} {Science}\ }\textbf {\bibinfo {volume} {336}},\ \bibinfo
  {pages} {1280} (\bibinfo {year} {2012})}\BibitemShut {NoStop}%
\bibitem [{\citenamefont {Saeedi}\ \emph {et~al.}(2013)\citenamefont {Saeedi},
  \citenamefont {Simmons}, \citenamefont {Salvail}, \citenamefont {Dluhy},
  \citenamefont {Riemann}, \citenamefont {Abrosimov}, \citenamefont {Becker},
  \citenamefont {Pohl}, \citenamefont {Morton},\ and\ \citenamefont
  {Thewalt}}]{saeedi2013room}%
  \BibitemOpen
  \bibfield  {author} {\bibinfo {author} {\bibfnamefont {K.}~\bibnamefont
  {Saeedi}}, \bibinfo {author} {\bibfnamefont {S.}~\bibnamefont {Simmons}},
  \bibinfo {author} {\bibfnamefont {J.~Z.}\ \bibnamefont {Salvail}}, \bibinfo
  {author} {\bibfnamefont {P.}~\bibnamefont {Dluhy}}, \bibinfo {author}
  {\bibfnamefont {H.}~\bibnamefont {Riemann}}, \bibinfo {author} {\bibfnamefont
  {N.~V.}\ \bibnamefont {Abrosimov}}, \bibinfo {author} {\bibfnamefont
  {P.}~\bibnamefont {Becker}}, \bibinfo {author} {\bibfnamefont {H.-J.}\
  \bibnamefont {Pohl}}, \bibinfo {author} {\bibfnamefont {J.~J.~L.}\
  \bibnamefont {Morton}},\ and\ \bibinfo {author} {\bibfnamefont {M.~L.~W.}\
  \bibnamefont {Thewalt}},\ }\bibfield  {title} {\bibinfo {title}
  {Room-temperature quantum bit storage exceeding 39 minutes using ionized
  donors in silicon-28},\ }\href@noop {} {\bibfield  {journal} {\bibinfo
  {journal} {Science}\ }\textbf {\bibinfo {volume} {342}},\ \bibinfo {pages}
  {830} (\bibinfo {year} {2013})}\BibitemShut {NoStop}%
\bibitem [{\citenamefont {Kane}(1998)}]{kane1998silicon}%
  \BibitemOpen
  \bibfield  {author} {\bibinfo {author} {\bibfnamefont {B.~E.}\ \bibnamefont
  {Kane}},\ }\bibfield  {title} {\bibinfo {title} {A silicon-based nuclear spin
  quantum computer},\ }\href@noop {} {\bibfield  {journal} {\bibinfo  {journal}
  {nature}\ }\textbf {\bibinfo {volume} {393}},\ \bibinfo {pages} {133}
  (\bibinfo {year} {1998})}\BibitemShut {NoStop}%
\bibitem [{\citenamefont {Loss}\ and\ \citenamefont
  {DiVincenzo}(1998)}]{loss1998quantum}%
  \BibitemOpen
  \bibfield  {author} {\bibinfo {author} {\bibfnamefont {D.}~\bibnamefont
  {Loss}}\ and\ \bibinfo {author} {\bibfnamefont {D.~P.}\ \bibnamefont
  {DiVincenzo}},\ }\bibfield  {title} {\bibinfo {title} {Quantum computation
  with quantum dots},\ }\href@noop {} {\bibfield  {journal} {\bibinfo
  {journal} {Phys.~Rev.~A}\ }\textbf {\bibinfo {volume} {57}},\ \bibinfo
  {pages} {120} (\bibinfo {year} {1998})}\BibitemShut {NoStop}%
\bibitem [{\citenamefont {Pla}\ \emph {et~al.}(2013)\citenamefont {Pla},
  \citenamefont {Tan}, \citenamefont {Dehollain}, \citenamefont {Lim},
  \citenamefont {Morton}, \citenamefont {Zwanenburg}, \citenamefont {Jamieson},
  \citenamefont {Dzurak},\ and\ \citenamefont {Morello}}]{pla2013high}%
  \BibitemOpen
  \bibfield  {author} {\bibinfo {author} {\bibfnamefont {J.~J.}\ \bibnamefont
  {Pla}}, \bibinfo {author} {\bibfnamefont {K.~Y.}\ \bibnamefont {Tan}},
  \bibinfo {author} {\bibfnamefont {J.~P.}\ \bibnamefont {Dehollain}}, \bibinfo
  {author} {\bibfnamefont {W.~H.}\ \bibnamefont {Lim}}, \bibinfo {author}
  {\bibfnamefont {J.~J.~L.}\ \bibnamefont {Morton}}, \bibinfo {author}
  {\bibfnamefont {F.~A.}\ \bibnamefont {Zwanenburg}}, \bibinfo {author}
  {\bibfnamefont {D.~N.}\ \bibnamefont {Jamieson}}, \bibinfo {author}
  {\bibfnamefont {A.~S.}\ \bibnamefont {Dzurak}},\ and\ \bibinfo {author}
  {\bibfnamefont {A.}~\bibnamefont {Morello}},\ }\bibfield  {title} {\bibinfo
  {title} {High-fidelity readout and control of a nuclear spin qubit in
  silicon},\ }\href@noop {} {\bibfield  {journal} {\bibinfo  {journal}
  {Nature}\ }\textbf {\bibinfo {volume} {496}},\ \bibinfo {pages} {334}
  (\bibinfo {year} {2013})}\BibitemShut {NoStop}%
\bibitem [{\citenamefont {Fricke}\ \emph {et~al.}(2021)\citenamefont {Fricke},
  \citenamefont {Hile}, \citenamefont {Kranz}, \citenamefont {Chung},
  \citenamefont {He}, \citenamefont {Pakkiam}, \citenamefont {House},
  \citenamefont {Keizer},\ and\ \citenamefont {Simmons}}]{fricke2021coherent}%
  \BibitemOpen
  \bibfield  {author} {\bibinfo {author} {\bibfnamefont {L.}~\bibnamefont
  {Fricke}}, \bibinfo {author} {\bibfnamefont {S.~J.}\ \bibnamefont {Hile}},
  \bibinfo {author} {\bibfnamefont {L.}~\bibnamefont {Kranz}}, \bibinfo
  {author} {\bibfnamefont {Y.}~\bibnamefont {Chung}}, \bibinfo {author}
  {\bibfnamefont {Y.}~\bibnamefont {He}}, \bibinfo {author} {\bibfnamefont
  {P.}~\bibnamefont {Pakkiam}}, \bibinfo {author} {\bibfnamefont {M.~G.}\
  \bibnamefont {House}}, \bibinfo {author} {\bibfnamefont {J.~G.}\ \bibnamefont
  {Keizer}},\ and\ \bibinfo {author} {\bibfnamefont {M.~Y.}\ \bibnamefont
  {Simmons}},\ }\bibfield  {title} {\bibinfo {title} {Coherent control of a
  donor-molecule electron spin qubit in silicon},\ }\href@noop {} {\bibfield
  {journal} {\bibinfo  {journal} {Nat.~Commun.}\ }\textbf {\bibinfo {volume}
  {12}},\ \bibinfo {pages} {3323} (\bibinfo {year} {2021})}\BibitemShut
  {NoStop}%
\bibitem [{\citenamefont {M{\k{a}}dzik}\ \emph {et~al.}(2022)\citenamefont
  {M{\k{a}}dzik}, \citenamefont {Asaad}, \citenamefont {Youssry}, \citenamefont
  {Joecker}, \citenamefont {Rudinger}, \citenamefont {Nielsen}, \citenamefont
  {Young}, \citenamefont {Proctor}, \citenamefont {Baczewski}, \citenamefont
  {Laucht} \emph {et~al.}}]{mkadzik2022precision}%
  \BibitemOpen
  \bibfield  {author} {\bibinfo {author} {\bibfnamefont {M.~T.}\ \bibnamefont
  {M{\k{a}}dzik}}, \bibinfo {author} {\bibfnamefont {S.}~\bibnamefont {Asaad}},
  \bibinfo {author} {\bibfnamefont {A.}~\bibnamefont {Youssry}}, \bibinfo
  {author} {\bibfnamefont {B.}~\bibnamefont {Joecker}}, \bibinfo {author}
  {\bibfnamefont {K.~M.}\ \bibnamefont {Rudinger}}, \bibinfo {author}
  {\bibfnamefont {E.}~\bibnamefont {Nielsen}}, \bibinfo {author} {\bibfnamefont
  {K.~C.}\ \bibnamefont {Young}}, \bibinfo {author} {\bibfnamefont {T.~J.}\
  \bibnamefont {Proctor}}, \bibinfo {author} {\bibfnamefont {A.~D.}\
  \bibnamefont {Baczewski}}, \bibinfo {author} {\bibfnamefont {A.}~\bibnamefont
  {Laucht}}, \emph {et~al.},\ }\bibfield  {title} {\bibinfo {title} {Precision
  tomography of a three-qubit donor quantum processor in silicon},\ }\href@noop
  {} {\bibfield  {journal} {\bibinfo  {journal} {Nature}\ }\textbf {\bibinfo
  {volume} {601}},\ \bibinfo {pages} {348} (\bibinfo {year}
  {2022})}\BibitemShut {NoStop}%
\bibitem [{\citenamefont {Reiner}\ \emph {et~al.}(2024)\citenamefont {Reiner},
  \citenamefont {Chung}, \citenamefont {Misha}, \citenamefont {Lehner},
  \citenamefont {Moehle}, \citenamefont {Poulos}, \citenamefont {Monir},
  \citenamefont {Charde}, \citenamefont {Macha}, \citenamefont {Kranz} \emph
  {et~al.}}]{reiner2024high}%
  \BibitemOpen
  \bibfield  {author} {\bibinfo {author} {\bibfnamefont {J.}~\bibnamefont
  {Reiner}}, \bibinfo {author} {\bibfnamefont {Y.}~\bibnamefont {Chung}},
  \bibinfo {author} {\bibfnamefont {S.~H.}\ \bibnamefont {Misha}}, \bibinfo
  {author} {\bibfnamefont {C.}~\bibnamefont {Lehner}}, \bibinfo {author}
  {\bibfnamefont {C.}~\bibnamefont {Moehle}}, \bibinfo {author} {\bibfnamefont
  {D.}~\bibnamefont {Poulos}}, \bibinfo {author} {\bibfnamefont
  {S.}~\bibnamefont {Monir}}, \bibinfo {author} {\bibfnamefont {K.~J.}\
  \bibnamefont {Charde}}, \bibinfo {author} {\bibfnamefont {P.}~\bibnamefont
  {Macha}}, \bibinfo {author} {\bibfnamefont {L.}~\bibnamefont {Kranz}}, \emph
  {et~al.},\ }\bibfield  {title} {\bibinfo {title} {High-fidelity
  initialization and control of electron and nuclear spins in a four-qubit
  register},\ }\href@noop {} {\bibfield  {journal} {\bibinfo  {journal}
  {Nat.~Nanotechnol.}\ }\textbf {\bibinfo {volume} {19}},\ \bibinfo {pages}
  {605} (\bibinfo {year} {2024})}\BibitemShut {NoStop}%
\bibitem [{\citenamefont {Asaad}\ \emph {et~al.}(2020)\citenamefont {Asaad},
  \citenamefont {Mourik}, \citenamefont {Joecker}, \citenamefont {Johnson},
  \citenamefont {Baczewski}, \citenamefont {Firgau}, \citenamefont
  {M{\k{a}}dzik}, \citenamefont {Schmitt}, \citenamefont {Pla}, \citenamefont
  {Hudson} \emph {et~al.}}]{asaad2020coherent}%
  \BibitemOpen
  \bibfield  {author} {\bibinfo {author} {\bibfnamefont {S.}~\bibnamefont
  {Asaad}}, \bibinfo {author} {\bibfnamefont {V.}~\bibnamefont {Mourik}},
  \bibinfo {author} {\bibfnamefont {B.}~\bibnamefont {Joecker}}, \bibinfo
  {author} {\bibfnamefont {M.~A.~I.}\ \bibnamefont {Johnson}}, \bibinfo
  {author} {\bibfnamefont {A.~D.}\ \bibnamefont {Baczewski}}, \bibinfo {author}
  {\bibfnamefont {H.~R.}\ \bibnamefont {Firgau}}, \bibinfo {author}
  {\bibfnamefont {M.~T.}\ \bibnamefont {M{\k{a}}dzik}}, \bibinfo {author}
  {\bibfnamefont {V.}~\bibnamefont {Schmitt}}, \bibinfo {author} {\bibfnamefont
  {J.~J.}\ \bibnamefont {Pla}}, \bibinfo {author} {\bibfnamefont {F.~E.}\
  \bibnamefont {Hudson}}, \emph {et~al.},\ }\bibfield  {title} {\bibinfo
  {title} {Coherent electrical control of a single high-spin nucleus in
  silicon},\ }\href@noop {} {\bibfield  {journal} {\bibinfo  {journal}
  {Nature}\ }\textbf {\bibinfo {volume} {579}},\ \bibinfo {pages} {205}
  (\bibinfo {year} {2020})}\BibitemShut {NoStop}%
\bibitem [{\citenamefont {Fern{\'a}ndez~de Fuentes}\ \emph
  {et~al.}(2024)\citenamefont {Fern{\'a}ndez~de Fuentes}, \citenamefont
  {Botzem}, \citenamefont {Johnson}, \citenamefont {Vaartjes}, \citenamefont
  {Asaad}, \citenamefont {Mourik}, \citenamefont {Hudson}, \citenamefont
  {Itoh}, \citenamefont {Johnson}, \citenamefont {Jakob} \emph
  {et~al.}}]{fernandez2024navigating}%
  \BibitemOpen
  \bibfield  {author} {\bibinfo {author} {\bibfnamefont {I.}~\bibnamefont
  {Fern{\'a}ndez~de Fuentes}}, \bibinfo {author} {\bibfnamefont
  {T.}~\bibnamefont {Botzem}}, \bibinfo {author} {\bibfnamefont {M.~A.~I.}\
  \bibnamefont {Johnson}}, \bibinfo {author} {\bibfnamefont {A.}~\bibnamefont
  {Vaartjes}}, \bibinfo {author} {\bibfnamefont {S.}~\bibnamefont {Asaad}},
  \bibinfo {author} {\bibfnamefont {V.}~\bibnamefont {Mourik}}, \bibinfo
  {author} {\bibfnamefont {F.~E.}\ \bibnamefont {Hudson}}, \bibinfo {author}
  {\bibfnamefont {K.~M.}\ \bibnamefont {Itoh}}, \bibinfo {author}
  {\bibfnamefont {B.~C.}\ \bibnamefont {Johnson}}, \bibinfo {author}
  {\bibfnamefont {A.~M.}\ \bibnamefont {Jakob}}, \emph {et~al.},\ }\bibfield
  {title} {\bibinfo {title} {Navigating the 16-dimensional {H}ilbert space of a
  high-spin donor qudit with electric and magnetic fields},\ }\href@noop {}
  {\bibfield  {journal} {\bibinfo  {journal} {Nat.~Commun.}\ }\textbf {\bibinfo
  {volume} {15}},\ \bibinfo {pages} {1380} (\bibinfo {year}
  {2024})}\BibitemShut {NoStop}%
\bibitem [{\citenamefont {Yu}\ \emph {et~al.}(2025)\citenamefont {Yu},
  \citenamefont {Wilhelm}, \citenamefont {Holmes}, \citenamefont {Vaartjes},
  \citenamefont {Schwienbacher}, \citenamefont {Nurizzo}, \citenamefont
  {Kringh{\o}j}, \citenamefont {Blankenstein}, \citenamefont {Jakob},
  \citenamefont {Gupta} \emph {et~al.}}]{yu2025schrodinger}%
  \BibitemOpen
  \bibfield  {author} {\bibinfo {author} {\bibfnamefont {X.}~\bibnamefont
  {Yu}}, \bibinfo {author} {\bibfnamefont {B.}~\bibnamefont {Wilhelm}},
  \bibinfo {author} {\bibfnamefont {D.}~\bibnamefont {Holmes}}, \bibinfo
  {author} {\bibfnamefont {A.}~\bibnamefont {Vaartjes}}, \bibinfo {author}
  {\bibfnamefont {D.}~\bibnamefont {Schwienbacher}}, \bibinfo {author}
  {\bibfnamefont {M.}~\bibnamefont {Nurizzo}}, \bibinfo {author} {\bibfnamefont
  {A.}~\bibnamefont {Kringh{\o}j}}, \bibinfo {author} {\bibfnamefont
  {M.~R.~v.}\ \bibnamefont {Blankenstein}}, \bibinfo {author} {\bibfnamefont
  {A.~M.}\ \bibnamefont {Jakob}}, \bibinfo {author} {\bibfnamefont
  {P.}~\bibnamefont {Gupta}}, \emph {et~al.},\ }\bibfield  {title} {\bibinfo
  {title} {Schr{\"o}dinger cat states of a nuclear spin qudit in silicon},\
  }\href@noop {} {\bibfield  {journal} {\bibinfo  {journal} {Nat.~Phys.}\
  }\textbf {\bibinfo {volume} {21}},\ \bibinfo {pages} {362} (\bibinfo {year}
  {2025})}\BibitemShut {NoStop}%
\bibitem [{\citenamefont {Mourik}\ \emph {et~al.}(2018)\citenamefont {Mourik},
  \citenamefont {Asaad}, \citenamefont {Firgau}, \citenamefont {Pla},
  \citenamefont {Holmes}, \citenamefont {Milburn}, \citenamefont {McCallum},\
  and\ \citenamefont {Morello}}]{mourik2018exploring}%
  \BibitemOpen
  \bibfield  {author} {\bibinfo {author} {\bibfnamefont {V.}~\bibnamefont
  {Mourik}}, \bibinfo {author} {\bibfnamefont {S.}~\bibnamefont {Asaad}},
  \bibinfo {author} {\bibfnamefont {H.}~\bibnamefont {Firgau}}, \bibinfo
  {author} {\bibfnamefont {J.~J.}\ \bibnamefont {Pla}}, \bibinfo {author}
  {\bibfnamefont {C.}~\bibnamefont {Holmes}}, \bibinfo {author} {\bibfnamefont
  {G.~J.}\ \bibnamefont {Milburn}}, \bibinfo {author} {\bibfnamefont {J.~C.}\
  \bibnamefont {McCallum}},\ and\ \bibinfo {author} {\bibfnamefont
  {A.}~\bibnamefont {Morello}},\ }\bibfield  {title} {\bibinfo {title}
  {Exploring quantum chaos with a single nuclear spin},\ }\href@noop {}
  {\bibfield  {journal} {\bibinfo  {journal} {Phys.~Rev.~E}\ }\textbf {\bibinfo
  {volume} {98}},\ \bibinfo {pages} {042206} (\bibinfo {year}
  {2018})}\BibitemShut {NoStop}%
\bibitem [{\citenamefont {Vaartjes}\ \emph {et~al.}(2025)\citenamefont
  {Vaartjes}, \citenamefont {Nurizzo}, \citenamefont {Zaw}, \citenamefont
  {Wilhelm}, \citenamefont {Yu}, \citenamefont {Holmes}, \citenamefont
  {Schwienbacher}, \citenamefont {Kringh{\o}j}, \citenamefont {van
  Blankenstein}, \citenamefont {Jakob} \emph
  {et~al.}}]{vaartjes2025certifying}%
  \BibitemOpen
  \bibfield  {author} {\bibinfo {author} {\bibfnamefont {A.}~\bibnamefont
  {Vaartjes}}, \bibinfo {author} {\bibfnamefont {M.}~\bibnamefont {Nurizzo}},
  \bibinfo {author} {\bibfnamefont {L.~H.}\ \bibnamefont {Zaw}}, \bibinfo
  {author} {\bibfnamefont {B.}~\bibnamefont {Wilhelm}}, \bibinfo {author}
  {\bibfnamefont {X.}~\bibnamefont {Yu}}, \bibinfo {author} {\bibfnamefont
  {D.}~\bibnamefont {Holmes}}, \bibinfo {author} {\bibfnamefont
  {D.}~\bibnamefont {Schwienbacher}}, \bibinfo {author} {\bibfnamefont
  {A.}~\bibnamefont {Kringh{\o}j}}, \bibinfo {author} {\bibfnamefont {M.~R.}\
  \bibnamefont {van Blankenstein}}, \bibinfo {author} {\bibfnamefont {A.~M.}\
  \bibnamefont {Jakob}}, \emph {et~al.},\ }\bibfield  {title} {\bibinfo {title}
  {Certifying the quantumness of a nuclear spin qudit through its uniform
  precession},\ }\href@noop {} {\bibfield  {journal} {\bibinfo  {journal}
  {Newton}\ }\textbf {\bibinfo {volume} {1}} (\bibinfo {year}
  {2025})}\BibitemShut {NoStop}%
\bibitem [{\citenamefont {Morello}\ \emph {et~al.}(2020)\citenamefont
  {Morello}, \citenamefont {Pla}, \citenamefont {Bertet},\ and\ \citenamefont
  {Jamieson}}]{morello2020donor}%
  \BibitemOpen
  \bibfield  {author} {\bibinfo {author} {\bibfnamefont {A.}~\bibnamefont
  {Morello}}, \bibinfo {author} {\bibfnamefont {J.~J.}\ \bibnamefont {Pla}},
  \bibinfo {author} {\bibfnamefont {P.}~\bibnamefont {Bertet}},\ and\ \bibinfo
  {author} {\bibfnamefont {D.~N.}\ \bibnamefont {Jamieson}},\ }\bibfield
  {title} {\bibinfo {title} {Donor spins in silicon for quantum technologies},\
  }\href@noop {} {\bibfield  {journal} {\bibinfo  {journal} {Adv.~Quantum
  Technol.}\ }\textbf {\bibinfo {volume} {3}},\ \bibinfo {pages} {2000005}
  (\bibinfo {year} {2020})}\BibitemShut {NoStop}%
\bibitem [{\citenamefont {Gross}(2021)}]{gross2021designing}%
  \BibitemOpen
  \bibfield  {author} {\bibinfo {author} {\bibfnamefont {J.~A.}\ \bibnamefont
  {Gross}},\ }\bibfield  {title} {\bibinfo {title} {Designing codes around
  interactions: {T}he case of a spin},\ }\href@noop {} {\bibfield  {journal}
  {\bibinfo  {journal} {Phys.~Rev.~Lett.}\ }\textbf {\bibinfo {volume} {127}},\
  \bibinfo {pages} {010504} (\bibinfo {year} {2021})}\BibitemShut {NoStop}%
\bibitem [{\citenamefont {Leuenberger}\ and\ \citenamefont
  {Loss}(2003)}]{leuenberger2003grover}%
  \BibitemOpen
  \bibfield  {author} {\bibinfo {author} {\bibfnamefont {M.~N.}\ \bibnamefont
  {Leuenberger}}\ and\ \bibinfo {author} {\bibfnamefont {D.}~\bibnamefont
  {Loss}},\ }\bibfield  {title} {\bibinfo {title} {Grover algorithm for large
  nuclear spins in semiconductors},\ }\href@noop {} {\bibfield  {journal}
  {\bibinfo  {journal} {Phys.~Rev.~B}\ }\textbf {\bibinfo {volume} {68}},\
  \bibinfo {pages} {165317} (\bibinfo {year} {2003})}\BibitemShut {NoStop}%
\bibitem [{\citenamefont {Muhonen}\ \emph {et~al.}(2014)\citenamefont
  {Muhonen}, \citenamefont {Dehollain}, \citenamefont {Laucht}, \citenamefont
  {Hudson}, \citenamefont {Kalra}, \citenamefont {Sekiguchi}, \citenamefont
  {Itoh}, \citenamefont {Jamieson}, \citenamefont {McCallum}, \citenamefont
  {Dzurak} \emph {et~al.}}]{muhonen2014storing}%
  \BibitemOpen
  \bibfield  {author} {\bibinfo {author} {\bibfnamefont {J.~T.}\ \bibnamefont
  {Muhonen}}, \bibinfo {author} {\bibfnamefont {J.~P.}\ \bibnamefont
  {Dehollain}}, \bibinfo {author} {\bibfnamefont {A.}~\bibnamefont {Laucht}},
  \bibinfo {author} {\bibfnamefont {F.~E.}\ \bibnamefont {Hudson}}, \bibinfo
  {author} {\bibfnamefont {R.}~\bibnamefont {Kalra}}, \bibinfo {author}
  {\bibfnamefont {T.}~\bibnamefont {Sekiguchi}}, \bibinfo {author}
  {\bibfnamefont {K.~M.}\ \bibnamefont {Itoh}}, \bibinfo {author}
  {\bibfnamefont {D.~N.}\ \bibnamefont {Jamieson}}, \bibinfo {author}
  {\bibfnamefont {J.~C.}\ \bibnamefont {McCallum}}, \bibinfo {author}
  {\bibfnamefont {A.~S.}\ \bibnamefont {Dzurak}}, \emph {et~al.},\ }\bibfield
  {title} {\bibinfo {title} {Storing quantum information for 30 seconds in a
  nanoelectronic device},\ }\href@noop {} {\bibfield  {journal} {\bibinfo
  {journal} {Nat.~Nanotechnol.}\ }\textbf {\bibinfo {volume} {9}},\ \bibinfo
  {pages} {986} (\bibinfo {year} {2014})}\BibitemShut {NoStop}%
\bibitem [{\citenamefont {Longdell}\ \emph {et~al.}(2006)\citenamefont
  {Longdell}, \citenamefont {Alexander},\ and\ \citenamefont
  {Sellars}}]{longdell2006characterization}%
  \BibitemOpen
  \bibfield  {author} {\bibinfo {author} {\bibfnamefont {J.~J.}\ \bibnamefont
  {Longdell}}, \bibinfo {author} {\bibfnamefont {A.~L.}\ \bibnamefont
  {Alexander}},\ and\ \bibinfo {author} {\bibfnamefont {M.}~\bibnamefont
  {Sellars}},\ }\bibfield  {title} {\bibinfo {title} {Characterization of the
  hyperfine interaction in europium-doped yttrium orthosilicate and europium
  chloride hexahydrate},\ }\href@noop {} {\bibfield  {journal} {\bibinfo
  {journal} {Phys.~Rev.~B}\ }\textbf {\bibinfo {volume} {74}},\ \bibinfo
  {pages} {195101} (\bibinfo {year} {2006})}\BibitemShut {NoStop}%
\bibitem [{\citenamefont {McAuslan}\ \emph {et~al.}(2012)\citenamefont
  {McAuslan}, \citenamefont {Bartholomew}, \citenamefont {Sellars},\ and\
  \citenamefont {Longdell}}]{mcauslan2012reducing}%
  \BibitemOpen
  \bibfield  {author} {\bibinfo {author} {\bibfnamefont {D.~L.}\ \bibnamefont
  {McAuslan}}, \bibinfo {author} {\bibfnamefont {J.~G.}\ \bibnamefont
  {Bartholomew}}, \bibinfo {author} {\bibfnamefont {M.~J.}\ \bibnamefont
  {Sellars}},\ and\ \bibinfo {author} {\bibfnamefont {J.~J.}\ \bibnamefont
  {Longdell}},\ }\bibfield  {title} {\bibinfo {title} {Reducing decoherence in
  optical and spin transitions in rare-earth-metal-ion--doped materials},\
  }\href@noop {} {\bibfield  {journal} {\bibinfo  {journal} {Phys.~Rev.~A}\
  }\textbf {\bibinfo {volume} {85}},\ \bibinfo {pages} {032339} (\bibinfo
  {year} {2012})}\BibitemShut {NoStop}%
\bibitem [{\citenamefont {Wolfowicz}\ \emph {et~al.}(2013)\citenamefont
  {Wolfowicz}, \citenamefont {Tyryshkin}, \citenamefont {George}, \citenamefont
  {Riemann}, \citenamefont {Abrosimov}, \citenamefont {Becker}, \citenamefont
  {Pohl}, \citenamefont {Thewalt}, \citenamefont {Lyon},\ and\ \citenamefont
  {Morton}}]{wolfowicz2013atomic}%
  \BibitemOpen
  \bibfield  {author} {\bibinfo {author} {\bibfnamefont {G.}~\bibnamefont
  {Wolfowicz}}, \bibinfo {author} {\bibfnamefont {A.~M.}\ \bibnamefont
  {Tyryshkin}}, \bibinfo {author} {\bibfnamefont {R.~E.}\ \bibnamefont
  {George}}, \bibinfo {author} {\bibfnamefont {H.}~\bibnamefont {Riemann}},
  \bibinfo {author} {\bibfnamefont {N.~V.}\ \bibnamefont {Abrosimov}}, \bibinfo
  {author} {\bibfnamefont {P.}~\bibnamefont {Becker}}, \bibinfo {author}
  {\bibfnamefont {H.-J.}\ \bibnamefont {Pohl}}, \bibinfo {author}
  {\bibfnamefont {M.~L.~W.}\ \bibnamefont {Thewalt}}, \bibinfo {author}
  {\bibfnamefont {S.~A.}\ \bibnamefont {Lyon}},\ and\ \bibinfo {author}
  {\bibfnamefont {J.~J.~L.}\ \bibnamefont {Morton}},\ }\bibfield  {title}
  {\bibinfo {title} {Atomic clock transitions in silicon-based spin qubits},\
  }\href@noop {} {\bibfield  {journal} {\bibinfo  {journal}
  {Nat.~Nanotechnol.}\ }\textbf {\bibinfo {volume} {8}},\ \bibinfo {pages}
  {561} (\bibinfo {year} {2013})}\BibitemShut {NoStop}%
\bibitem [{\citenamefont {Tosi}\ \emph {et~al.}(2017)\citenamefont {Tosi},
  \citenamefont {Mohiyaddin}, \citenamefont {Schmitt}, \citenamefont {Tenberg},
  \citenamefont {Rahman}, \citenamefont {Klimeck},\ and\ \citenamefont
  {Morello}}]{tosi2017silicon}%
  \BibitemOpen
  \bibfield  {author} {\bibinfo {author} {\bibfnamefont {G.}~\bibnamefont
  {Tosi}}, \bibinfo {author} {\bibfnamefont {F.~A.}\ \bibnamefont
  {Mohiyaddin}}, \bibinfo {author} {\bibfnamefont {V.}~\bibnamefont {Schmitt}},
  \bibinfo {author} {\bibfnamefont {S.}~\bibnamefont {Tenberg}}, \bibinfo
  {author} {\bibfnamefont {R.}~\bibnamefont {Rahman}}, \bibinfo {author}
  {\bibfnamefont {G.}~\bibnamefont {Klimeck}},\ and\ \bibinfo {author}
  {\bibfnamefont {A.}~\bibnamefont {Morello}},\ }\bibfield  {title} {\bibinfo
  {title} {Silicon quantum processor with robust long-distance qubit
  couplings},\ }\href@noop {} {\bibfield  {journal} {\bibinfo  {journal}
  {Nat.~Commun.}\ }\textbf {\bibinfo {volume} {8}},\ \bibinfo {pages} {450}
  (\bibinfo {year} {2017})}\BibitemShut {NoStop}%
\bibitem [{\citenamefont {Savytskyy}\ \emph {et~al.}(2023)\citenamefont
  {Savytskyy}, \citenamefont {Botzem}, \citenamefont {Fernandez~de Fuentes},
  \citenamefont {Joecker}, \citenamefont {Pla}, \citenamefont {Hudson},
  \citenamefont {Itoh}, \citenamefont {Jakob}, \citenamefont {Johnson},
  \citenamefont {Jamieson} \emph {et~al.}}]{savytskyy2023electrically}%
  \BibitemOpen
  \bibfield  {author} {\bibinfo {author} {\bibfnamefont {R.}~\bibnamefont
  {Savytskyy}}, \bibinfo {author} {\bibfnamefont {T.}~\bibnamefont {Botzem}},
  \bibinfo {author} {\bibfnamefont {I.}~\bibnamefont {Fernandez~de Fuentes}},
  \bibinfo {author} {\bibfnamefont {B.}~\bibnamefont {Joecker}}, \bibinfo
  {author} {\bibfnamefont {J.~J.}\ \bibnamefont {Pla}}, \bibinfo {author}
  {\bibfnamefont {F.~E.}\ \bibnamefont {Hudson}}, \bibinfo {author}
  {\bibfnamefont {K.~M.}\ \bibnamefont {Itoh}}, \bibinfo {author}
  {\bibfnamefont {A.~M.}\ \bibnamefont {Jakob}}, \bibinfo {author}
  {\bibfnamefont {B.~C.}\ \bibnamefont {Johnson}}, \bibinfo {author}
  {\bibfnamefont {D.~N.}\ \bibnamefont {Jamieson}}, \emph {et~al.},\ }\bibfield
   {title} {\bibinfo {title} {An electrically driven single-atom
  “flip-flop” qubit},\ }\href@noop {} {\bibfield  {journal} {\bibinfo
  {journal} {Sci.~Adv.}\ }\textbf {\bibinfo {volume} {9}},\ \bibinfo {pages}
  {eadd9408} (\bibinfo {year} {2023})}\BibitemShut {NoStop}%
\bibitem [{\citenamefont {Puri}\ \emph {et~al.}(2017)\citenamefont {Puri},
  \citenamefont {Boutin},\ and\ \citenamefont {Blais}}]{puri2017engineering}%
  \BibitemOpen
  \bibfield  {author} {\bibinfo {author} {\bibfnamefont {S.}~\bibnamefont
  {Puri}}, \bibinfo {author} {\bibfnamefont {S.}~\bibnamefont {Boutin}},\ and\
  \bibinfo {author} {\bibfnamefont {A.}~\bibnamefont {Blais}},\ }\bibfield
  {title} {\bibinfo {title} {Engineering the quantum states of light in a
  {K}err-nonlinear resonator by two-photon driving},\ }\href@noop {} {\bibfield
   {journal} {\bibinfo  {journal} {npj Quantum Inf.}\ }\textbf {\bibinfo
  {volume} {3}},\ \bibinfo {pages} {18} (\bibinfo {year} {2017})}\BibitemShut
  {NoStop}%
\bibitem [{\citenamefont {Grimm}\ \emph {et~al.}(2020)\citenamefont {Grimm},
  \citenamefont {Frattini}, \citenamefont {Puri}, \citenamefont {Mundhada},
  \citenamefont {Touzard}, \citenamefont {Mirrahimi}, \citenamefont {Girvin},
  \citenamefont {Shankar},\ and\ \citenamefont
  {Devoret}}]{grimm2020stabilization}%
  \BibitemOpen
  \bibfield  {author} {\bibinfo {author} {\bibfnamefont {A.}~\bibnamefont
  {Grimm}}, \bibinfo {author} {\bibfnamefont {N.~E.}\ \bibnamefont {Frattini}},
  \bibinfo {author} {\bibfnamefont {S.}~\bibnamefont {Puri}}, \bibinfo {author}
  {\bibfnamefont {S.~O.}\ \bibnamefont {Mundhada}}, \bibinfo {author}
  {\bibfnamefont {S.}~\bibnamefont {Touzard}}, \bibinfo {author} {\bibfnamefont
  {M.}~\bibnamefont {Mirrahimi}}, \bibinfo {author} {\bibfnamefont {S.~M.}\
  \bibnamefont {Girvin}}, \bibinfo {author} {\bibfnamefont {S.}~\bibnamefont
  {Shankar}},\ and\ \bibinfo {author} {\bibfnamefont {M.~H.}\ \bibnamefont
  {Devoret}},\ }\bibfield  {title} {\bibinfo {title} {Stabilization and
  operation of a {K}err-cat qubit},\ }\href@noop {} {\bibfield  {journal}
  {\bibinfo  {journal} {Nature}\ }\textbf {\bibinfo {volume} {584}},\ \bibinfo
  {pages} {205} (\bibinfo {year} {2020})}\BibitemShut {NoStop}%
\bibitem [{\citenamefont {Gross}\ \emph {et~al.}(2024)\citenamefont {Gross},
  \citenamefont {Godfrin}, \citenamefont {Blais},\ and\ \citenamefont
  {Dupont-Ferrier}}]{gross2024hardware}%
  \BibitemOpen
  \bibfield  {author} {\bibinfo {author} {\bibfnamefont {J.~A.}\ \bibnamefont
  {Gross}}, \bibinfo {author} {\bibfnamefont {C.}~\bibnamefont {Godfrin}},
  \bibinfo {author} {\bibfnamefont {A.}~\bibnamefont {Blais}},\ and\ \bibinfo
  {author} {\bibfnamefont {E.}~\bibnamefont {Dupont-Ferrier}},\ }\bibfield
  {title} {\bibinfo {title} {Hardware-efficient error-correcting codes for
  large nuclear spins},\ }\href@noop {} {\bibfield  {journal} {\bibinfo
  {journal} {Phys.~Rev.~Appl.}\ }\textbf {\bibinfo {volume} {22}},\ \bibinfo
  {pages} {014006} (\bibinfo {year} {2024})}\BibitemShut {NoStop}%
\bibitem [{\citenamefont {Omanakuttan}\ \emph {et~al.}(2024)\citenamefont
  {Omanakuttan}, \citenamefont {Buchemmavari}, \citenamefont {Gross},
  \citenamefont {Deutsch},\ and\ \citenamefont
  {Marvian}}]{omanakuttan2024fault}%
  \BibitemOpen
  \bibfield  {author} {\bibinfo {author} {\bibfnamefont {S.}~\bibnamefont
  {Omanakuttan}}, \bibinfo {author} {\bibfnamefont {V.}~\bibnamefont
  {Buchemmavari}}, \bibinfo {author} {\bibfnamefont {J.~A.}\ \bibnamefont
  {Gross}}, \bibinfo {author} {\bibfnamefont {I.~H.}\ \bibnamefont {Deutsch}},\
  and\ \bibinfo {author} {\bibfnamefont {M.}~\bibnamefont {Marvian}},\
  }\bibfield  {title} {\bibinfo {title} {Fault-tolerant quantum computation
  using large spin-cat codes},\ }\href@noop {} {\bibfield  {journal} {\bibinfo
  {journal} {PRX Quantum}\ }\textbf {\bibinfo {volume} {5}},\ \bibinfo {pages}
  {020355} (\bibinfo {year} {2024})}\BibitemShut {NoStop}%
\bibitem [{\citenamefont {Gupta}\ \emph {et~al.}(2024)\citenamefont {Gupta},
  \citenamefont {Vaartjes}, \citenamefont {Yu}, \citenamefont {Morello},\ and\
  \citenamefont {Sanders}}]{gupta2024robust}%
  \BibitemOpen
  \bibfield  {author} {\bibinfo {author} {\bibfnamefont {P.}~\bibnamefont
  {Gupta}}, \bibinfo {author} {\bibfnamefont {A.}~\bibnamefont {Vaartjes}},
  \bibinfo {author} {\bibfnamefont {X.}~\bibnamefont {Yu}}, \bibinfo {author}
  {\bibfnamefont {A.}~\bibnamefont {Morello}},\ and\ \bibinfo {author}
  {\bibfnamefont {B.~C.}\ \bibnamefont {Sanders}},\ }\bibfield  {title}
  {\bibinfo {title} {Robust macroscopic {S}chr{\"o}dinger's cat on a nucleus},\
  }\href@noop {} {\bibfield  {journal} {\bibinfo  {journal} {Phys.~Rev.~Res.}\
  }\textbf {\bibinfo {volume} {6}},\ \bibinfo {pages} {013101} (\bibinfo {year}
  {2024})}\BibitemShut {NoStop}%
\bibitem [{\citenamefont {Cohen}\ and\ \citenamefont
  {Reif}(1957)}]{cohen1957quadrupole}%
  \BibitemOpen
  \bibfield  {author} {\bibinfo {author} {\bibfnamefont {M.~H.}\ \bibnamefont
  {Cohen}}\ and\ \bibinfo {author} {\bibfnamefont {F.}~\bibnamefont {Reif}},\
  }\bibfield  {title} {\bibinfo {title} {Quadrupole effects in nuclear magnetic
  resonance studies of solids},\ }in\ \href@noop {} {\emph {\bibinfo
  {booktitle} {Solid State Physics}}},\ Vol.~\bibinfo {volume} {5}\ (\bibinfo
  {publisher} {Elsevier},\ \bibinfo {year} {1957})\ pp.\ \bibinfo {pages}
  {321--438}\BibitemShut {NoStop}%
\bibitem [{\citenamefont {Franke}\ \emph {et~al.}(2015)\citenamefont {Franke},
  \citenamefont {Hrubesch}, \citenamefont {K{\"u}nzl}, \citenamefont {Itoh},
  \citenamefont {Stutzmann}, \citenamefont {Hoehne}, \citenamefont {Dreher},\
  and\ \citenamefont {Brandt}}]{franke2015interaction}%
  \BibitemOpen
  \bibfield  {author} {\bibinfo {author} {\bibfnamefont {D.~P.}\ \bibnamefont
  {Franke}}, \bibinfo {author} {\bibfnamefont {F.~M.}\ \bibnamefont
  {Hrubesch}}, \bibinfo {author} {\bibfnamefont {M.}~\bibnamefont {K{\"u}nzl}},
  \bibinfo {author} {\bibfnamefont {K.~M.}\ \bibnamefont {Itoh}}, \bibinfo
  {author} {\bibfnamefont {M.}~\bibnamefont {Stutzmann}}, \bibinfo {author}
  {\bibfnamefont {F.}~\bibnamefont {Hoehne}}, \bibinfo {author} {\bibfnamefont
  {L.}~\bibnamefont {Dreher}},\ and\ \bibinfo {author} {\bibfnamefont {M.~S.}\
  \bibnamefont {Brandt}},\ }\bibfield  {title} {\bibinfo {title} {Interaction
  of strain and nuclear spins in silicon: {Q}uadrupolar effects on ionized
  donors},\ }\href@noop {} {\bibfield  {journal} {\bibinfo  {journal}
  {Phys.~Rev.~Lett.}\ }\textbf {\bibinfo {volume} {115}},\ \bibinfo {pages}
  {057601} (\bibinfo {year} {2015})}\BibitemShut {NoStop}%
\bibitem [{\citenamefont {Pla}\ \emph {et~al.}(2018)\citenamefont {Pla},
  \citenamefont {Bienfait}, \citenamefont {Pica}, \citenamefont {Mansir},
  \citenamefont {Mohiyaddin}, \citenamefont {Zeng}, \citenamefont {Niquet},
  \citenamefont {Morello}, \citenamefont {Schenkel}, \citenamefont {Morton}
  \emph {et~al.}}]{pla2018strain}%
  \BibitemOpen
  \bibfield  {author} {\bibinfo {author} {\bibfnamefont {J.~J.}\ \bibnamefont
  {Pla}}, \bibinfo {author} {\bibfnamefont {A.}~\bibnamefont {Bienfait}},
  \bibinfo {author} {\bibfnamefont {G.}~\bibnamefont {Pica}}, \bibinfo {author}
  {\bibfnamefont {J.}~\bibnamefont {Mansir}}, \bibinfo {author} {\bibfnamefont
  {F.~A.}\ \bibnamefont {Mohiyaddin}}, \bibinfo {author} {\bibfnamefont
  {Z.}~\bibnamefont {Zeng}}, \bibinfo {author} {\bibfnamefont {Y.-M.}\
  \bibnamefont {Niquet}}, \bibinfo {author} {\bibfnamefont {A.}~\bibnamefont
  {Morello}}, \bibinfo {author} {\bibfnamefont {T.}~\bibnamefont {Schenkel}},
  \bibinfo {author} {\bibfnamefont {J.~J.~L.}\ \bibnamefont {Morton}}, \emph
  {et~al.},\ }\bibfield  {title} {\bibinfo {title} {Strain-induced
  spin-resonance shifts in silicon devices},\ }\href@noop {} {\bibfield
  {journal} {\bibinfo  {journal} {Phys.~Rev.~Appl.}\ }\textbf {\bibinfo
  {volume} {9}},\ \bibinfo {pages} {044014} (\bibinfo {year}
  {2018})}\BibitemShut {NoStop}%
\bibitem [{\citenamefont {Bloembergen}(1961)}]{bloembergen1961linear}%
  \BibitemOpen
  \bibfield  {author} {\bibinfo {author} {\bibfnamefont {N.}~\bibnamefont
  {Bloembergen}},\ }\bibfield  {title} {\bibinfo {title} {Linear {S}tark effect
  in magnetic resonance spectra},\ }\href@noop {} {\bibfield  {journal}
  {\bibinfo  {journal} {Science}\ }\textbf {\bibinfo {volume} {133}},\ \bibinfo
  {pages} {1363} (\bibinfo {year} {1961})}\BibitemShut {NoStop}%
\bibitem [{\citenamefont {Ono}\ \emph {et~al.}(2013)\citenamefont {Ono},
  \citenamefont {Ishihara}, \citenamefont {Sato}, \citenamefont {Ohno},\ and\
  \citenamefont {Ohno}}]{ono2013coherent}%
  \BibitemOpen
  \bibfield  {author} {\bibinfo {author} {\bibfnamefont {M.}~\bibnamefont
  {Ono}}, \bibinfo {author} {\bibfnamefont {J.}~\bibnamefont {Ishihara}},
  \bibinfo {author} {\bibfnamefont {G.}~\bibnamefont {Sato}}, \bibinfo {author}
  {\bibfnamefont {Y.}~\bibnamefont {Ohno}},\ and\ \bibinfo {author}
  {\bibfnamefont {H.}~\bibnamefont {Ohno}},\ }\bibfield  {title} {\bibinfo
  {title} {Coherent manipulation of nuclear spins in semiconductors with an
  electric field},\ }\href@noop {} {\bibfield  {journal} {\bibinfo  {journal}
  {Appl.~Phys.~Express}\ }\textbf {\bibinfo {volume} {6}},\ \bibinfo {pages}
  {033002} (\bibinfo {year} {2013})}\BibitemShut {NoStop}%
\bibitem [{\citenamefont {Slichter}(2013)}]{slichter2013principles}%
  \BibitemOpen
  \bibfield  {author} {\bibinfo {author} {\bibfnamefont {C.~P.}\ \bibnamefont
  {Slichter}},\ }\href@noop {} {\emph {\bibinfo {title} {Principles of
  {M}agnetic {R}esonance}}},\ Vol.~\bibinfo {volume} {1}\ (\bibinfo
  {publisher} {Springer Science \& Business Media},\ \bibinfo {year}
  {2013})\BibitemShut {NoStop}%
\bibitem [{\citenamefont {Pezze}\ \emph {et~al.}(2018)\citenamefont {Pezze},
  \citenamefont {Smerzi}, \citenamefont {Oberthaler}, \citenamefont {Schmied},\
  and\ \citenamefont {Treutlein}}]{pezze2018quantum}%
  \BibitemOpen
  \bibfield  {author} {\bibinfo {author} {\bibfnamefont {L.}~\bibnamefont
  {Pezze}}, \bibinfo {author} {\bibfnamefont {A.}~\bibnamefont {Smerzi}},
  \bibinfo {author} {\bibfnamefont {M.~K.}\ \bibnamefont {Oberthaler}},
  \bibinfo {author} {\bibfnamefont {R.}~\bibnamefont {Schmied}},\ and\ \bibinfo
  {author} {\bibfnamefont {P.}~\bibnamefont {Treutlein}},\ }\bibfield  {title}
  {\bibinfo {title} {Quantum metrology with nonclassical states of atomic
  ensembles},\ }\href@noop {} {\bibfield  {journal} {\bibinfo  {journal}
  {Rev.~Mod.~Phys.}\ }\textbf {\bibinfo {volume} {90}},\ \bibinfo {pages}
  {035005} (\bibinfo {year} {2018})}\BibitemShut {NoStop}%
\bibitem [{\citenamefont {Leuenberger}\ and\ \citenamefont
  {Loss}(2001)}]{leuenberger2001spin}%
  \BibitemOpen
  \bibfield  {author} {\bibinfo {author} {\bibfnamefont {M.~N.}\ \bibnamefont
  {Leuenberger}}\ and\ \bibinfo {author} {\bibfnamefont {D.}~\bibnamefont
  {Loss}},\ }\bibfield  {title} {\bibinfo {title} {Spin tunneling and
  topological selection rules for integer spins},\ }\href
  {https://doi.org/10.1103/PhysRevB.63.054414} {\bibfield  {journal} {\bibinfo
  {journal} {Phys.~Rev.~B}\ }\textbf {\bibinfo {volume} {63}},\ \bibinfo
  {pages} {054414} (\bibinfo {year} {2001})}\BibitemShut {NoStop}%
\bibitem [{\citenamefont {Leuenberger}\ \emph {et~al.}(2002)\citenamefont
  {Leuenberger}, \citenamefont {Loss}, \citenamefont {Poggio},\ and\
  \citenamefont {Awschalom}}]{leuenberger2002quantum}%
  \BibitemOpen
  \bibfield  {author} {\bibinfo {author} {\bibfnamefont {M.~N.}\ \bibnamefont
  {Leuenberger}}, \bibinfo {author} {\bibfnamefont {D.}~\bibnamefont {Loss}},
  \bibinfo {author} {\bibfnamefont {M.}~\bibnamefont {Poggio}},\ and\ \bibinfo
  {author} {\bibfnamefont {D.~D.}\ \bibnamefont {Awschalom}},\ }\bibfield
  {title} {\bibinfo {title} {Quantum information processing with large nuclear
  spins in {G}a{A}s semiconductors},\ }\href@noop {} {\bibfield  {journal}
  {\bibinfo  {journal} {Phys.~Rev.~Lett.}\ }\textbf {\bibinfo {volume} {89}},\
  \bibinfo {pages} {207601} (\bibinfo {year} {2002})}\BibitemShut {NoStop}%
\bibitem [{\citenamefont {Morley}\ \emph {et~al.}(2010)\citenamefont {Morley},
  \citenamefont {Warner}, \citenamefont {Stoneham}, \citenamefont {Greenland},
  \citenamefont {Van~Tol}, \citenamefont {Kay},\ and\ \citenamefont
  {Aeppli}}]{morley2010initialization}%
  \BibitemOpen
  \bibfield  {author} {\bibinfo {author} {\bibfnamefont {G.~W.}\ \bibnamefont
  {Morley}}, \bibinfo {author} {\bibfnamefont {M.}~\bibnamefont {Warner}},
  \bibinfo {author} {\bibfnamefont {A.~M.}\ \bibnamefont {Stoneham}}, \bibinfo
  {author} {\bibfnamefont {P.~T.}\ \bibnamefont {Greenland}}, \bibinfo {author}
  {\bibfnamefont {J.}~\bibnamefont {Van~Tol}}, \bibinfo {author} {\bibfnamefont
  {C.~W.~M.}\ \bibnamefont {Kay}},\ and\ \bibinfo {author} {\bibfnamefont
  {G.}~\bibnamefont {Aeppli}},\ }\bibfield  {title} {\bibinfo {title} {The
  initialization and manipulation of quantum information stored in silicon by
  bismuth dopants},\ }\href@noop {} {\bibfield  {journal} {\bibinfo  {journal}
  {Nat.~Mater.}\ }\textbf {\bibinfo {volume} {9}},\ \bibinfo {pages} {725}
  (\bibinfo {year} {2010})}\BibitemShut {NoStop}%
\bibitem [{\citenamefont {Radcliffe}(1971)}]{radcliffe1971some}%
  \BibitemOpen
  \bibfield  {author} {\bibinfo {author} {\bibfnamefont {J.~M.}\ \bibnamefont
  {Radcliffe}},\ }\bibfield  {title} {\bibinfo {title} {Some properties of
  coherent spin states},\ }\href@noop {} {\bibfield  {journal} {\bibinfo
  {journal} {J.~Phys.~A: Gen.~Phys.}\ }\textbf {\bibinfo {volume} {4}},\
  \bibinfo {pages} {313} (\bibinfo {year} {1971})}\BibitemShut {NoStop}%
\bibitem [{\citenamefont {Arecchi}\ \emph {et~al.}(1972)\citenamefont
  {Arecchi}, \citenamefont {Courtens}, \citenamefont {Gilmore},\ and\
  \citenamefont {Thomas}}]{arecchi1972atomic}%
  \BibitemOpen
  \bibfield  {author} {\bibinfo {author} {\bibfnamefont {F.~T.}\ \bibnamefont
  {Arecchi}}, \bibinfo {author} {\bibfnamefont {E.}~\bibnamefont {Courtens}},
  \bibinfo {author} {\bibfnamefont {R.}~\bibnamefont {Gilmore}},\ and\ \bibinfo
  {author} {\bibfnamefont {H.}~\bibnamefont {Thomas}},\ }\bibfield  {title}
  {\bibinfo {title} {Atomic coherent states in quantum optics},\ }\href@noop {}
  {\bibfield  {journal} {\bibinfo  {journal} {Phys.~Rev.~A}\ }\textbf {\bibinfo
  {volume} {6}},\ \bibinfo {pages} {2211} (\bibinfo {year} {1972})}\BibitemShut
  {NoStop}%
\bibitem [{\citenamefont {Pirandola}\ \emph {et~al.}(2008)\citenamefont
  {Pirandola}, \citenamefont {Mancini}, \citenamefont {Braunstein},\ and\
  \citenamefont {Vitali}}]{pirandola2008minimal}%
  \BibitemOpen
  \bibfield  {author} {\bibinfo {author} {\bibfnamefont {S.}~\bibnamefont
  {Pirandola}}, \bibinfo {author} {\bibfnamefont {S.}~\bibnamefont {Mancini}},
  \bibinfo {author} {\bibfnamefont {S.~L.}\ \bibnamefont {Braunstein}},\ and\
  \bibinfo {author} {\bibfnamefont {D.}~\bibnamefont {Vitali}},\ }\bibfield
  {title} {\bibinfo {title} {Minimal qudit code for a qubit in the
  phase-damping channel},\ }\href@noop {} {\bibfield  {journal} {\bibinfo
  {journal} {Phys.~Rev.~A}\ }\textbf {\bibinfo {volume} {77}},\ \bibinfo
  {pages} {032309} (\bibinfo {year} {2008})}\BibitemShut {NoStop}%
\bibitem [{\citenamefont {Chiesa}\ \emph {et~al.}(2020)\citenamefont {Chiesa},
  \citenamefont {Macaluso}, \citenamefont {Petiziol}, \citenamefont
  {Wimberger}, \citenamefont {Santini},\ and\ \citenamefont
  {Carretta}}]{chiesa2020molecular}%
  \BibitemOpen
  \bibfield  {author} {\bibinfo {author} {\bibfnamefont {A.}~\bibnamefont
  {Chiesa}}, \bibinfo {author} {\bibfnamefont {E.}~\bibnamefont {Macaluso}},
  \bibinfo {author} {\bibfnamefont {F.}~\bibnamefont {Petiziol}}, \bibinfo
  {author} {\bibfnamefont {S.}~\bibnamefont {Wimberger}}, \bibinfo {author}
  {\bibfnamefont {P.}~\bibnamefont {Santini}},\ and\ \bibinfo {author}
  {\bibfnamefont {S.}~\bibnamefont {Carretta}},\ }\bibfield  {title} {\bibinfo
  {title} {Molecular nanomagnets as qubits with embedded quantum-error
  correction},\ }\href@noop {} {\bibfield  {journal} {\bibinfo  {journal}
  {J.~Phys.~Chem.~Lett.}\ }\textbf {\bibinfo {volume} {11}},\ \bibinfo {pages}
  {8610} (\bibinfo {year} {2020})}\BibitemShut {NoStop}%
\bibitem [{\citenamefont {Chiesa}\ \emph {et~al.}(2021)\citenamefont {Chiesa},
  \citenamefont {Petiziol}, \citenamefont {Macaluso}, \citenamefont
  {Wimberger}, \citenamefont {Santini},\ and\ \citenamefont
  {Carretta}}]{chiesa2021embedded}%
  \BibitemOpen
  \bibfield  {author} {\bibinfo {author} {\bibfnamefont {A.}~\bibnamefont
  {Chiesa}}, \bibinfo {author} {\bibfnamefont {F.}~\bibnamefont {Petiziol}},
  \bibinfo {author} {\bibfnamefont {E.}~\bibnamefont {Macaluso}}, \bibinfo
  {author} {\bibfnamefont {S.}~\bibnamefont {Wimberger}}, \bibinfo {author}
  {\bibfnamefont {P.}~\bibnamefont {Santini}},\ and\ \bibinfo {author}
  {\bibfnamefont {S.}~\bibnamefont {Carretta}},\ }\bibfield  {title} {\bibinfo
  {title} {Embedded quantum-error correction and controlled-phase gate for
  molecular spin qubits},\ }\href@noop {} {\bibfield  {journal} {\bibinfo
  {journal} {AIP Advances}\ }\textbf {\bibinfo {volume} {11}} (\bibinfo {year}
  {2021})}\BibitemShut {NoStop}%
\bibitem [{\citenamefont {Lim}\ \emph {et~al.}(2025)\citenamefont {Lim},
  \citenamefont {Vaganov}, \citenamefont {Liu},\ and\ \citenamefont
  {Ardavan}}]{lim2025demonstrating}%
  \BibitemOpen
  \bibfield  {author} {\bibinfo {author} {\bibfnamefont {S.}~\bibnamefont
  {Lim}}, \bibinfo {author} {\bibfnamefont {M.~V.}\ \bibnamefont {Vaganov}},
  \bibinfo {author} {\bibfnamefont {J.}~\bibnamefont {Liu}},\ and\ \bibinfo
  {author} {\bibfnamefont {A.}~\bibnamefont {Ardavan}},\ }\bibfield  {title}
  {\bibinfo {title} {Demonstrating experimentally the encoding and dynamics of
  an error-correctable logical qubit on a hyperfine-coupled nuclear spin
  qudit},\ }\href@noop {} {\bibfield  {journal} {\bibinfo  {journal}
  {Phys.~Rev.~Lett.}\ }\textbf {\bibinfo {volume} {134}},\ \bibinfo {pages}
  {070603} (\bibinfo {year} {2025})}\BibitemShut {NoStop}%
\bibitem [{\citenamefont {Lim}\ \emph {et~al.}(2023)\citenamefont {Lim},
  \citenamefont {Liu},\ and\ \citenamefont {Ardavan}}]{lim2023fault}%
  \BibitemOpen
  \bibfield  {author} {\bibinfo {author} {\bibfnamefont {S.}~\bibnamefont
  {Lim}}, \bibinfo {author} {\bibfnamefont {J.}~\bibnamefont {Liu}},\ and\
  \bibinfo {author} {\bibfnamefont {A.}~\bibnamefont {Ardavan}},\ }\bibfield
  {title} {\bibinfo {title} {Fault-tolerant qubit encoding using a spin-7/2
  qudit},\ }\href@noop {} {\bibfield  {journal} {\bibinfo  {journal}
  {Phys.~Rev.~A}\ }\textbf {\bibinfo {volume} {108}},\ \bibinfo {pages}
  {062403} (\bibinfo {year} {2023})}\BibitemShut {NoStop}%
\bibitem [{\citenamefont {Darmawan}\ \emph {et~al.}(2021)\citenamefont
  {Darmawan}, \citenamefont {Brown}, \citenamefont {Grimsmo}, \citenamefont
  {Tuckett},\ and\ \citenamefont {Puri}}]{darmawan2021practical}%
  \BibitemOpen
  \bibfield  {author} {\bibinfo {author} {\bibfnamefont {A.~S.}\ \bibnamefont
  {Darmawan}}, \bibinfo {author} {\bibfnamefont {B.~J.}\ \bibnamefont {Brown}},
  \bibinfo {author} {\bibfnamefont {A.~L.}\ \bibnamefont {Grimsmo}}, \bibinfo
  {author} {\bibfnamefont {D.~K.}\ \bibnamefont {Tuckett}},\ and\ \bibinfo
  {author} {\bibfnamefont {S.}~\bibnamefont {Puri}},\ }\bibfield  {title}
  {\bibinfo {title} {Practical quantum error correction with the {XZZX} code
  and {K}err-cat qubits},\ }\href@noop {} {\bibfield  {journal} {\bibinfo
  {journal} {PRX Quantum}\ }\textbf {\bibinfo {volume} {2}},\ \bibinfo {pages}
  {030345} (\bibinfo {year} {2021})}\BibitemShut {NoStop}%
\bibitem [{\citenamefont {Putterman}\ \emph {et~al.}(2025)\citenamefont
  {Putterman}, \citenamefont {Noh}, \citenamefont {Hann}, \citenamefont
  {MacCabe}, \citenamefont {Aghaeimeibodi}, \citenamefont {Patel},
  \citenamefont {Lee}, \citenamefont {Jones}, \citenamefont {Moradinejad},
  \citenamefont {Rodriguez} \emph {et~al.}}]{putterman2025hardware}%
  \BibitemOpen
  \bibfield  {author} {\bibinfo {author} {\bibfnamefont {H.}~\bibnamefont
  {Putterman}}, \bibinfo {author} {\bibfnamefont {K.}~\bibnamefont {Noh}},
  \bibinfo {author} {\bibfnamefont {C.~T.}\ \bibnamefont {Hann}}, \bibinfo
  {author} {\bibfnamefont {G.~S.}\ \bibnamefont {MacCabe}}, \bibinfo {author}
  {\bibfnamefont {S.}~\bibnamefont {Aghaeimeibodi}}, \bibinfo {author}
  {\bibfnamefont {R.~N.}\ \bibnamefont {Patel}}, \bibinfo {author}
  {\bibfnamefont {M.}~\bibnamefont {Lee}}, \bibinfo {author} {\bibfnamefont
  {W.~M.}\ \bibnamefont {Jones}}, \bibinfo {author} {\bibfnamefont
  {H.}~\bibnamefont {Moradinejad}}, \bibinfo {author} {\bibfnamefont
  {R.}~\bibnamefont {Rodriguez}}, \emph {et~al.},\ }\bibfield  {title}
  {\bibinfo {title} {Hardware-efficient quantum error correction via
  concatenated bosonic qubits},\ }\href@noop {} {\bibfield  {journal} {\bibinfo
   {journal} {Nature}\ }\textbf {\bibinfo {volume} {638}},\ \bibinfo {pages}
  {927} (\bibinfo {year} {2025})}\BibitemShut {NoStop}%
\bibitem [{\citenamefont {Stratonovich}(1957)}]{stratonovich1957distributions}%
  \BibitemOpen
  \bibfield  {author} {\bibinfo {author} {\bibfnamefont {R.~L.}\ \bibnamefont
  {Stratonovich}},\ }\bibfield  {title} {\bibinfo {title} {On distributions in
  representation space},\ }\href@noop {} {\bibfield  {journal} {\bibinfo
  {journal} {Sov.~Phys.~JETP}\ }\textbf {\bibinfo {volume} {4}},\ \bibinfo
  {pages} {891} (\bibinfo {year} {1957})}\BibitemShut {NoStop}%
\bibitem [{\citenamefont {Davis}\ \emph {et~al.}(2021)\citenamefont {Davis},
  \citenamefont {Kumari}, \citenamefont {Mann},\ and\ \citenamefont
  {Ghose}}]{davis2021wigner}%
  \BibitemOpen
  \bibfield  {author} {\bibinfo {author} {\bibfnamefont {J.}~\bibnamefont
  {Davis}}, \bibinfo {author} {\bibfnamefont {M.}~\bibnamefont {Kumari}},
  \bibinfo {author} {\bibfnamefont {R.~B.}\ \bibnamefont {Mann}},\ and\
  \bibinfo {author} {\bibfnamefont {S.}~\bibnamefont {Ghose}},\ }\bibfield
  {title} {\bibinfo {title} {Wigner negativity in spin-j systems},\ }\href@noop
  {} {\bibfield  {journal} {\bibinfo  {journal} {Phys.~Rev.~Res.}\ }\textbf
  {\bibinfo {volume} {3}},\ \bibinfo {pages} {033134} (\bibinfo {year}
  {2021})}\BibitemShut {NoStop}%
\bibitem [{\citenamefont {Franke}\ \emph {et~al.}(2017)\citenamefont {Franke},
  \citenamefont {Pfl{\"u}ger}, \citenamefont {Itoh},\ and\ \citenamefont
  {Brandt}}]{franke2017multiple}%
  \BibitemOpen
  \bibfield  {author} {\bibinfo {author} {\bibfnamefont {D.~P.}\ \bibnamefont
  {Franke}}, \bibinfo {author} {\bibfnamefont {M.~P.~D.}\ \bibnamefont
  {Pfl{\"u}ger}}, \bibinfo {author} {\bibfnamefont {K.~M.}\ \bibnamefont
  {Itoh}},\ and\ \bibinfo {author} {\bibfnamefont {M.~S.}\ \bibnamefont
  {Brandt}},\ }\bibfield  {title} {\bibinfo {title} {Multiple-quantum
  transitions and charge-induced decoherence of donor nuclear spins in
  silicon},\ }\href@noop {} {\bibfield  {journal} {\bibinfo  {journal}
  {Phys.~Rev.~Lett.}\ }\textbf {\bibinfo {volume} {118}},\ \bibinfo {pages}
  {246401} (\bibinfo {year} {2017})}\BibitemShut {NoStop}%
\bibitem [{\citenamefont {Armstrong}\ \emph {et~al.}(1961)\citenamefont
  {Armstrong}, \citenamefont {Bloembergen},\ and\ \citenamefont
  {Gill}}]{armstrong1961linear}%
  \BibitemOpen
  \bibfield  {author} {\bibinfo {author} {\bibfnamefont {J.}~\bibnamefont
  {Armstrong}}, \bibinfo {author} {\bibfnamefont {N.}~\bibnamefont
  {Bloembergen}},\ and\ \bibinfo {author} {\bibfnamefont {D.}~\bibnamefont
  {Gill}},\ }\bibfield  {title} {\bibinfo {title} {Linear effect of applied
  electric field on nuclear quadrupole resonance},\ }\href@noop {} {\bibfield
  {journal} {\bibinfo  {journal} {Phys.~Rev.~Lett.}\ }\textbf {\bibinfo
  {volume} {7}},\ \bibinfo {pages} {11} (\bibinfo {year} {1961})}\BibitemShut
  {NoStop}%
\bibitem [{\citenamefont {Golding}\ \emph {et~al.}(1979)\citenamefont
  {Golding}, \citenamefont {Schickfus}, \citenamefont {Hunklinger},\ and\
  \citenamefont {Dransfeld}}]{golding1979intrinsic}%
  \BibitemOpen
  \bibfield  {author} {\bibinfo {author} {\bibfnamefont {B.}~\bibnamefont
  {Golding}}, \bibinfo {author} {\bibfnamefont {M.~v.}\ \bibnamefont
  {Schickfus}}, \bibinfo {author} {\bibfnamefont {S.}~\bibnamefont
  {Hunklinger}},\ and\ \bibinfo {author} {\bibfnamefont {K.}~\bibnamefont
  {Dransfeld}},\ }\bibfield  {title} {\bibinfo {title} {Intrinsic electric
  dipole moment of tunneling systems in silica glasses},\ }\href@noop {}
  {\bibfield  {journal} {\bibinfo  {journal} {Phys.~Rev.~Lett.}\ }\textbf
  {\bibinfo {volume} {43}},\ \bibinfo {pages} {1817} (\bibinfo {year}
  {1979})}\BibitemShut {NoStop}%
\bibitem [{\citenamefont {Martinis}\ \emph {et~al.}(2005)\citenamefont
  {Martinis}, \citenamefont {Cooper}, \citenamefont {McDermott}, \citenamefont
  {Steffen}, \citenamefont {Ansmann}, \citenamefont {Osborn}, \citenamefont
  {Cicak}, \citenamefont {Oh}, \citenamefont {Pappas}, \citenamefont {Simmonds}
  \emph {et~al.}}]{martinis2005decoherence}%
  \BibitemOpen
  \bibfield  {author} {\bibinfo {author} {\bibfnamefont {J.~M.}\ \bibnamefont
  {Martinis}}, \bibinfo {author} {\bibfnamefont {K.~B.}\ \bibnamefont
  {Cooper}}, \bibinfo {author} {\bibfnamefont {R.}~\bibnamefont {McDermott}},
  \bibinfo {author} {\bibfnamefont {M.}~\bibnamefont {Steffen}}, \bibinfo
  {author} {\bibfnamefont {M.}~\bibnamefont {Ansmann}}, \bibinfo {author}
  {\bibfnamefont {K.~D.}\ \bibnamefont {Osborn}}, \bibinfo {author}
  {\bibfnamefont {K.}~\bibnamefont {Cicak}}, \bibinfo {author} {\bibfnamefont
  {S.}~\bibnamefont {Oh}}, \bibinfo {author} {\bibfnamefont {D.~P.}\
  \bibnamefont {Pappas}}, \bibinfo {author} {\bibfnamefont {R.~W.}\
  \bibnamefont {Simmonds}}, \emph {et~al.},\ }\bibfield  {title} {\bibinfo
  {title} {Decoherence in {J}osephson qubits from dielectric loss},\
  }\href@noop {} {\bibfield  {journal} {\bibinfo  {journal} {Phys.~Rev.~Lett.}\
  }\textbf {\bibinfo {volume} {95}},\ \bibinfo {pages} {210503} (\bibinfo
  {year} {2005})}\BibitemShut {NoStop}%
\bibitem [{\citenamefont {Adshead}\ \emph {et~al.}(2025)\citenamefont
  {Adshead}, \citenamefont {Coke}, \citenamefont {Tillotson}, \citenamefont
  {Li}, \citenamefont {Sullivan-Allsop}, \citenamefont {Egoavil}, \citenamefont
  {Thornley}, \citenamefont {Cui}, \citenamefont {Gourlay}, \citenamefont
  {Moore} \emph {et~al.}}]{adshead2025isotopically}%
  \BibitemOpen
  \bibfield  {author} {\bibinfo {author} {\bibfnamefont {M.}~\bibnamefont
  {Adshead}}, \bibinfo {author} {\bibfnamefont {M.}~\bibnamefont {Coke}},
  \bibinfo {author} {\bibfnamefont {E.}~\bibnamefont {Tillotson}}, \bibinfo
  {author} {\bibfnamefont {K.}~\bibnamefont {Li}}, \bibinfo {author}
  {\bibfnamefont {S.}~\bibnamefont {Sullivan-Allsop}}, \bibinfo {author}
  {\bibfnamefont {R.}~\bibnamefont {Egoavil}}, \bibinfo {author} {\bibfnamefont
  {W.}~\bibnamefont {Thornley}}, \bibinfo {author} {\bibfnamefont
  {Y.}~\bibnamefont {Cui}}, \bibinfo {author} {\bibfnamefont {C.~M.}\
  \bibnamefont {Gourlay}}, \bibinfo {author} {\bibfnamefont {K.~L.}\
  \bibnamefont {Moore}}, \emph {et~al.},\ }\bibfield  {title} {\bibinfo {title}
  {Isotopically selected single antimony molecule doping},\ }\href@noop {}
  {\bibfield  {journal} {\bibinfo  {journal} {arXiv preprint arXiv:2509.03243}\
  } (\bibinfo {year} {2025})}\BibitemShut {NoStop}%
\bibitem [{\citenamefont {Elsayed}\ \emph {et~al.}(2024)\citenamefont
  {Elsayed}, \citenamefont {Shehata}, \citenamefont {Godfrin}, \citenamefont
  {Kubicek}, \citenamefont {Massar}, \citenamefont {Canvel}, \citenamefont
  {Jussot}, \citenamefont {Simion}, \citenamefont {Mongillo}, \citenamefont
  {Wan} \emph {et~al.}}]{elsayed2024low}%
  \BibitemOpen
  \bibfield  {author} {\bibinfo {author} {\bibfnamefont {A.}~\bibnamefont
  {Elsayed}}, \bibinfo {author} {\bibfnamefont {M.}~\bibnamefont {Shehata}},
  \bibinfo {author} {\bibfnamefont {C.}~\bibnamefont {Godfrin}}, \bibinfo
  {author} {\bibfnamefont {S.}~\bibnamefont {Kubicek}}, \bibinfo {author}
  {\bibfnamefont {S.}~\bibnamefont {Massar}}, \bibinfo {author} {\bibfnamefont
  {Y.}~\bibnamefont {Canvel}}, \bibinfo {author} {\bibfnamefont
  {J.}~\bibnamefont {Jussot}}, \bibinfo {author} {\bibfnamefont
  {G.}~\bibnamefont {Simion}}, \bibinfo {author} {\bibfnamefont
  {M.}~\bibnamefont {Mongillo}}, \bibinfo {author} {\bibfnamefont
  {D.}~\bibnamefont {Wan}}, \emph {et~al.},\ }\bibfield  {title} {\bibinfo
  {title} {Low charge noise quantum dots with industrial {CMOS}
  manufacturing},\ }\href@noop {} {\bibfield  {journal} {\bibinfo  {journal}
  {npj Quantum Inf.}\ }\textbf {\bibinfo {volume} {10}},\ \bibinfo {pages} {70}
  (\bibinfo {year} {2024})}\BibitemShut {NoStop}%
\bibitem [{\citenamefont {Cywi{\'n}ski}\ \emph {et~al.}(2008)\citenamefont
  {Cywi{\'n}ski}, \citenamefont {Lutchyn}, \citenamefont {Nave},\ and\
  \citenamefont {Das~Sarma}}]{cywinski2008enhance}%
  \BibitemOpen
  \bibfield  {author} {\bibinfo {author} {\bibfnamefont {{\L}.}~\bibnamefont
  {Cywi{\'n}ski}}, \bibinfo {author} {\bibfnamefont {R.~M.}\ \bibnamefont
  {Lutchyn}}, \bibinfo {author} {\bibfnamefont {C.~P.}\ \bibnamefont {Nave}},\
  and\ \bibinfo {author} {\bibfnamefont {S.}~\bibnamefont {Das~Sarma}},\
  }\bibfield  {title} {\bibinfo {title} {How to enhance dephasing time in
  superconducting qubits},\ }\href@noop {} {\bibfield  {journal} {\bibinfo
  {journal} {Phys.~Rev.~B}\ }\textbf {\bibinfo {volume} {77}},\ \bibinfo
  {pages} {174509} (\bibinfo {year} {2008})}\BibitemShut {NoStop}%
\bibitem [{\citenamefont {Lemke}\ and\ \citenamefont
  {Haneman}(1978)}]{lemke1978dangling}%
  \BibitemOpen
  \bibfield  {author} {\bibinfo {author} {\bibfnamefont {B.~P.}\ \bibnamefont
  {Lemke}}\ and\ \bibinfo {author} {\bibfnamefont {D.}~\bibnamefont
  {Haneman}},\ }\bibfield  {title} {\bibinfo {title} {Dangling bonds on
  silicon},\ }\href@noop {} {\bibfield  {journal} {\bibinfo  {journal}
  {Phys.~Rev.~B}\ }\textbf {\bibinfo {volume} {17}},\ \bibinfo {pages} {1893}
  (\bibinfo {year} {1978})}\BibitemShut {NoStop}%
\bibitem [{\citenamefont {De~Sousa}(2007)}]{de2007dangling}%
  \BibitemOpen
  \bibfield  {author} {\bibinfo {author} {\bibfnamefont {R.}~\bibnamefont
  {De~Sousa}},\ }\bibfield  {title} {\bibinfo {title} {Dangling-bond spin
  relaxation and magnetic 1/f noise from the amorphous-semiconductor/oxide
  interface: {T}heory},\ }\href@noop {} {\bibfield  {journal} {\bibinfo
  {journal} {Phys.~Rev.~B}\ }\textbf {\bibinfo {volume} {76}},\ \bibinfo
  {pages} {245306} (\bibinfo {year} {2007})}\BibitemShut {NoStop}%
\bibitem [{\citenamefont {Galperin}\ \emph {et~al.}(2006)\citenamefont
  {Galperin}, \citenamefont {Altshuler}, \citenamefont {Bergli},\ and\
  \citenamefont {Shantsev}}]{galperin2006non}%
  \BibitemOpen
  \bibfield  {author} {\bibinfo {author} {\bibfnamefont {Y.~M.}\ \bibnamefont
  {Galperin}}, \bibinfo {author} {\bibfnamefont {B.}~\bibnamefont {Altshuler}},
  \bibinfo {author} {\bibfnamefont {J.}~\bibnamefont {Bergli}},\ and\ \bibinfo
  {author} {\bibfnamefont {D.~V.}\ \bibnamefont {Shantsev}},\ }\bibfield
  {title} {\bibinfo {title} {Non-{G}aussian low-frequency noise as a source of
  qubit decoherence},\ }\href@noop {} {\bibfield  {journal} {\bibinfo
  {journal} {Phys.~Rev.~Lett.}\ }\textbf {\bibinfo {volume} {96}},\ \bibinfo
  {pages} {097009} (\bibinfo {year} {2006})}\BibitemShut {NoStop}%
\bibitem [{\citenamefont {Schriefl}\ \emph {et~al.}(2006)\citenamefont
  {Schriefl}, \citenamefont {Makhlin}, \citenamefont {Shnirman},\ and\
  \citenamefont {Sch{\"o}n}}]{schriefl2006decoherence}%
  \BibitemOpen
  \bibfield  {author} {\bibinfo {author} {\bibfnamefont {J.}~\bibnamefont
  {Schriefl}}, \bibinfo {author} {\bibfnamefont {Y.}~\bibnamefont {Makhlin}},
  \bibinfo {author} {\bibfnamefont {A.}~\bibnamefont {Shnirman}},\ and\
  \bibinfo {author} {\bibfnamefont {G.}~\bibnamefont {Sch{\"o}n}},\ }\bibfield
  {title} {\bibinfo {title} {Decoherence from ensembles of two-level
  fluctuators},\ }\href@noop {} {\bibfield  {journal} {\bibinfo  {journal} {New
  J.~Phys.}\ }\textbf {\bibinfo {volume} {8}},\ \bibinfo {pages} {1} (\bibinfo
  {year} {2006})}\BibitemShut {NoStop}%
\bibitem [{\citenamefont {Paladino}\ \emph {et~al.}(2014)\citenamefont
  {Paladino}, \citenamefont {Galperin}, \citenamefont {Falci},\ and\
  \citenamefont {Altshuler}}]{paladino20141}%
  \BibitemOpen
  \bibfield  {author} {\bibinfo {author} {\bibfnamefont {E.}~\bibnamefont
  {Paladino}}, \bibinfo {author} {\bibfnamefont {Y.~M.}\ \bibnamefont
  {Galperin}}, \bibinfo {author} {\bibfnamefont {G.}~\bibnamefont {Falci}},\
  and\ \bibinfo {author} {\bibfnamefont {B.~L.}\ \bibnamefont {Altshuler}},\
  }\bibfield  {title} {\bibinfo {title} {1/f noise: {I}mplications for
  solid-state quantum information},\ }\href@noop {} {\bibfield  {journal}
  {\bibinfo  {journal} {Rev.~Mod.~Phys.}\ }\textbf {\bibinfo {volume} {86}},\
  \bibinfo {pages} {361} (\bibinfo {year} {2014})}\BibitemShut {NoStop}%
\bibitem [{\citenamefont {Makhlin}\ \emph {et~al.}(2004)\citenamefont
  {Makhlin}, \citenamefont {Sch{\"o}n},\ and\ \citenamefont
  {Shnirman}}]{makhlin2004dissipative}%
  \BibitemOpen
  \bibfield  {author} {\bibinfo {author} {\bibfnamefont {Y.}~\bibnamefont
  {Makhlin}}, \bibinfo {author} {\bibfnamefont {G.}~\bibnamefont {Sch{\"o}n}},\
  and\ \bibinfo {author} {\bibfnamefont {A.}~\bibnamefont {Shnirman}},\
  }\bibfield  {title} {\bibinfo {title} {Dissipative effects in {J}osephson
  qubits},\ }\href@noop {} {\bibfield  {journal} {\bibinfo  {journal} {Chemical
  Physics}\ }\textbf {\bibinfo {volume} {296}},\ \bibinfo {pages} {315}
  (\bibinfo {year} {2004})}\BibitemShut {NoStop}%
\bibitem [{\citenamefont {Fehse}\ \emph {et~al.}(2023)\citenamefont {Fehse},
  \citenamefont {David}, \citenamefont {Pioro-Ladriere},\ and\ \citenamefont
  {Coish}}]{fehse2023generalized}%
  \BibitemOpen
  \bibfield  {author} {\bibinfo {author} {\bibfnamefont {F.}~\bibnamefont
  {Fehse}}, \bibinfo {author} {\bibfnamefont {M.}~\bibnamefont {David}},
  \bibinfo {author} {\bibfnamefont {M.}~\bibnamefont {Pioro-Ladriere}},\ and\
  \bibinfo {author} {\bibfnamefont {W.~A.}\ \bibnamefont {Coish}},\ }\bibfield
  {title} {\bibinfo {title} {Generalized fast quasiadiabatic population
  transfer for improved qubit readout, shuttling, and noise mitigation},\
  }\href@noop {} {\bibfield  {journal} {\bibinfo  {journal} {Phys.~Rev.~B}\
  }\textbf {\bibinfo {volume} {107}},\ \bibinfo {pages} {245303} (\bibinfo
  {year} {2023})}\BibitemShut {NoStop}%
\bibitem [{\citenamefont {Yang}\ \emph {et~al.}(2019)\citenamefont {Yang},
  \citenamefont {Coppersmith},\ and\ \citenamefont
  {Friesen}}]{yang2019achieving}%
  \BibitemOpen
  \bibfield  {author} {\bibinfo {author} {\bibfnamefont {Y.-C.}\ \bibnamefont
  {Yang}}, \bibinfo {author} {\bibfnamefont {S.~N.}\ \bibnamefont
  {Coppersmith}},\ and\ \bibinfo {author} {\bibfnamefont {M.}~\bibnamefont
  {Friesen}},\ }\bibfield  {title} {\bibinfo {title} {Achieving high-fidelity
  single-qubit gates in a strongly driven charge qubit with 1/f charge noise},\
  }\href@noop {} {\bibfield  {journal} {\bibinfo  {journal} {npj Quantum Inf.}\
  }\textbf {\bibinfo {volume} {5}},\ \bibinfo {pages} {12} (\bibinfo {year}
  {2019})}\BibitemShut {NoStop}%
\bibitem [{\citenamefont {McKay}\ \emph {et~al.}(2017)\citenamefont {McKay},
  \citenamefont {Wood}, \citenamefont {Sheldon}, \citenamefont {Chow},\ and\
  \citenamefont {Gambetta}}]{mckay2017efficient}%
  \BibitemOpen
  \bibfield  {author} {\bibinfo {author} {\bibfnamefont {D.~C.}\ \bibnamefont
  {McKay}}, \bibinfo {author} {\bibfnamefont {C.~J.}\ \bibnamefont {Wood}},
  \bibinfo {author} {\bibfnamefont {S.}~\bibnamefont {Sheldon}}, \bibinfo
  {author} {\bibfnamefont {J.~M.}\ \bibnamefont {Chow}},\ and\ \bibinfo
  {author} {\bibfnamefont {J.~M.}\ \bibnamefont {Gambetta}},\ }\bibfield
  {title} {\bibinfo {title} {Efficient {Z} gates for quantum computing},\
  }\href@noop {} {\bibfield  {journal} {\bibinfo  {journal} {Phys.~Rev.~A}\
  }\textbf {\bibinfo {volume} {96}},\ \bibinfo {pages} {022330} (\bibinfo
  {year} {2017})}\BibitemShut {NoStop}%
\bibitem [{\citenamefont {Morello}\ \emph {et~al.}(2010)\citenamefont
  {Morello}, \citenamefont {Pla}, \citenamefont {Zwanenburg}, \citenamefont
  {Chan}, \citenamefont {Tan}, \citenamefont {Huebl}, \citenamefont
  {M{\"o}tt{\"o}nen}, \citenamefont {Nugroho}, \citenamefont {Yang},
  \citenamefont {Van~Donkelaar} \emph {et~al.}}]{morello2010single}%
  \BibitemOpen
  \bibfield  {author} {\bibinfo {author} {\bibfnamefont {A.}~\bibnamefont
  {Morello}}, \bibinfo {author} {\bibfnamefont {J.~J.}\ \bibnamefont {Pla}},
  \bibinfo {author} {\bibfnamefont {F.~A.}\ \bibnamefont {Zwanenburg}},
  \bibinfo {author} {\bibfnamefont {K.~W.}\ \bibnamefont {Chan}}, \bibinfo
  {author} {\bibfnamefont {K.~Y.}\ \bibnamefont {Tan}}, \bibinfo {author}
  {\bibfnamefont {H.}~\bibnamefont {Huebl}}, \bibinfo {author} {\bibfnamefont
  {M.}~\bibnamefont {M{\"o}tt{\"o}nen}}, \bibinfo {author} {\bibfnamefont
  {C.~D.}\ \bibnamefont {Nugroho}}, \bibinfo {author} {\bibfnamefont
  {C.}~\bibnamefont {Yang}}, \bibinfo {author} {\bibfnamefont {J.~A.}\
  \bibnamefont {Van~Donkelaar}}, \emph {et~al.},\ }\bibfield  {title} {\bibinfo
  {title} {Single-shot readout of an electron spin in silicon},\ }\href@noop {}
  {\bibfield  {journal} {\bibinfo  {journal} {Nature}\ }\textbf {\bibinfo
  {volume} {467}},\ \bibinfo {pages} {687} (\bibinfo {year}
  {2010})}\BibitemShut {NoStop}%
\bibitem [{\citenamefont {Geng}\ \emph {et~al.}(2025)\citenamefont {Geng},
  \citenamefont {Kiczynski}, \citenamefont {Timofeev}, \citenamefont {Osika},
  \citenamefont {Keith}, \citenamefont {Rowlands}, \citenamefont {Kranz},
  \citenamefont {Rahman}, \citenamefont {Chung}, \citenamefont {Keizer} \emph
  {et~al.}}]{geng2025high}%
  \BibitemOpen
  \bibfield  {author} {\bibinfo {author} {\bibfnamefont {H.}~\bibnamefont
  {Geng}}, \bibinfo {author} {\bibfnamefont {M.}~\bibnamefont {Kiczynski}},
  \bibinfo {author} {\bibfnamefont {A.~V.}\ \bibnamefont {Timofeev}}, \bibinfo
  {author} {\bibfnamefont {E.~N.}\ \bibnamefont {Osika}}, \bibinfo {author}
  {\bibfnamefont {D.}~\bibnamefont {Keith}}, \bibinfo {author} {\bibfnamefont
  {J.}~\bibnamefont {Rowlands}}, \bibinfo {author} {\bibfnamefont
  {L.}~\bibnamefont {Kranz}}, \bibinfo {author} {\bibfnamefont
  {R.}~\bibnamefont {Rahman}}, \bibinfo {author} {\bibfnamefont
  {Y.}~\bibnamefont {Chung}}, \bibinfo {author} {\bibfnamefont {J.~G.}\
  \bibnamefont {Keizer}}, \emph {et~al.},\ }\bibfield  {title} {\bibinfo
  {title} {High-fidelity sub-microsecond single-shot electron spin readout
  above 3.5 {K}},\ }\href@noop {} {\bibfield  {journal} {\bibinfo  {journal}
  {Nat.~Commun.}\ }\textbf {\bibinfo {volume} {16}},\ \bibinfo {pages} {3382}
  (\bibinfo {year} {2025})}\BibitemShut {NoStop}%
\bibitem [{\citenamefont {Mielke}\ \emph {et~al.}(2021)\citenamefont {Mielke},
  \citenamefont {Petta},\ and\ \citenamefont {Burkard}}]{mielke2021nuclear}%
  \BibitemOpen
  \bibfield  {author} {\bibinfo {author} {\bibfnamefont {J.}~\bibnamefont
  {Mielke}}, \bibinfo {author} {\bibfnamefont {J.~R.}\ \bibnamefont {Petta}},\
  and\ \bibinfo {author} {\bibfnamefont {G.}~\bibnamefont {Burkard}},\
  }\bibfield  {title} {\bibinfo {title} {Nuclear spin readout in a
  cavity-coupled hybrid quantum dot-donor system},\ }\href@noop {} {\bibfield
  {journal} {\bibinfo  {journal} {PRX Quantum}\ }\textbf {\bibinfo {volume}
  {2}},\ \bibinfo {pages} {020347} (\bibinfo {year} {2021})}\BibitemShut
  {NoStop}%
\bibitem [{\citenamefont {Unseld}\ \emph {et~al.}(2025)\citenamefont {Unseld},
  \citenamefont {Undseth}, \citenamefont {Raymenants}, \citenamefont
  {Matsumoto}, \citenamefont {de~Snoo}, \citenamefont {Karwal}, \citenamefont
  {Pietx-Casas}, \citenamefont {Ivlev}, \citenamefont {Meyer}, \citenamefont
  {Sammak} \emph {et~al.}}]{unseld2025baseband}%
  \BibitemOpen
  \bibfield  {author} {\bibinfo {author} {\bibfnamefont {F.~K.}\ \bibnamefont
  {Unseld}}, \bibinfo {author} {\bibfnamefont {B.}~\bibnamefont {Undseth}},
  \bibinfo {author} {\bibfnamefont {E.}~\bibnamefont {Raymenants}}, \bibinfo
  {author} {\bibfnamefont {Y.}~\bibnamefont {Matsumoto}}, \bibinfo {author}
  {\bibfnamefont {S.~L.}\ \bibnamefont {de~Snoo}}, \bibinfo {author}
  {\bibfnamefont {S.}~\bibnamefont {Karwal}}, \bibinfo {author} {\bibfnamefont
  {O.}~\bibnamefont {Pietx-Casas}}, \bibinfo {author} {\bibfnamefont {A.~S.}\
  \bibnamefont {Ivlev}}, \bibinfo {author} {\bibfnamefont {M.}~\bibnamefont
  {Meyer}}, \bibinfo {author} {\bibfnamefont {A.}~\bibnamefont {Sammak}}, \emph
  {et~al.},\ }\bibfield  {title} {\bibinfo {title} {Baseband control of
  single-electron silicon spin qubits in two dimensions},\ }\href@noop {}
  {\bibfield  {journal} {\bibinfo  {journal} {Nat.~Commun.}\ }\textbf {\bibinfo
  {volume} {16}},\ \bibinfo {pages} {5605} (\bibinfo {year}
  {2025})}\BibitemShut {NoStop}%
\bibitem [{\citenamefont {Sun}\ \emph {et~al.}(2014)\citenamefont {Sun},
  \citenamefont {Petrenko}, \citenamefont {Leghtas}, \citenamefont {Vlastakis},
  \citenamefont {Kirchmair}, \citenamefont {Sliwa}, \citenamefont {Narla},
  \citenamefont {Hatridge}, \citenamefont {Shankar}, \citenamefont {Blumoff}
  \emph {et~al.}}]{sun2014tracking}%
  \BibitemOpen
  \bibfield  {author} {\bibinfo {author} {\bibfnamefont {L.}~\bibnamefont
  {Sun}}, \bibinfo {author} {\bibfnamefont {A.}~\bibnamefont {Petrenko}},
  \bibinfo {author} {\bibfnamefont {Z.}~\bibnamefont {Leghtas}}, \bibinfo
  {author} {\bibfnamefont {B.}~\bibnamefont {Vlastakis}}, \bibinfo {author}
  {\bibfnamefont {G.}~\bibnamefont {Kirchmair}}, \bibinfo {author}
  {\bibfnamefont {K.~M.}\ \bibnamefont {Sliwa}}, \bibinfo {author}
  {\bibfnamefont {A.}~\bibnamefont {Narla}}, \bibinfo {author} {\bibfnamefont
  {M.}~\bibnamefont {Hatridge}}, \bibinfo {author} {\bibfnamefont
  {S.}~\bibnamefont {Shankar}}, \bibinfo {author} {\bibfnamefont
  {J.}~\bibnamefont {Blumoff}}, \emph {et~al.},\ }\bibfield  {title} {\bibinfo
  {title} {Tracking photon jumps with repeated quantum non-demolition parity
  measurements},\ }\href@noop {} {\bibfield  {journal} {\bibinfo  {journal}
  {Nature}\ }\textbf {\bibinfo {volume} {511}},\ \bibinfo {pages} {444}
  (\bibinfo {year} {2014})}\BibitemShut {NoStop}%
\bibitem [{\citenamefont {Rosenblum}\ \emph {et~al.}(2018)\citenamefont
  {Rosenblum}, \citenamefont {Reinhold}, \citenamefont {Mirrahimi},
  \citenamefont {Jiang}, \citenamefont {Frunzio},\ and\ \citenamefont
  {Schoelkopf}}]{rosenblum2018fault}%
  \BibitemOpen
  \bibfield  {author} {\bibinfo {author} {\bibfnamefont {S.}~\bibnamefont
  {Rosenblum}}, \bibinfo {author} {\bibfnamefont {P.}~\bibnamefont {Reinhold}},
  \bibinfo {author} {\bibfnamefont {M.}~\bibnamefont {Mirrahimi}}, \bibinfo
  {author} {\bibfnamefont {L.}~\bibnamefont {Jiang}}, \bibinfo {author}
  {\bibfnamefont {L.}~\bibnamefont {Frunzio}},\ and\ \bibinfo {author}
  {\bibfnamefont {R.~J.}\ \bibnamefont {Schoelkopf}},\ }\bibfield  {title}
  {\bibinfo {title} {Fault-tolerant detection of a quantum error},\ }\href@noop
  {} {\bibfield  {journal} {\bibinfo  {journal} {Science}\ }\textbf {\bibinfo
  {volume} {361}},\ \bibinfo {pages} {266} (\bibinfo {year}
  {2018})}\BibitemShut {NoStop}%
\bibitem [{\citenamefont {De~Smet}\ \emph {et~al.}(2025)\citenamefont
  {De~Smet}, \citenamefont {Matsumoto}, \citenamefont {Zwerver}, \citenamefont
  {Tryputen}, \citenamefont {de~Snoo}, \citenamefont {Amitonov}, \citenamefont
  {Katiraee-Far}, \citenamefont {Sammak}, \citenamefont {Samkharadze},
  \citenamefont {G{\"u}l} \emph {et~al.}}]{de2025high}%
  \BibitemOpen
  \bibfield  {author} {\bibinfo {author} {\bibfnamefont {M.}~\bibnamefont
  {De~Smet}}, \bibinfo {author} {\bibfnamefont {Y.}~\bibnamefont {Matsumoto}},
  \bibinfo {author} {\bibfnamefont {A.-M.~J.}\ \bibnamefont {Zwerver}},
  \bibinfo {author} {\bibfnamefont {L.}~\bibnamefont {Tryputen}}, \bibinfo
  {author} {\bibfnamefont {S.~L.}\ \bibnamefont {de~Snoo}}, \bibinfo {author}
  {\bibfnamefont {S.~V.}\ \bibnamefont {Amitonov}}, \bibinfo {author}
  {\bibfnamefont {S.~R.}\ \bibnamefont {Katiraee-Far}}, \bibinfo {author}
  {\bibfnamefont {A.}~\bibnamefont {Sammak}}, \bibinfo {author} {\bibfnamefont
  {N.}~\bibnamefont {Samkharadze}}, \bibinfo {author} {\bibfnamefont
  {{\"O}.}~\bibnamefont {G{\"u}l}}, \emph {et~al.},\ }\bibfield  {title}
  {\bibinfo {title} {High-fidelity single-spin shuttling in silicon},\
  }\href@noop {} {\bibfield  {journal} {\bibinfo  {journal} {Nature
  Nanotechnology}\ ,\ \bibinfo {pages} {1}} (\bibinfo {year}
  {2025})}\BibitemShut {NoStop}%
\bibitem [{\citenamefont {Connors}\ \emph {et~al.}(2020)\citenamefont
  {Connors}, \citenamefont {Nelson},\ and\ \citenamefont
  {Nichol}}]{connors2020rapid}%
  \BibitemOpen
  \bibfield  {author} {\bibinfo {author} {\bibfnamefont {E.~J.}\ \bibnamefont
  {Connors}}, \bibinfo {author} {\bibfnamefont {J.~J.}\ \bibnamefont
  {Nelson}},\ and\ \bibinfo {author} {\bibfnamefont {J.~M.}\ \bibnamefont
  {Nichol}},\ }\bibfield  {title} {\bibinfo {title} {Rapid high-fidelity
  spin-state readout in {S}i/{S}i-{G}e quantum dots via {RF} reflectometry},\
  }\href@noop {} {\bibfield  {journal} {\bibinfo  {journal} {Phys.~Rev.~Appl.}\
  }\textbf {\bibinfo {volume} {13}},\ \bibinfo {pages} {024019} (\bibinfo
  {year} {2020})}\BibitemShut {NoStop}%
\bibitem [{\citenamefont {Elzerman}\ \emph {et~al.}(2004)\citenamefont
  {Elzerman}, \citenamefont {Hanson}, \citenamefont {Willems~van Beveren},
  \citenamefont {Witkamp}, \citenamefont {Vandersypen},\ and\ \citenamefont
  {Kouwenhoven}}]{elzerman2004single}%
  \BibitemOpen
  \bibfield  {author} {\bibinfo {author} {\bibfnamefont {J.~M.}\ \bibnamefont
  {Elzerman}}, \bibinfo {author} {\bibfnamefont {R.}~\bibnamefont {Hanson}},
  \bibinfo {author} {\bibfnamefont {L.~H.}\ \bibnamefont {Willems~van
  Beveren}}, \bibinfo {author} {\bibfnamefont {B.}~\bibnamefont {Witkamp}},
  \bibinfo {author} {\bibfnamefont {L.~M.~K.}\ \bibnamefont {Vandersypen}},\
  and\ \bibinfo {author} {\bibfnamefont {L.~P.}\ \bibnamefont {Kouwenhoven}},\
  }\bibfield  {title} {\bibinfo {title} {Single-shot read-out of an individual
  electron spin in a quantum dot},\ }\href@noop {} {\bibfield  {journal}
  {\bibinfo  {journal} {Nature}\ }\textbf {\bibinfo {volume} {430}},\ \bibinfo
  {pages} {431} (\bibinfo {year} {2004})}\BibitemShut {NoStop}%
\bibitem [{\citenamefont {Vaartjes}\ \emph {et~al.}(2026)\citenamefont
  {Vaartjes}, \citenamefont {Su}, \citenamefont {O’Neill}, \citenamefont
  {Steinacker}, \citenamefont {Goenka}, \citenamefont {van Blankenstein},
  \citenamefont {Yu}, \citenamefont {Wilhelm}, \citenamefont {Jakob},
  \citenamefont {Hudson} \emph {et~al.}}]{vaartjes2026maximizing}%
  \BibitemOpen
  \bibfield  {author} {\bibinfo {author} {\bibfnamefont {A.}~\bibnamefont
  {Vaartjes}}, \bibinfo {author} {\bibfnamefont {R.~Y.}\ \bibnamefont {Su}},
  \bibinfo {author} {\bibfnamefont {L.~A.}\ \bibnamefont {O’Neill}}, \bibinfo
  {author} {\bibfnamefont {P.}~\bibnamefont {Steinacker}}, \bibinfo {author}
  {\bibfnamefont {G.}~\bibnamefont {Goenka}}, \bibinfo {author} {\bibfnamefont
  {M.~R.}\ \bibnamefont {van Blankenstein}}, \bibinfo {author} {\bibfnamefont
  {X.}~\bibnamefont {Yu}}, \bibinfo {author} {\bibfnamefont {B.}~\bibnamefont
  {Wilhelm}}, \bibinfo {author} {\bibfnamefont {A.~M.}\ \bibnamefont {Jakob}},
  \bibinfo {author} {\bibfnamefont {F.~E.}\ \bibnamefont {Hudson}}, \emph
  {et~al.},\ }\bibfield  {title} {\bibinfo {title} {Maximizing the
  nondemolition nature of a quantum measurement via an adaptive readout
  protocol},\ }\href@noop {} {\bibfield  {journal} {\bibinfo  {journal} {PRX
  Quantum}\ }\textbf {\bibinfo {volume} {7}},\ \bibinfo {pages} {020330}
  (\bibinfo {year} {2026})}\BibitemShut {NoStop}%
\bibitem [{Note1()}]{Note1}%
  \BibitemOpen
  \bibinfo {note} {Since the magnetic field in Refs.~\cite
  {asaad2020coherent,fernandez2024navigating, yu2025schrodinger} is not aligned
  with the quadrupole tensor’s principal axis, the quadrupole splittings
  reported in these works are not exactly equal to $Q$. We nonetheless take the
  values reported here as reasonable order-of-magnitude estimates for
  $Q$.}\BibitemShut {Stop}%
\bibitem [{\citenamefont {Dreher}\ \emph {et~al.}(2011)\citenamefont {Dreher},
  \citenamefont {Hilker}, \citenamefont {Brandlmaier}, \citenamefont
  {Goennenwein}, \citenamefont {Huebl}, \citenamefont {Stutzmann},\ and\
  \citenamefont {Brandt}}]{dreher2011electroelastic}%
  \BibitemOpen
  \bibfield  {author} {\bibinfo {author} {\bibfnamefont {L.}~\bibnamefont
  {Dreher}}, \bibinfo {author} {\bibfnamefont {T.~A.}\ \bibnamefont {Hilker}},
  \bibinfo {author} {\bibfnamefont {A.}~\bibnamefont {Brandlmaier}}, \bibinfo
  {author} {\bibfnamefont {S.~T.~B.}\ \bibnamefont {Goennenwein}}, \bibinfo
  {author} {\bibfnamefont {H.}~\bibnamefont {Huebl}}, \bibinfo {author}
  {\bibfnamefont {M.}~\bibnamefont {Stutzmann}},\ and\ \bibinfo {author}
  {\bibfnamefont {M.~S.}\ \bibnamefont {Brandt}},\ }\bibfield  {title}
  {\bibinfo {title} {Electroelastic hyperfine tuning of phosphorus donors in
  silicon},\ }\href@noop {} {\bibfield  {journal} {\bibinfo  {journal}
  {Phys.~Rev.~Lett.}\ }\textbf {\bibinfo {volume} {106}},\ \bibinfo {pages}
  {037601} (\bibinfo {year} {2011})}\BibitemShut {NoStop}%
\bibitem [{\citenamefont {Bosco}\ and\ \citenamefont
  {Loss}(2021)}]{bosco2021fully}%
  \BibitemOpen
  \bibfield  {author} {\bibinfo {author} {\bibfnamefont {S.}~\bibnamefont
  {Bosco}}\ and\ \bibinfo {author} {\bibfnamefont {D.}~\bibnamefont {Loss}},\
  }\bibfield  {title} {\bibinfo {title} {Fully tunable hyperfine interactions
  of hole spin qubits in {S}i and {G}e quantum dots},\ }\href@noop {}
  {\bibfield  {journal} {\bibinfo  {journal} {Phys.~Rev.~Lett.}\ }\textbf
  {\bibinfo {volume} {127}},\ \bibinfo {pages} {190501} (\bibinfo {year}
  {2021})}\BibitemShut {NoStop}%
\bibitem [{\citenamefont {Bassi}\ \emph {et~al.}(2025)\citenamefont {Bassi},
  \citenamefont {Rodr{\i}guez-Mena}, \citenamefont {Brun}, \citenamefont
  {Zihlmann}, \citenamefont {Nguyen}, \citenamefont {Champain}, \citenamefont
  {Abadillo-Uriel}, \citenamefont {Bertrand}, \citenamefont {Niebojewski},
  \citenamefont {Maurand} \emph {et~al.}}]{bassi2025optimal}%
  \BibitemOpen
  \bibfield  {author} {\bibinfo {author} {\bibfnamefont {M.}~\bibnamefont
  {Bassi}}, \bibinfo {author} {\bibfnamefont {E.-A.}\ \bibnamefont
  {Rodr{\i}guez-Mena}}, \bibinfo {author} {\bibfnamefont {B.}~\bibnamefont
  {Brun}}, \bibinfo {author} {\bibfnamefont {S.}~\bibnamefont {Zihlmann}},
  \bibinfo {author} {\bibfnamefont {T.}~\bibnamefont {Nguyen}}, \bibinfo
  {author} {\bibfnamefont {V.}~\bibnamefont {Champain}}, \bibinfo {author}
  {\bibfnamefont {J.~C.}\ \bibnamefont {Abadillo-Uriel}}, \bibinfo {author}
  {\bibfnamefont {B.}~\bibnamefont {Bertrand}}, \bibinfo {author}
  {\bibfnamefont {H.}~\bibnamefont {Niebojewski}}, \bibinfo {author}
  {\bibfnamefont {R.}~\bibnamefont {Maurand}}, \emph {et~al.},\ }\bibfield
  {title} {\bibinfo {title} {Optimal operation of hole spin qubits},\
  }\href@noop {} {\bibfield  {journal} {\bibinfo  {journal} {Nat.~Phys.}\
  }\textbf {\bibinfo {volume} {22}},\ \bibinfo {pages} {75} (\bibinfo {year}
  {2025})}\BibitemShut {NoStop}%
\bibitem [{\citenamefont {Carballido}\ \emph {et~al.}(2025)\citenamefont
  {Carballido}, \citenamefont {Svab}, \citenamefont {Eggli}, \citenamefont
  {Patlatiuk}, \citenamefont {Chevalier~Kwon}, \citenamefont {Schuff},
  \citenamefont {Kaiser}, \citenamefont {Camenzind}, \citenamefont {Li},
  \citenamefont {Ares} \emph {et~al.}}]{carballido2025compromise}%
  \BibitemOpen
  \bibfield  {author} {\bibinfo {author} {\bibfnamefont {M.~J.}\ \bibnamefont
  {Carballido}}, \bibinfo {author} {\bibfnamefont {S.}~\bibnamefont {Svab}},
  \bibinfo {author} {\bibfnamefont {R.~S.}\ \bibnamefont {Eggli}}, \bibinfo
  {author} {\bibfnamefont {T.}~\bibnamefont {Patlatiuk}}, \bibinfo {author}
  {\bibfnamefont {P.}~\bibnamefont {Chevalier~Kwon}}, \bibinfo {author}
  {\bibfnamefont {J.}~\bibnamefont {Schuff}}, \bibinfo {author} {\bibfnamefont
  {R.~M.}\ \bibnamefont {Kaiser}}, \bibinfo {author} {\bibfnamefont {L.~C.}\
  \bibnamefont {Camenzind}}, \bibinfo {author} {\bibfnamefont {A.}~\bibnamefont
  {Li}}, \bibinfo {author} {\bibfnamefont {N.}~\bibnamefont {Ares}}, \emph
  {et~al.},\ }\bibfield  {title} {\bibinfo {title} {Compromise-free scaling of
  qubit speed and coherence},\ }\href@noop {} {\bibfield  {journal} {\bibinfo
  {journal} {Nat.~Commun.}\ }\textbf {\bibinfo {volume} {16}},\ \bibinfo
  {pages} {7616} (\bibinfo {year} {2025})}\BibitemShut {NoStop}%
\end{thebibliography}
\end{document}